\documentclass[11pt]{article}
\usepackage[margin=.8in,left=.8in]{geometry}
\usepackage{amsmath}
\usepackage{amsfonts}
\usepackage{amssymb}
\usepackage{amsthm}
\usepackage{mathtools}
\usepackage{subcaption}
\usepackage{float}
\usepackage{graphicx}
\usepackage{color}
\usepackage{xcolor}
\usepackage{xfrac}
\usepackage{stmaryrd}
\usepackage[utf8]{inputenc}
\usepackage{hyperref}
\usepackage[all]{xy}

\usepackage{tikz}
\usepackage{tikz-cd}
\usepackage{lipsum}
\usepackage{adjustbox}

\usepackage{multirow}

\usepackage{stmaryrd}    

\usepackage{enumerate} 

\usepackage{helvet}   

\usepackage{color, colortbl}
\definecolor{Gray}{gray}{0.9}
\definecolor{lightgray}{rgb}{0.9,0.9,0.9}

\definecolor{darkblue}{rgb}{0.05,0.25,0.65}
\definecolor{greenii}{RGB}{20,140,10}

\definecolor{orangeii}{RGB}{200,100,5}

\usepackage{mathptmx}
\usepackage{amsmath}

\usepackage{floatflt}  
\usepackage{wrapfig} 
\usepackage{array}     
\newcolumntype{L}[1]{>{\raggedright\let\newline\\\arraybackslash\hspace{0pt}}m{#1}}
\newcolumntype{C}[1]{>{\centering\let\newline\\\arraybackslash\hspace{0pt}}m{#1}}
\newcolumntype{R}[1]{>{\raggedleft\let\newline\\\arraybackslash\hspace{0pt}}m{#1}}

\usepackage{diagbox}  

\makeatletter
\newcommand{\raisemath}[1]{\mathpalette{\raisem@th{#1}}}
\newcommand{\raisem@th}[3]{\raisebox{#1}{$#2#3$}}
\makeatother

\usepackage{tabularx}   

\usepackage[new]{old-arrows}   

\newdir{> }{{}*!/10pt/@{>}}

\newcommand{\dslash}{/\!\!/}

\makeatletter
\newif\if@sup
\newtoks\@sups
\def\append@sup#1{\edef\act{\noexpand\@sups={\the\@sups #1}}\act}%
\def\reset@sup{\@supfalse\@sups={}}%
\def\mk@scripts#1#2{\if #2/ \if@sup ^{\the\@sups}\fi \else%
  \ifx #1_ \if@sup ^{\the\@sups}\reset@sup \fi {}_{#2}%
  \else \append@sup#2 \@suptrue \fi%
  \expandafter\mk@scripts\fi}
\def\tensor#1#2{\reset@sup#1\mk@scripts#2_/}
\def\multiscripts#1#2#3{\reset@sup{}\mk@scripts#1_/#2%
  \reset@sup\mk@scripts#3_/}
\makeatother

\makeatletter
\newbox\slashbox \setbox\slashbox=\hbox{$/$}
\def\itex@pslash#1{\setbox\@tempboxa=\hbox{$#1$}
  \@tempdima=0.5\wd\slashbox \advance\@tempdima 0.5\wd\@tempboxa
  \copy\slashbox \kern-\@tempdima \box\@tempboxa}
\def\slash{\protect\itex@pslash}
\makeatother

\def\clap#1{\hbox to 0pt{\hss#1\hss}}

\def\mathrlap{\mathpalette\mathrlapinternal}

\def\mathrlapinternal#1#2{\rlap{$\mathsurround=0pt#1{#2}$}}

\let\oldroot\root
\def\root#1#2{\oldroot #1 \of{#2}}
\renewcommand{\sqrt}[2][]{\oldroot #1 \of{#2}}

\DeclareSymbolFont{symbolsC}{U}{txsyc}{m}{n}
\SetSymbolFont{symbolsC}{bold}{U}{txsyc}{bx}{n}
\DeclareFontSubstitution{U}{txsyc}{m}{n}

\DeclareSymbolFont{stmry}{U}{stmry}{m}{n}
\SetSymbolFont{stmry}{bold}{U}{stmry}{b}{n}

\DeclareFontFamily{OMX}{MnSymbolE}{}
\DeclareSymbolFont{mnomx}{OMX}{MnSymbolE}{m}{n}
\SetSymbolFont{mnomx}{bold}{OMX}{MnSymbolE}{b}{n}
\DeclareFontShape{OMX}{MnSymbolE}{m}{n}{
    <-6>  MnSymbolE5
   <6-7>  MnSymbolE6
   <7-8>  MnSymbolE7
   <8-9>  MnSymbolE8
   <9-10> MnSymbolE9
  <10-12> MnSymbolE10
  <12->   MnSymbolE12}{}

\makeatletter
\def\Decl@Mn@Delim#1#2#3#4{%
  \if\relax\noexpand#1%
    \let#1\undefined
  \fi
  \DeclareMathDelimiter{#1}{#2}{#3}{#4}{#3}{#4}}
\def\Decl@Mn@Open#1#2#3{\Decl@Mn@Delim{#1}{\mathopen}{#2}{#3}}
\def\Decl@Mn@Close#1#2#3{\Decl@Mn@Delim{#1}{\mathclose}{#2}{#3}}
\Decl@Mn@Open{\llangle}{mnomx}{'164}
\Decl@Mn@Close{\rrangle}{mnomx}{'171}
\Decl@Mn@Open{\lmoustache}{mnomx}{'245}
\Decl@Mn@Close{\rmoustache}{mnomx}{'244}
\makeatother

\makeatletter
\DeclareRobustCommand\widecheck[1]{{\mathpalette\@widecheck{#1}}}
\def\@widecheck#1#2{%
    \setbox\z@\hbox{\m@th$#1#2$}%
    \setbox\tw@\hbox{\m@th$#1%
       \widehat{%
          \vrule\@width\z@\@height\ht\z@
          \vrule\@height\z@\@width\wd\z@}$}%
    \dp\tw@-\ht\z@
    \@tempdima\ht\z@ \advance\@tempdima2\ht\tw@ \divide\@tempdima\thr@@
    \setbox\tw@\hbox{%
       \raise\@tempdima\hbox{\scalebox{1}[-1]{\lower\@tempdima\box
\tw@}}}%
    {\ooalign{\box\tw@ \cr \box\z@}}}
\makeatother

\makeatletter
\def\udots{\mathinner{\mkern2mu\raise\p@\hbox{.}
\mkern2mu\raise4\p@\hbox{.}\mkern1mu
\raise7\p@\vbox{\kern7\p@\hbox{.}}\mkern1mu}}
\makeatother

\newcommand{\mathfr}{\mathfrak}

\newcommand{\g}{\mathfrak{g}}
\newcommand{\h}{\mathfrak{h}}

\newcommand{\R}{\ensuremath{\mathbb R}}
\newcommand{\Z}{\ensuremath{\mathbb Z}}
\newcommand{\Q}{\ensuremath{\mathbb Q}}

\newcommand{\cE}{\ensuremath{\mathcal E}}
\newcommand{\cH}{\ensuremath{\mathcal H}}
\newcommand{\cK}{\ensuremath{\mathcal K}}
\newcommand{\cD}{\ensuremath{\mathcal D}}
\newcommand{\cC}{\ensuremath{\mathcal C}}

\renewcommand{\(}{\begin{equation}}
\renewcommand{\)}{\end{equation}}
\newcommand{\bea}{\begin{eqnarray*}}
\newcommand{\eea}{\end{eqnarray*}}

\newcommand{\mc}[1]{\mathcal{#1}}
\newcommand{\abs}[1]{\lvert #1 \rvert}
\newcommand{\on}[1]{\operatorname{#1}}

\def\BB{\mathbb{B}}
\def\CC{\mathbb{C}}
\def\GG{\mathbb{G}}
\def\HH{\mathbb{H}}

\def\PP{\mathbb{P}}
\def\QQ{\mathbb{Q}}
\def\RR{\mathbb{R}}
\def\ZZ{\mathbb{Z}}

\DeclareMathOperator{\Aut}{Aut}

\DeclareMathOperator{\Der}{Der}

\DeclareMathOperator{\GL}{GL}
\DeclareMathOperator{\Hom}{Hom}
\DeclareMathOperator{\id}{id}
\DeclareMathOperator{\im}{im}
\DeclareMathOperator{\Lie}{Lie}
\DeclareMathOperator{\Map}{Map}

\DeclareMathOperator{\Pic}{Pic}
\DeclareMathOperator{\rank}{rank}
\DeclareMathOperator{\Spec}{Spec}

\usepackage{cleveref}

\crefformat{section}{\S#2#1#3} 
\crefformat{subsection}{\S#2#1#3}
\crefformat{subsubsection}{\S#2#1#3}

\newtheorem{theorem}{Theorem}[section]

\theoremstyle{definition}

\newtheorem{example}[theorem]{Example}

\newtheorem{note[theorem]}{Note}

\usepackage{amsfonts}

\definecolor{mygreen}{RGB}{80,150,10}
\newcommand{\old}[1]{}

\begin{document}

\title{Mysterious Triality and M-Theory}

\author{Hisham Sati, \; Alexander A. Voronov}

\maketitle
  

\vspace{-3mm}
\bigskip 

\begin{abstract}
In a previous paper \cite{SV1}, we introduced Mysterious Triality as 
an extension, via algebraic topology in the form of rational homotopy theory,  of Mysterious 
Duality by Iqbal, Neitzke, and Vafa
\cite{INV}, which provides connections between physics, in the form of dimensional reduction of M-theory, and algebraic geometry, in the form of intersection theory on del Pezzo surfaces. 
The starting point for that connection to rational homotopy theory is the description of M-theory dynamics using the 4-sphere, via 
Hypothesis H.
This progresses to dimensional reduction of M-theory on tori $T^k$ with its dynamics described via cyclic loop spaces of the 4-sphere $\mathcal{L}_c^k S^4$, producing a series of data analogous to that given by the del Pezzo surfaces
$\mathbb{B}_k$, for $k=0, \dots, 8$.

\smallskip
With the mathematical constructions established in \cite{SV1}, 
in this companion physics paper we present novel connections to M-theory that enhance
the triality, including those strengthening the duality. This uncovers interesting ties between algebraic geometry, algebraic topology, and M-theory and provides 
tantalizing links.
We further expand on the extension of the duality and triality to the Kac-Moody setting. 
 \end{abstract}

\medskip

\tableofcontents


\section{Introduction}
\label{intro}

Mysterious Duality has been discovered by Iqbal, Neitzke, and Vafa \cite{INV} as a remarkable, yet mysterious 
correspondence between certain symmetry patterns in compactifications of M-theory
on tori $T^k$ and del Pezzo surfaces  ${\mathbb B}_k$, both governed by the root system corresponding to 
the exceptional series $E_k$, $k\leq 8$.

$$
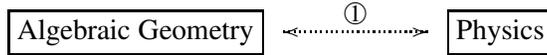
\xymatrix@R=.1em{
 \fbox{\text{Algebraic Geometry}} \;\; \ar@{<..>}[rr]^-{\footnotesize
 \raisebox{.5pt}{\textcircled{\raisebox{-.9pt} {1}}}} 
 &  & \;\; 
 \fbox{\text{Physics}}
}
$$
The moduli space of M-theory compactified on (reduced on, or simply on) $T^k$ is 
 usually taken to be the double quotient $G(\ZZ) \backslash G / K$, where $G$ is the U-duality group, which is the (split real form of the) Lie group $E_k$, $G(\ZZ)$ is its integral form, and $K$ is the maximal compact subgroup of $G$
 \cite{HT}\cite{OP}. 
 In \cite{INV}, for ``flat tori with no C-field",  this was taken to be of the form
$
\mathcal{M}_k := A/W,
$
where $A$ is the maximal split torus of $G$ and $W$ is the Weyl group. This, as we explained in \cite{SV1}, can be thought of as abelianization:
\vspace{-2mm}
\(
\label{abelianization} 
K \backslash G / G(\ZZ) = (AN) /G(\ZZ) \xymatrix{\ar@{~>}[rr]^{\rm \bf \color{darkblue} abelianize}&&}  
\mathcal{M}_k = A/W\;.
\)  

The duality highlights relations of del Pezzo geometry to $p$-branes in M-theory (M-branes) as well as to string theory upon dimensional 
reductions and application of dualities (D-branes). 
The main features include \cite{INV}:

\vspace{-2mm} 
\begin{enumerate}[{\bf (i)}]
\setlength\itemsep{-2pt}
\item Rational curves $\cC$ on $\BB_k$ with volume $\omega (\cC)$ and degree $d_\cC=(p+1)$ 
correspond to $\tfrac{1}{2}$BPS $p$-brane states with tension $T_p=2\pi \exp \omega(\cC)$;

\item Automorphisms of $\mathbb{B}_k$ and $H^2 (\BB_k; \ZZ)$
correspond to the U-duality 
transformations of M-theory on $T^k$.
The U-duality group of M-theory on $T^k$, 
which for rectangular compactifications
with no $C$-field is given by the Weyl group $W$ of $E_k$,  
is mapped to a subgroup of the automorphism group of 
$H^2(\mathbb{B}_k; \ZZ)$ which preserves the intersection form and canonical class $\cK_k$;

\item The moduli space $A/W$ of compactified M-theory corresponds to the moduli space $H^2 (\BB_k; \RR)/W$ of generalized K\"{a}hler forms $\omega$ on $\BB_k$;

 
\item Two classes of rational curves $\cC_1$ and $\cC_2$ related as $\cC_1 + \cC_2= - \cK_k$ on $\BB_k$
correspond to two ${\rm D}p_1$-branes 
with ${\rm D}p_2$-branes  with $p_1 + p_2=7-k$, expressing electric-magnetic duality;

\item The $p$-branes of type IIA and type IIB string theory in 10 dimensions correspond to
classes of rational curves on $\mathbb{B}_1$ and $\CC \PP^1 \times \CC \PP^1$, respectively. 
Further explicit realizations of these are given in \cite{AHRS}. 

\end{enumerate}

\vspace{-2mm} 
Studying the cohomology of the del Pezzo surfaces, the authors of
\cite{HLJP1}\cite{HLJP2} managed to extract the algebras found in
\cite{CJLP2} as Borcherds algebras with a truncated
set of their positive roots corresponding to the generators of the
potentials.

\smallskip 
Rational homotopy theory  of the 4-sphere $S^4$ captures the dynamics of 
the fields in M-theory, as proposed in \cite{Sati13}, and developed further under
the name {\it Hypothesis H} in \cite{FSS17}\cite{FSS19b}\cite{FSS-WZW}\cite{GS21} (and applied in \cite{Roberts}). Many consistency checks have been made, correctly  yielding
various subtle effects of charge quantization (see \cite{FSS19b}\cite{FSS-WZW}\cite{SS19a}\cite{SS19b}\cite{MF-theory}).

In the companion mathematical paper \cite{SV1}, we provided an explicit description of the fields and their dynamics in  M-theory compactified on tori, using the power of algebraic topology, through which many detailed aspects can be captured. 
We explained there how 
bringing in topology,
in the form of (rational) homotopy theory,
leads to a triality
extending the above Mysterious Duality picture ${\footnotesize \raisebox{.5pt}{\textcircled{\raisebox{-.9pt} {1}}}}$: 
\vspace{0mm} 
\begin{figure}[htb]
$$
\xymatrix@R=.1em{
 \fbox{\text{Algebraic Geometry}} \;\; \ar@{<..>}[rr]^-{\footnotesize
 \raisebox{.5pt}{\textcircled{\raisebox{-.9pt} {3}}}} 
 \ar@{<..>}[dr]_-{\footnotesize \raisebox{.5pt}{\textcircled{\raisebox{-.9pt} {1}}}}&  & \;\; 
 \fbox{\text{Algebraic Topology}}
\\
& \fbox{\text{Physics}} \ar@{<->}[ur]_-{\footnotesize \raisebox{.5pt}{\textcircled{\raisebox{-.9pt} {2}}}}& 
}
$$

\vspace{-5mm} 
\caption{\footnotesize {\it Mysterious Triality}. The solid arrow 
${\footnotesize \raisebox{.5pt}{\textcircled{\raisebox{-.9pt} {2}}}}$
stands for a relationship between physics and algebraic topology which is made explicit in \cite{SV1}, while the dotted arrows 
${\footnotesize \raisebox{.5pt}{\textcircled{\raisebox{-.9pt} {1}}}}$
and
${\footnotesize \raisebox{.5pt}{\textcircled{\raisebox{-.9pt} {3}}}}$
denote still mysterious correspondences 
based on combinatorial coincidences. It is enough to solidify one of the dotted arrows to remove the mystery veil off the face of Triality.}
\label{MT}
\end{figure}

While in \cite{SV1} we focused on building the duality ${\footnotesize \raisebox{.5pt}{\textcircled{\raisebox{-.9pt} {2}}}}$, our contributions here are of three types:
\begin{enumerate}[\bf (i)]
\setlength\itemsep{-2pt}
\vspace{-2mm} 
\item those that confirm Mysterious Duality ${\footnotesize \raisebox{.5pt}{\textcircled{\raisebox{-.9pt} {3}}}}$;
\item those that expand the duality to have further features;
\item and those that extend the duality to the triality. 
\end{enumerate}

\vspace{2mm} 
There are two main models for capturing  rational homotopy via algebra, which are dual in some sense (Koszul duality) \cite{Ta83}\cite{Maj}\cite{felix-halperin-thomas}:
\vspace{-1mm} 
\begin{enumerate}[{\bf (I)}] 
\setlength\itemsep{-2pt}
\item 
The {\it Sullivan model}:  
Sullivan models \cite{Su77} (see also \cite{felix-halperin-thomas}\cite{FOT08}\cite{GM13}\cite{felix-halperin})
in rational homotopy theory (see \cref{smm})
bring out structures that are visible in 
supergravity 
(see \cite{FSS17}\cite{FSS-pbranes}\cite{FSS-L00} \cite{FSS19a})
and which standard homotopy theory might otherwise obscure. 
The Sullivan minimal model $M(S^4)$ describes the dynamics of the 
form fields, expressed as differential forms on spacetime. 
A gentle introduction to the relationship between 
M-theory and rational homotopy theory
is given in \cite{FSS19a}.

\item The {\it Quillen model:} 
Quillen models \cite{Quillen} (see also \cite{felix-halperin-thomas}\cite{GM13}\cite{BFMT20}) capture the algebra of gauge transformations, or alternatively that of the branes.
We showed in \cite{SV1} that the algebra of gauge transformations for these fields is captured 
by the Quillen model.  
Remarkably, we found that the degree on del Pezzo matches the
degree arising from the Quillen model.

\end{enumerate}

\vspace{-3mm} 
\(
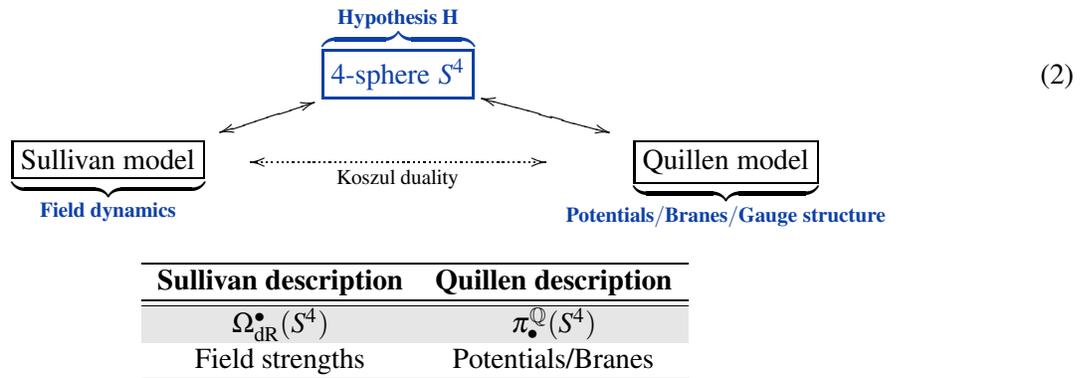
\label{SulQuil}
\hspace{9mm}
\xymatrix@R=-8pt{
& \protect\overbrace{\fbox{\text{4-sphere $S^4$}}}^{\rm \bf \color{darkblue} Hypothesis \; H} \ar@{<->}[dr] \ar@{<->}[dl]& 
\\
 \protect\underbrace{\fbox{\text{Sullivan model}}}_{\rm \bf \color{darkblue} Field \; dynamics} \;\; \; \; \; \; \ar@{<..>}[rr]_-{\rm Koszul \; duality}&  & \;\; 
 \protect\underbrace{\fbox{\text{Quillen model}}}_{\rm \bf \color{darkblue} Potentials / Branes / Gauge \; structure}
}
\)

\vspace{-2mm} 
\begin{center}
    \begin{tabular}{cc}
    \hline 
 {\bf   Sullivan description} & {\bf Quillen description}   
    \\
    \hline 
    \hline 
\rowcolor{lightgray}   $\Omega^\bullet_{\rm dR}(S^4)$ & $\pi^\QQ_\bullet (S^4)$
    \\
    Field strengths & Potentials/Branes 
     \\
    \hline 
    \end{tabular}
    \end{center}

The Lie groups and Lie algebras that we extract from the topological data $M = M(\mathcal{L}_c^k S^4)$
of cyclic loop spaces $\mathcal{L}_c^k S^4$, arising in the dimensional reduction of M-theory, take the form 

\(
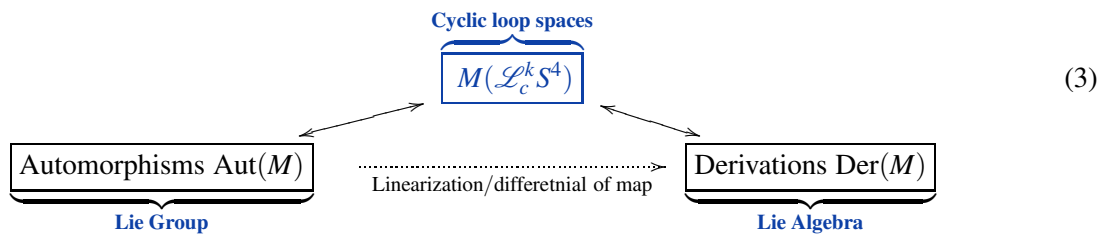
\label{AutDer}
\hspace{9mm}
\xymatrix@R=-8pt{
& \protect\overbrace{\fbox{\text{ $M(\mathcal{L}_c^k S^4)$}}}^{\rm \bf \color{darkblue} Cyclic\; loop \; spaces } \ar@{<->}[dr] \ar@{<->}[dl]& 
\\
 \protect\underbrace{\fbox{\text{Automorphisms Aut$(M)$}}}_{\rm \bf \color{darkblue} Lie \; Group} \;\; \; \; \; \; \ar@{..>}[rr]_-{\rm Linearization/ differetnial \; of \; map}&  & \;\; 
 \protect\underbrace{\fbox{\text{Derivations Der$(M)$}}}_{\rm \bf \color{darkblue} Lie \; Algebra}
}
\)

\vspace{-3mm} 
\paragraph{Coefficients.} Rational homotopy theory is about taking homotopy groups $\pi_\bullet$
over the rational numbers $\mathbb{Q}$: $\pi^\QQ_\bullet = \pi_\bullet \otimes \QQ$. The theory works equally well over the real numbers 
$\mathbb{R}$, and indeed, here as in \cite{SV1} we consider rational homotopy theory over $\mathbb{R}$ 
(as in \cite{GM13}). This is the version in
which rational (more precisely, real) homotopy theory connects to differential geometry (e.g. \cite{FOT08}), 
since the smooth de Rham complex is not defined over $\mathbb{Q}$  but over
$\mathbb{R}$. This is favorable when dealing with differential forms $\Omega^\bullet_{\rm dR}$ 
and physical fields
-- see \cite{FSS-Chern} for an extensive discussion and applications. 
Still, we do at times 
consider rational and integral coefficients, for instance
when dealing with algebraic (Lie) groups. 
We also get the root lattice, though that sits in a real vector
space, indirectly through combinatorial data. So, in this sense, we
connect to the $\mathbb{Z}$-forms of the U-duality groups \cite{SV1}.

\paragraph{The cyclified loop space.} 
We found  in \cite{SV1} a striking matching between the equations
  for the Sullivan minimal models of the iterated cyclic loop space (\emph{cyclification}) ${\mc{L}_c^k S^4}$  of the
  four-sphere and the equations of motion (EOMs) of toroidal
  compactifications of M-theory to $11-k$ dimensions.
  This is deeply rooted in the fact that the
  process of compactification of supergravity on a circle $S^1$ is akin to the process
  of working out the Sullivan minimal model of the cyclic loop space.
 This matching also implies the following general philosophy \cite{SV1} (extending the $k=1$ case in \cite{FSS-pbranes}\cite{FSS-L00}):
\begin{quote}
{\it Any feature of or statement about the Sullivan minimal model of an iterated cyclic
loop space $\mc{L}_c^k S^4$ (or the real homotopy type thereof)
may be translated into a feature of or statement about the
compactification of M-theory on the $k$-torus.}
\end{quote}

Explicit examples of these features are the root system $E_k$, its Weyl group
$W(E_k)$ and U-duality, the split real torus action and 
trombone and rescaling symmetry, etc.\ (see \Cref{Sec-TorSymm,Sullmm}). 
We view this as providing a strengthening of the Mysterious Duality, as 
having an explicit relation was the essence of 
the mystery behind the duality proposed by \cite{INV}.

\medskip 
 The action of the integral form $G(\mathbb{Z})$ on $G/K$ (see \eqref{abelianization}), 
 hence orbits of discrete U--duality groups, are studied in 
\cite{CMS} \cite{BDDFMR}. 
 Note that the above are symmetries of the equations of motion and
not of the action,\footnote{
These symmetries can be implemented at the level of the action, but this
involves a Legendre transformation of the fields.}
hence it is appropriate that here and in \cite{SV1} we consider these symmetries from 
studying dynamics given by the EOMs and Bianchi identities arising from the corresponding Sullivan models.

\paragraph{New perspectives and further contributions.} In addition 
to extending the duality 
${\footnotesize \raisebox{.5pt}{\textcircled{\raisebox{-.9pt} {1}}}}$
to a triality involving rational homotopy theory via 
${\footnotesize \raisebox{.5pt}{\textcircled{\raisebox{-.9pt} {2}}}}$
and
${\footnotesize \raisebox{.5pt}{\textcircled{\raisebox{-.9pt} {3}}}}$ in \Cref{MT}, 
here
we 
provide the following perspectives:
\vspace{-2mm} 
\begin{enumerate}[{\bf (i)}] 
\setlength\itemsep{-2pt}

\item \emph{Including the C-field universally via $S^4$}: This 
  brings in the $C$-field and its (Hodge) dual at the level of nontrivial field strengths, 
  modeled dynamically through $S^4$ as a universal
  nonlinear space \cite{Sati13}. Previously these fields were introduced as  algebraic coefficients at the level of potentials in \cite{HLJP1}\cite{HLJP2}.
 
\item \emph{Cohomological nontriviality}: 
Closed differential forms on flat spacetime are exact, but the Sullivan and Quillen 
models we consider, by construction, reflect 
the nontrivial (co)homology of the universal space $S^4$ and its cyclifications. 
We correspondingly take the field strengths to
not necessarily be exact, extending what is traditional in the supergravity literature.
 
\item \emph{Nonlinearity}: Instead of working with  (co)homology which is intrinsically linear,
we bring in nonlinear spaces, with a rich structure encoding dualities and dynamics. 
The setting in \cite{INV} is kinematic and linear, while we will see that ours is dynamic and 
both linear (in regards to weights/roots and Weyl group actions) as well 
as nonlinear (in terms of generators/field strengths and their equations of motion). 
 
\item \emph{Nontrivialilty of torus bundles}: Instead of having trivial torus bundles,
we have nontrivial
ones, each equipped with 
an Euler class or a first Chern class that enters crucially into the dynamics leading to the uncovering of the 
underlying homotopy structure.
 
\item \emph{Global dimensional reduction}: 
We already provided in \cite{SV1} a systematic global reduction on $T^k$ of the
duality-symmetric theories. 
We used cyclic loop spaces for reduction on a circle 
as in  \cite{FSS-pbranes}  and for reductions on tori (as instances of more general Lie groups) as in \cite{BSS}. 
We found iterative reduction on circles 
 via ``cyclification" ${\mc{L}_c^k S^4}$ to be most relevant. 
 Here we contrast this with reducing directly on tori
at once via ``toroidification'' $\mathbb{T}^k(S^4)$ (see \cref{Sec-toroid}).

\item \emph{Moduli space}:
We introduced in \cite{SV1} a new moduli space, similar to the moduli space $A/W$ of compactified M-theory 
(see \eqref{abelianization}) and the moduli space $H^2 (\BB_k; \RR)/W$ of generalized K\"{a}hler forms $\omega$ on the del Pezzo surface \cite{INV}, in the context of rational homotopy theory as the moduli space of metric data added to the rational homotopy type. 
  
\item \emph{Probing into the nonabelian part of the moduli space}:
We are, in a sense, supplying the topological content of the 
nonabelian unipotent part $N$ of \eqref{abelianization} by considering fields, which carry nontrivial weights of the toroidal part $A$.
We plan to expand on this further in \cite{SV3}. 

 \item \emph{The role of the Weyl group}:
Bringing in both dynamical/nontrivial tori, as the fibers  of
iterated circle/torus bundles as above, allows us to 
explain and clarify the exact role the Weyl group plays from physical and topological points of view.

 \item \emph{Data beyond that of Lie groups}: Our approach allows for extending Mysterious Duality 
 and Triality 
 to the rank 
 $k \geq 9$ case, taking us beyond Lie (to Kac-Moody) groups on the group/root-system side,
and beyond del Pezzo (to non-Fano and general-type) surfaces on the algebraic geometry side in \cref{Sec-KM}.
The context of Triality allows for an interpretation of the 
imaginary roots of the Kac-Moody algebras, see \Cref{sec-blow9,sec-blow10}.

 \item \emph{Real vs.\ complex}: We provide a careful treatment of real 
 structures appearing in M-theory and U-duality vs.\ complex forms that
 appear 
 on the algebraic geometry side. This also allows for a natural transition between 
 $\CC \PP^2$ and $S^4$ initiating link  
${\footnotesize \raisebox{.5pt}{\textcircled{\raisebox{-.9pt} {3}}}}$ for $k = 0$,
 as corresponding to modding out by complex conjugation. 
 
 \item \emph{Compatibility with T-duality and S-duality}:
 Our approach and extension are compatible with both S-duality in type IIB string 
 and T-duality between type IIA and type IIB string theories, and same at the 
 dynamical level from the physical and topological points of view. 
 See \cref{Sec-IIBS} and \cref{IIB}.
 
 \item \emph{Modeling classical statements in algebraic geometry}: 
 Adopting this perspective allows us to provide remarkable topological and  physical 
 manifestations of various famous statements in classical algebraic geometry, including
  27 lines on a cubic, etc.\ in \cref{Sec-27}. 
  For instance, we showed in \cite{SV1} that $\pi_2^\RR (\mc{L}_c^6 S^4)$ is 27-dimensional and splits 
  canonically into 27 lines $\CC\PP^1 \to \mc{L}_c^6 S^4$. Here we provide further connections
  in \cref{Sec-M}. 
  
  \item \emph{$E_k$-symmetry structure on rational homotopy groups}: The $E_k$ symmetry gives a way to organize and handle the computation of rational homotopy groups of cyclic loop spaces $\mc{L}_c^k S^4$, 
 arising in the dimensional reduction on $T^k$.
 We also expand on this further in \cite{SV3}. 
\end{enumerate}

\paragraph{The duality-symmetric approach.} 
In the systematic dimensional reduction presented in \cite{SV1}, we
treat the fields and their $*$-duals as independent, 
with all equations of motion and the duality relations between the doubled fields accounted 
for, as in the 
duality-symmetric (doubled field) formulation of $D=11$ supergravity
 \cite{BBS98}, of type IIB $D=10$ supergravity \cite{DLS97}\cite{DLT98},
and of type IIA $D=10$ supergravity \cite{BNS04}, the last obtained 
by dimensionally reducing the first. 
The global dimensional reduction leading to the latter is performed in 
\cite{MaS}\cite{FOS03}.
This duality-symmetric formulation also allows for treating higher-order corrections, as in \cite{ST17}.
A higher cohomological duality-symmetric formulation was initiated in \cite{Sati-Form} and highlighted in \cite{tcu}. 
Such a duality-symmetric formulation, when pursued in the context of $S^4$, 
allows for remarkable cancellation of all prominent anomalies in M-theory \cite{FSS19b} and for compatibility with the K-theoretic formulations in types IIA/IIB 
\cite{FSS-pbranes}\cite{FSS-L00}\cite{BSS}.

\vspace{-2mm} 
\paragraph{Bridging the gap between $\CC \PP^2$ and $S^4$.} 
As above, 
the topological side of the 
Triality 
uses the (rational homotopy type of the) 4-sphere $S^4$ as a starting point. This is a departure from the setting of the complex projective plane $\CC \PP^2$ on the algebraic-geometric side. 
Aside from this being a shift in perspective, can one find a mechanism for going from one to the other? 
We propose that this is generally a transition from the complex setting
to the real setting. Beyond the general conceptual level, we offer 
two tantalizing routes which support adopting this perspective:

\vspace{-2mm} 
\begin{enumerate}[{\bf (i)}] 
\setlength\itemsep{-2pt}
\item Via modding out by {\it complex conjugation} on the algebraic geometry side (cf.\ item \textbf{(x)} above).
We discuss this in \cref{Subsec-RvsC}. This uses the Arnold-Kuipers-Massey theorem 
(\cite{Arnold88}\cite{Ku74}\cite{Letizia}\cite{Lawson}\cite{Massey}\cite{Marin}\linebreak[0]\cite{HH11}).
We also establish the compatibility of this process with 
S-duality in type IIB string theory in \cref{Sec-IIBS}:
\vspace{0mm} 
$$
\hspace{-4mm}
\xymatrix{ 
\fbox{{Type IIB via} 
$\CC \PP^1 \times \CC \PP^1$ } 
\ar[rrr]^-{\rm \bf  \color{darkblue} quotient\; by}_-{\rm \bf  \color{darkblue} S-duality} 
&&&
\fbox{{M-theory via} $\mathbb{C}\PP^2$} 
\ar[rrr]^-{\rm \bf  \color{darkblue} quotient\; by}_-{\rm \bf  \color{darkblue} complex \; conjugation} &&&
\fbox{{M-theory via} $S^4$}
}
$$

\item Via {\it Hopf fibrations}: The transition is from the complex Hopf fibration,
which can be viewed as part of the process of blowing up, 
to the quaternionic Hopf fibration 
(viewed as a topological, i.e., in some sense, real fibration):\footnote{Real and
quaternionic are closely related and  can occur together in 
topology, much more so than with the complex case; e.g. in the topology of the orthogonal group.
M-theory has prominent real and quaternionic aspects, but not as many complex ones.} 
\vspace{-2mm} 
$$
\hspace{-.6cm} 
\fbox{\footnotesize \color{darkblue} \bf Complex algebraic  geometry/branes}
\;\;
\xymatrix@=1em{
S^1 \ar[r] & S^3 \ar[d]^{h_{\CC}} &  \ar@{~>}[rrrrrr]^{\rm complex \; to \; real/quaternionic} &&&&&&
\\ & S^2
}
\xymatrix@=1em{
S^3 \ar[r] & S^7 \ar[d]^{h_{\HH}}    
\\ & S^4 &&&
}
\hspace{-1.2cm} 
\fbox{\footnotesize \color{darkblue} \bf Algebraic topology/fields}
$$

\vspace{-1mm} 
\noindent We discuss this in \cref{Subsec-Hopf}. See \cite{Sati13}\cite{FSS-WZW}\cite{FSS19b}\cite{MF-theory} for extensive discussion about the importance of the quaternionic Hopf fibration in the context of M-theory. \end{enumerate}

\paragraph{Automorphisms.} 
The Weyl group acts on cohomology and, equivalently, homology of del Pezzo surfaces. 
Viewing the perspective in \cite{INV} schematically as:
\begin{equation} 
\label{lin-dual-coh} 
\xymatrix@C=8em{
\text{\color{darkblue} \bf \footnotesize Weyl group} \ \rotatebox[origin=c]{-90}{$\circlearrowright$}~~
H^2(\mathbb{B}_k; \R) = \widehat{\mathcal{M}}_k
\ar@{<->}[r]^{\rm linear}_{\rm duality}  
& 
\widehat{\mathcal{M}}^*_k \cong  H_2(\mathbb{B}_k; \R) 
\ \rotatebox[origin=c]{-270}{$\circlearrowright$}~~ \text{\color{darkblue} \bf \footnotesize Weyl group},
}
\end{equation} 
the picture we have on the topological side is considering, in a certain sense, 
a nonlinear version of this;
schematically:
\begin{equation} 
\label{dual-models}
\xymatrix@C=8em{
\text{\color{darkblue} \bf \footnotesize automorphisms} \ \rotatebox[origin=c]{-90}{$\circlearrowright$}~~
\fbox{Sullivan model} \ar@{<->}[r]^{\rm Koszul}_{\rm duality} &
\fbox{Quillen model} 
\ \rotatebox[origin=c]{-270}{$\circlearrowright$}~~ \text{\color{darkblue}
\bf \footnotesize automorphisms}.
}
\end{equation} 
There are two features to highlight here, as demonstrated in \cite{SV1}:

\vspace{-2mm} 
\begin{enumerate}[{\bf (a)}] 
\setlength\itemsep{-2pt}
\item 
 We are replacing linear entities, namely cohomology and homology, with nonlinear entities, 
namely rational models. The Sullivan model is nonlinear in two respects: it has products of 
elements and a nonlinear differential. The Quillen model
combines these two 
nonlinearities into one (upon dualizing) captured by the Lie bracket or, more generally, the $L_\infty$-brackets.

\item The structures are
secondary, in the sense of ``gauge transformations of gauge
transformations:''
$$
\text{Weyl group} \quad  \rotatebox[origin=c]{-90}{$\circlearrowright$} \quad \text{Torus} \quad 
\rotatebox[origin=c]{-90}{$\circlearrowright$} \quad  M(\mc{L}_c^k S^4) \;.
$$
The first two, i.e., the Weyl group and the torus together with their actions, is what participates in the correspondence 
\eqref{lin-dual-coh}, while these acting in turn on the third is what is involved in 
the correspondence \eqref{dual-models}.
However, we emphasize that both correspondences share one and the same torus. 
\end{enumerate} 

\noindent Thus, the Weyl group, being an isometry group of the abelian Lie algebra $\h_k$ of infinitesimal symmetries of $M(\mc{L}_c^k S^4)$, is a ``second derived'' object with respect to $\mc{L}_c^k S^4$: $W(E_k)$ is the group of ``symmetries of symmetries'' of $\mc{L}_c^k S^4$.

\vspace{-2mm} 
\paragraph{Symmetry via torus actions.} 
We have seen in \cite{SV1} that topology emerges in our perspective on the scaling/toroidal 
symmetries  via self-maps of topological spaces at hand, namely the 4-sphere $S^4$ and the 
circle factors $S^1$ of the internal torus $T^k$, which 
involves various actions of the multiplicative group 
$\GG_m = \GL(1)$:
\vspace{-2mm} 
\begin{enumerate}[{\bf (i)}] 
\setlength\itemsep{-2pt}

\item
 An action of one factor $\GG_m$ already in eleven dimensions 
corresponds to the trombone symmetry of the C-field
(see \cref{Sec-TorSymm}). In turn, 
from the point of view
of cohomotopy (i.e., the generalized cohomology theory arising from mapping to 
the sphere)  and the 4-sphere as a universal target space
\cite{Sati13}, we find that this corresponds to the scaling symmetry
or self-maps of the 4-sphere.

\item Upon reduction on $T^k=S^1 \times \cdots \times S^1$, we get an
action of $\GG_m^k$, where the $i$th $\GG_m$ corresponds
to the scaling symmetry of the $i$th circle in $T^k$, i.e., 
self-maps $S^1 \to
S^1$   of the circle fibers. See \cref{Sec-IIAsym} and  \cref{symm-cycl-concrete}.
\end{enumerate}

\paragraph{Symmetry via Weyl group actions.} We 
consider
the following
aspects:

\vspace{-2mm} 
\begin{enumerate}[{\bf (i)}] 
\setlength\itemsep{-2pt}
    \item {\it Relation between the Weyl group $W(G)$ and the integral form  $G(\ZZ)$ of the 
    U-duality group $G$}:
For compactificatons of M-theory on 
rectangular tori with no 
$C$-field, 
the U-duality group acting on this class of compactifications is 
 realized as the Weyl group of $E_k$
\cite{EGKR98}\cite{BFM}. 
Likewise in supergravity, $W(G)$ is often assumed to be a subgroup of the discrete U-duality group $G(\ZZ)$. This is a subtle point, and
a 
discussion is needed here -- see the end of \cref{dP};

\item  {\it The Weyl group as a symmetry:} 
We explain how the Weyl group $W$ acts on the Picard group ${\rm Pic}(\mathbb{B}_k)$ 
as a symmetry. Hence it makes sense to mod out by it in the moduli space. 
Any two exceptional configurations on $\BB_k$ are  $W$-translates of one another,
provided the signs of the intersection forms are preserved 
 \cite{Nagata61}\cite{Har85}; see \cref{Sec-Ek};

\item {\it Weyl group `action' on field strengths}: 
From a physics perspective, the  action of $W(G)$ 
for  toroidal compactification for $k\leq 8$ has been considered in 
\cite{LPS96}, where the effect is captured by the axionic/dilatonic prefactors 
multiplying the field strengths in their dimensional reduction.\footnote{\label{barefields} This depends on what one calls a ``field strength". The topological 
perspective favors the bare ones, rather than the ones which 
are scaled by dilaton and 
axion vectors, and possibly additionally shifted  by composites of lower-degree
field strengths and potentials, with the understanding that the latter might
be favored in parts of the supergravity literature (see e.g. \cite{LP96}\cite{LLP98}).
} 
In particular, 
the  dilaton vectors transform in a similar fashion as 
weight vectors under a subgroup of the discrete U-duality group identified
as the Weyl group.
We parametrize the apparent mathematical difficulties encountered in \cite{LPS96} 
by clarifying the entities on which the Weyl groups act and how they do so: 
$W$ acts directly on the linear data arising
 from field strengths and not on the field strengths themselves. Nevertheless, 
 there  will be a correspondence of the form
  $$
\xymatrix@R=.3em{
& \protect{\fbox{\text{\bf \color{darkblue} Fields}}} \ar@{<->}[dr] \ar@{<->}[dl]& 
\\
\protect {\fbox{\text{\bf \color{darkblue} Root vectors in lattice}}} \;\; \ar@{<..>}[rr]^-{\rm dual}&  & \;\; 
\protect {\fbox{\text{\bf \color{darkblue} Divisors on del Pezzo}}}\; }
$$
See \cref{Sec-Ek} and \cref{dP}, especially the end of \cref{Sec-8lower};

     \item {\it Weyl orbits:} 
     With the relation between $W(G)$ and $G(\ZZ)$ clarified,
     the orbits of $W$ are related to the orbits of $G$, with the latter
being determined in \cite{CMS} (see also \cite{BDDFMR}). The Weyl orbits have also 
been studied in \cite{FRT} from the point of view of sigma models. 
Our formulation via moduli spaces clarifies the role of the Weyl group 
and the corresponding geometry (see \cref{Sec-Ek}); 

 \item {\it Weyl group and automorphisms of del Pezzo surfaces: }
 The two groups $W(G)$ and $\Aut(\mathbb{B}_k)$ are related but they 
 generally 
 do not coincide (depending on the rank), and we clarify the role of each
 and how they relate. This helps in understanding the relationship between
 U-duality and Mysterious Duality/Triality (see \cref{dP}). 
\end{enumerate}

\paragraph{The $E_k$ symmetry patterns.} 
The dimensional reduction of M-theory on 
a $k$-torus gives rise to a theory in $D = 11 - k$ dimensions
with  symmetry pattern \cite{INV} in the five columns in Table 
\ref{table1} below, matching the familiar pattern for del Pezzo surfaces \cite{Man}, 
to which we added in \cite{SV1} 
the 6th column for cyclic loop spaces (``cyclifications'' $\mc{L}_c^k S^4$ of $S^4$),
as well as the 7th column corresponding to torus symmetry. 
This highlights the interrelations among Lie theory (4th column), 
algebraic geometry (5th column), and topology/physics (6th column),
as appropriate by the trichotomy/triality in Figure \ref{MT}.

\begin{table}[H]
\centering
\begin{tabular}{ccccccc}
\hline
$D$ &  $k$ & {\bf Type of $E_k$} & {\bf Lie algebra $\mathfr{g}$} & {\bf del Pezzo} & 
{\bf Model} & {\bf Maximal Split Torus}  \\
\hline
 \hline
\rowcolor{lightgray} 11 & 0 & $A_{-1}$ & $\mathfr{sl}_0 = \varnothing$ & $\CC \mathbb{P}^2$
&  $S^4$       & $\GG_m$  \\
10 & 1 & $A_0$ & $\mathfr{sl}_1 = 0$ & $\mathbb{B}_1$ 
&  $\mathcal{L}_c S^4$ & $\GG_m \times \GG_m$ \\
10 & 1 & $A_1$ & $\mathfr{sl}_2$ & $\CC \mathbb{P}^1 \times \CC \mathbb{P}^1$ 
& $IIB$ & $\GG_m \times \GG_m$ \\
\rowcolor{lightgray} 9 & 2 & $A_{1}$ &$\mathfr{sl}_2$ & $\mathbb{B}_2$ 
&  $\mathcal{L}_c^2 S^4$ & $\GG_m^2 \times \GG_m$  \\
8 & 3 & $A_2 \times A_1$ &$\mathfr{sl}_3 \oplus \mathfr{sl}_2$ & $\mathbb{B}_3$ 
&  $\mathcal{L}_c^3 S^4$ & $\GG_m^3 \times \GG_m$  \\
\rowcolor{lightgray} 7 & 4 & $A_4$&$\mathfr{sl}_5$ & $\mathbb{B}_4$ 
&  $\mathcal{L}_c^4 S^4$ & $\GG_m^4 \times \GG_m$ \\
6 & 5 & $D_5$&$\mathfr{so}_{10}$  & $\mathbb{B}_5$ 
&  $\mathcal{L}_c^5 S^4$ & $\GG_m^5 \times \GG_m$  \\
\rowcolor{lightgray} 5 & 6 & $E_6$&$\mathfr{e}_6$   & $\mathbb{B}_6$ 
&  $\mathcal{L}_c^6 S^4$ & $\GG_m^6 \times \GG_m$  \\
4 & 7 & $E_7$&$\mathfr{e}_7$  & $\mathbb{B}_7$ 
&  $\mathcal{L}_c^7 S^4$ & $\GG_m^7 \times \GG_m$ \\
\rowcolor{lightgray} 3 & 8 & $E_8$&$\mathfr{e}_8$  & $\mathbb{B}_8$ 
&  $\mathcal{L}_c^8 S^4$ & $\GG_m^8 \times \GG_m$  \\
\hline
\end{tabular}
\vspace{-2mm} 
\caption{\label{table1} \footnotesize The $E_k$ pattern in Lie theory,
$(0 \leq k \leq 8)$ 
del Pezzo surfaces, and cyclifications of $S^4$.}
\end{table}

\vspace{-2mm} 
\paragraph{Extension to Kac-Moody.}
As indicated in \cite{SV1}, on which we expand further in \cref{Sec-KM} below,
our formulation allows us to extend the theory to the higher ranks, $k=9$, $10$, and $11$,
corresponding to infinite-dimensional cases. 
Indeed, since our discussion extends beyond the Lie setting to the  Kac-Moody 
setting, we 
have extensions of the Triality in Figure \ref{MT}  that go beyond 
the del Pezzo/Fano case on the algebraic side.
We also observe that cyclic loop spaces, when we go beyond $k=8$, undergo a transition analogous to that on the del Pezzo/root systems/Lie algebra side: the degree of the cyclification $\mathcal{L}_c^k S^4$ in the sense of \eqref{deg-L} ceases to be positive, the corresponding root system becomes infinite, and the metric on the $k$-dimensional real vector space holding the root system ceases to be Euclidean; see \cite[Remark 4.7]{SV1}. 
The surface $\mathbb{B}_9$ and cyclification $\mathcal{L}_c^9 S^4$ give rise to a rank-9 parabolic lattice, 
while $\mathbb{B}_{10}$ and $\mathcal{L}_c^{10} S^4$, and  $\mathbb{B}_{11}$ and $\mathcal{L}_c^{11} S^4$ correspond to rank-10 and rank-11 hyperbolic lattices, respectively. 

\begin{table}[H]
\centering
\renewcommand{\arraystretch}{1.2}
\begin{tabular}{ccccccc}
\hline
$D$ &  $k$ & {\bf Type of $E_k$} & {\bf Kac-Moody algebra $\mathfr{g}$} & {\bf Non-Fano Surface} & 
{\bf Model} & {\bf Maximal Split Torus}  \\
\hline
 \hline
\rowcolor{lightgray} 
2 & 9 &  $E_9= \widehat{E}_8$ & affine $\mathfr{e}_9=\widehat{\mathfr{e}}_8$  & 
  $\mathbb{B}_9$ 
&  $\mathcal{L}_c^9 S^4$ & $\GG_m^9 \times \GG_m$  \\
1 & 10 & $E_{10}$ & hyperbolic $\mathfr{e}_{10}$  & 
  $\mathbb{B}_{10}$ 
&  $\mathcal{L}_c^{10} S^4$ & $\GG_m^{10} \times \GG_m$  \\
\rowcolor{lightgray} 
0 & 11 & $E_{11}$ & Lorentzian $\mathfr{e}_{11}$  & 
  $\mathbb{B}_{11}$ 
&  $\mathcal{L}_c^{11} S^4$ & $\GG_m^{11} \times \GG_m$  \\
\hline 
\end{tabular}
\vspace{-2mm} 
\caption{\label{table2} \footnotesize The $E_k$ pattern in 
Kac-Moody theory 
$(k\geq 9)$, general surfaces, and cyclifications of $S^4$.}
\end{table}
 
\paragraph{Further proposals and outlook.} 
This paper, as well as its mathematical companion \cite{SV1}, 
can be viewed as an initiation of a research program. We  have established 
new connections
and provided correspondences in these two installments, but clearly there 
is a lot more that deserves to be investigated. Among these are the following: 

\vspace{-2mm} 
\begin{enumerate}[{\bf (i)}] 
\setlength\itemsep{-2.5pt}
\item We have obtained the rational model for type IIB 
directly via  minimal models in \cite{SV1}. 
It would be interesting to work
out a model at the topological space level, analogous 
(and certainly related) to the 4-sphere;

\item  Our discussion suggests a much bigger
  picture. For instance, we should take the Quillen model as a symmetry Lie algebra
  seriously. We expect that one would be able to work out general
  results on a gauge action of the Quillen model for any topological space. 
  Hence further Lie-theoretic aspects wait to be uncovered. 
  We will indeed discuss this in \cite{SV3}. 
  
\item Studying only toroidal, i.e., commuting symmetries of the real homotopy types of the 
space $\mc{L}_c^k S^4$ (and its Sullivan minimal model) for each $k$ already allows us to 
uncover the hidden $E_k$ root structure. It would be interesting to identify the full 
symmetry group of this real homotopy type. This would be a certain Lie group that contains 
the toroidal symmetry group $\GG_m^{k+1}$,  as the maximal split torus. We anticipate 
the answer to be some real algebraic group between $\GG_m^{k+1}$ and a toroidal extension 
of a real form of the Lie group $E_k$. In the latter case, we would have a manifest action 
of the U-duality group $E_k$ on the fields, coming from an action on the universal model $\mc{L}_c^k S^4$.
  
\item In order to encode geometric information, such as connections/potentials,
smooth structure, and (further)
metric data, we would need geometric/differential/smooth  refinements of our 
cyclification and toroidification (cf.\ the discussion in 
\cref{Sec-toroid}) models,
$\widehat{\mc{L}_c^k S^4}$ and $\widehat{\mathbb{T}^k(S^4)}$,
generally along the lines of 
\cite{FSS15}\cite{FSS17}\cite{GS21} for the special cases $k=0$ and $1$, i.e., 
M-theory and its reduction to type IIA string theory.

\item
A parametrized version
of the sphere model has been proposed in \cite{BSS}
as a way to include higher degree field strengths, $F_8$ and $F_{10}$
in type IIA and also $F_9$ in type IIB.
It would be interesting to see the relation
to del Pezzo surfaces (and beyond). Similarly, extending beyond rational homotopy theory 
via $S^4$ rather than $S^4_{\mathbb{R}}$, in the context of Hypothesis H (see \cref{smm}). 

\item Relating del Pezzo and cyclifications: extending the transition from $\CC \PP^2$ to $S^4$
to that of del Pezzo surfaces $\mathbb{B}_k$ to the cyclification spaces $\mc{L}_c^k S^4$.
 The correspondence for $k\geq 1$ seems to be subtler than
just modding out by complex conjugation:
\(
\label{Bk-conj} 
\xymatrix@R=1.7em{
\mathbb{B}'_1 = \CC \PP^1 \times \CC \PP^1 \ar[rr]^-{\color{darkblue}  \bf /S-duality} \ar@{~>}[ddrr]^{\color{orangeii}  \bf  blowup} 
&&  \mathbb{B}_0=\CC \PP^2 \ar[rrr]^-{\color{darkblue}  \bf / complex\; conjugation} \ar@{~>}[d]^{\color{orangeii}  \bf  blowup} 
&&& S^4 \ar@{~>}[d]^{\color{greenii}  \bf  cyclification}
\\
&&  \mathbb{B}_1  \ar@{<-->}[rrr] \ar@{~>}[d]^{\color{orangeii}  \bf  blowup} 
&&& \mathcal{L}_c S^4 \ar@{~>}[d]^{\color{greenii}  \bf cyclification} 
\\
&&  \mathbb{B}_2  \ar@{<-->}[rrr] \ar@{~>}[d]^{\color{orangeii}  \bf  blowup} 
&&& \mathcal{L}^2_c S^4 \ar@{~>}[d]^{\color{greenii}  \bf cyclification} 
\\
&&  \overset{\vdots}{\mathbb{B}_k} \ar@{<-->}[rrr] \ar@{~>}[d]^{\color{orangeii}  \bf  blowup} 
&&&  \overset{\vdots}{\mathcal{L}^k_c S^4} \ar@{~>}[d]^{\color{greenii}  \bf cyclification}
\\
&& \vdots 
&&& \vdots 
}
\)
\item Clarifying the role of Kac-Moody groups and algebras from a topological perspective: 
what is so special about $k=9$, $k=10$ or $k=11$ from the perspective of cyclic loop spaces? 

\end{enumerate}




\vspace{-5mm} 
\paragraph{\large Acknowledgments.}
The first author thanks the University of Minnesota, the Aspen Center for Physics, and 
the Park City Mathematics Institute (IAS) for hospitality during the work on this project,
and acknowledges the support by Tamkeen under the NYU Abu Dhabi Research Institute grant CG008.
The second author thanks NYU Abu Dhabi and Kavli IPMU for creating remarkable opportunities to initiate and work on this project.
His work
was also supported by World Premier International Research Center Initiative (WPI), MEXT, Japan, and a Collaboration Grant from the Simons Foundation (\#585720).

\section{From Mysterious Duality to Mysterious Triality via Hypothesis H} 
\label{Sec-S4cyc}

\subsection{Hypothesis H: M-theory dynamics via the
Sullivan model of $S^4$} 
 \label{smm}

As in \cite{SV1}, we replace the notion of a \emph{rational Sullivan minimal model}
of a topological space \cite{felix-halperin-thomas}\cite{FOT08} \cite{GM13}
with that of a \emph{real Sullivan minimal model}, given that real coefficients of physical
fields could be a bit more natural than rational ones (see the 
discussion in \cite{FSS-Chern}). We will therefore
assume that our algebraic models are defined over the reals $\RR$
(see \cite{BSzcz}\cite{GM13}).

\medskip 
We adopt the perspective proposed in \cite{Sati13} and known as \emph{Hypothesis H} of viewing the 4-sphere  $S^4$ as the 
universal space of form fields in M-theory. The  significance of this is that  $S^4$ encodes, entirely in its topology, the field $G_4$ and its dual $G_7$ as well as their 
{\it dynamics}.
This space is  viewed as a universal space in the sense that these field
configurations are given at the homotopy level by real homotopy classes of 
maps from spacetime $Y^{11}$ to $S^4$, and whenever geometry is included, one
would need to consider all maps; see \cite{FSS15}\cite{FSS17}\cite{GS21} (but 
here we will concentrate on topology).

\medskip 
In 11-dimensional supergravity, which is the low-energy limit of M-theory, the equations of motion (EOMs) are \cite{CJS}
 \(
 \label{M-EOM}
 d G_4 = 0, \qquad d*G_4 + \tfrac{1}{2}G_4 \wedge G_4=0\;.
 \)
When combined with the self-duality condition
\begin{equation}
  \label{C-fields}
G_7 := *G_4 \;,
\end{equation}
these may be rewritten as
\begin{equation}
\label{EOM}
d G_4 = 0, \qquad d G_7 + \tfrac{1}{2}G_4 \wedge G_4 = 0 \;,
\end{equation}
where the fields $G_4$ and $G_7$ are represented by differential forms of degree 4 and 7, respectively, on the 11-dimensional spacetime $Y^{11}$ 
of M-theory, and $*$ denotes the Hodge star operator, which captures the dependence on the metric on $Y^{11}$.
Note that {\it locally} we may write  
$$
G_4 = dC_3, \qquad G_7 = dC_6 - \tfrac{1}{2} C_3 \wedge G_4\;
$$ 
for some differential forms $C_3$ and $C_6$, viewed as the corresponding {\it potentials}.

\medskip 
However, we will use the duality-symmetric (doubled field) 
formulation, where $G_4$ and $G_7$ are treated 
independently \cite{BBS98} (see also \cite{MaS}\cite{Sati-Form}\cite{ST17}\cite{BSS}
for more global treatments). This will also 
suppress any explicit dependence on the metric, suitable for 
our topological perspective, which may be regarded as describing the topological background of the full story. To get the full picture at the level of fields, one simply adds metric data to spacetime and imposes the duality relation $* G_4 = G_7$. We have also found a way to add metric data to the universal real homotopy model of $S^4$ (and its cyclifications) via introducing moduli parameters \cite{SV1}; see also ``Geometric aspects of the moduli space\dots'' at the end of \Cref{Sec-Ek}.

\medskip
The topological aspects of the M-theory dynamics at the real homotopy level are captured by the real homotopy
theory description given by the Sullivan minimal model 
of $S^4$, \cite{Sati13},
which we denote $M(S^4)$:\footnote{We will use lowercase letters for universal elements and uppercase letters to denote
spacetime fields.
}  
\vspace{-1mm} 
\begin{gather}
 \label{S^4}
  M(S^4) = (\RR[g_4, g_7], d)\, \qquad
dg_4 = 0, \qquad dg_7 = -\tfrac{1}{2} g_4^2 \,,
\end{gather}
where the degree of each of the generators $g_4$ and $g_7$ is given by the 
corresponding subscript:
$\abs{g_4} = 4$, $\abs{g_7} = 7$.
Here choose to include the factor of $-\tfrac{1}{2}$ in the
equations of the model, as opposed to being absorbed by the generators
(see \cite[Ex.\ 3.3]{FSS17}).

\medskip 
The de Rham
cohomology of $S^4$ is generated by the volume form $d$vol of degree $n$. 
This gives a generator $g_4$ in degree 4, with $dg_4 = 0$, which generates a polynomial algebra. 
Since $M(S^4)$ must be free, the only way
to kill $g_4^2$ is to introduce an element $g_7$ of degree 7 with $dg_7 = g_4^2$ (or any scalar multiple of $g_4^2$). The resulting the DGCA,
\eqref{S^4},
is minimal. The map from $M(S^4)$ to the de Rham algebra (differential forms),
given by $\Omega^\bullet (S^4)$  by $g_4 \mapsto$ $d$vol and $g_7 \mapsto 0$,
defines a \emph{quasi-isomorphism}, \emph{i.e}., an algebra homomorphism which
induces an isomorphism on cohomology.

\medskip 
In order to include spacetime, comparing \eqref{EOM} with \eqref{S^4}, we see that there exists a differential graded (dg) algebra homomorphism
\begin{align}
\label{homo}
M(S^4) &\longrightarrow (\Omega^\bullet(Y), d),
\end{align}
identifying $g_4$ with  $G_4$
and 
$g_7$ with $G_7$,
where $(\Omega^\bullet(Y), d)$ is the de Rham algebra of the 11-dimensional spacetime $Y^{11}$. The de Rham algebra is, in fact, a real homotopy model of the manifold $Y^{11}$, and this model could be different from the Sullivan minimal model of $Y$.
Rational (or, actually, real \cite{FSS-Chern}) homotopy theory provides a canonical continuous map
\begin{equation}
\label{q-cohomotopy}
Y \longrightarrow  S^4_\R,
\end{equation}
where $S^4_\R$ is the \emph{rationalization over $\RR$}  
of $M(S^4)$, a certain universal topological space whose Sullivan minimal model $M(S^4_\R)$ is $M(S^4)$, such that the pullback map from the Sullivan minimal model of $S^4_\R$ to the de Rham model of $Y^{11}$ is given by \eqref{homo};
see \cite{BSzcz}\cite{FSS-Chern}.

\paragraph{Cohomotopy vs.\ cohomology.} 
Even though the above connection is at the rational level, 
 it was suggested in \cite{Sati13}, that there is actually a continuous map to 
 the honest-to-goodness 4-sphere $S^4$,
$Y \longrightarrow  S^4$,
which induces the identification between the rational model and differential forms. 
Indeed, a comparison of this target to its linearization, that is to say, the
Eilenberg-MacLane classifying space $K(\mathbb{Z}, 4)$,
which encodes the C-field as captured by a degree 4 class, with the 
4-sphere is presented in \cite{GS21} through a Postnikov tower analysis. 
 The nonabelian nature and the shift by Pontrjagin classes in the 
 quantization condition of the C-field are studied in \cite{FSS19b}.
As also indicated below in \cref{Subsec-Hopf}, the quaternionic 
Hopf fibration $h_{\mathbb{H}}$ over $S^4 \cong \HH \PP^1$ 
plays a central role in describing the field
strengths  on   the M-theory
side in  
\cite{Sati13}\cite{FSS-WZW}\cite{FSS19b},
while quaternionic orientation using $\HH \PP^\infty$ (and its 
finite-dimensional stabilizations in low dimensions) are 
central to the constructions in 
\cite{MF-theory}. 
Rationally, since $\pi_i(\HH \PP^\infty) \cong \pi_{i-1}(S^3)$ is torsion for 
$i>4$, we have a (weak) rational homotopy equivalence between 
$\HH \PP^\infty_{\QQ}$ and $K(\QQ, 4)$.
The Postnikov tower of $S^4$ can be related to $\mathbb{H}\PP^\infty$ 
(with stages described in \cite{GS21}):
\vspace{-2mm} 
 $$
 \xymatrix@C=2.5em{
&& S^4 \ar[r] \ar[d]
& \mathbb{H}\PP^\infty \ar[d] 
\\
 Y  \ar@{-->}[urr] \ar[rr] && K(\mathbb{Z}, 4) 
 \ar@{}[r]|{{\simeq}_{\dim \leq 4}} &
(\mathbb{H}\PP^\infty)_4
\ar@{}[r]|-{{\simeq}_{\dim \leq 15}}
& 
 BE_8.
}
$$
See \cite{SS19a}\cite{SS19b}\cite{MF-theory} for recent developments in that direction. Note that a map $Y \to K(\ZZ, 4)$ defines a cohomology class in $H^4(Y; \ZZ) \cong [Y, K(\ZZ, 4)]$, while $Y \to S^4$ defines a cohomotopy class in $[Y, S^4]$.


\paragraph{Postnikov tower and minimal model for $S^4$.}
The part of the Postnikov tower above can be viewed as a decomposition of $S^4$, and is intimately related to 
a dual decomposition of the corresponding Sullivan minimal model \eqref{S^4} (into successive cofibrations).
Generally, and noting that Postnikov towers can be constructed over any abelian group
including $\RR$ (see e.g. \cite[§ 3.5]{MayPonto}), 
we have the following scheme \cite{GM13} of connections:
$$
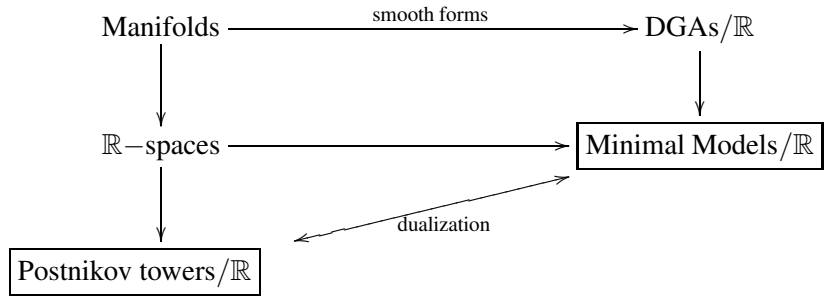
\xymatrix@C=4em{
{\rm Manifolds} \ar[rr]^{\rm smooth \; forms} \ar[d]   && {\rm DGAs}/\RR    \ar[d]  
\\
\RR{\rm -spaces} \ar[d]  \ar[rr] 
&& 
\fbox{$ {\rm Minimal\; Models} /\RR$}
\ar@{<->}[dll]^-{\rm dualization} 
\\
\fbox{${\rm Postnikov \; towers} /\RR$} \phantom{AAA} &&
    }
$$
 The starting point for both is the same. As in the case of Postnikov tower, where one starts with base space $K(\R, 4)$,
one can build the Sullivan model of $S^4$ by starting with the minimal model 
of $K(\R, 4)$, which is $\R[g_4]$. In order to build $M(S^4)$, we
adjoin the above data of $g_7, dg_7,$ and $z_8$ through two maps as follows:

 --   a map $i: \R[z_8] \to   \R [g_7, dg_7]$ taking 
   $z_8$ to $dg_7$. 

   -- a map $f: \R [z_8] \to   \R [ g_4 ]$ taking 
   $z_8$ to $g_4^2$ or a scalar multiple of it, such as $-\tfrac{1}{2} g_4^2$, which is used in this paper for better matching the EOMs.

\noindent These maps form the pushout square 
$$
\xymatrix{
\R[z_8]_{\phantom{A}}  \ar[r]^-f \ar@{^(->}[d]_{i}  \ar@{}[dr]^-{\ulcorner  \phantom{AAAAAAAAAAAAAAAAAAA} }& 
\R[g_4]
\ar[d]
\\
    \R[g_7, dg_7] \ar[r] & M(S^4),
    }
$$
which means we amalgamate the NE and SW pieces into the model in the SE corner in such a way that they agree in the 
NW corner. 
Dually (see \cite{GM13}), we have the first nontrivial Postnikov section  $(S^4_\R)_4 = K(\R, 4)$, whose minimal
model is $\R [g_4]$, and dually to the map $i$
we have the path fibration
$\pi: PK(\R, 8) \to K(\R, 8)$.
We get the following pullback diagram, where $k_2$ is the second $k$-invariant of the Postnikov tower for
$S^4_\R$, capturing the quadratic nature of the EOMs \eqref{M-EOM}
(see \cite{GS21} for the general integral case):
$$
\xymatrix{
& S^4_\R  \ar[r]  \ar[d] 
& 
PK(\R, 8)
\ar@{->>}[d]^\pi
\\
Y \ar[r] \ar@{-->}[ur] & (S^4_\R)_4 \simeq K(\R, 4)    \ar[r]^-{k_2} & K(\R, 8).
    }
$$

\subsection{M-theory gauge structure via the Quillen model of $S^4$}
\label{Sec-Quillen} 

 Quillen's rational homotopy theory allows us to pass back and forth from a simply connected 
 topological space $X$ to the Lie algebra of its loop space
$$
\fbox{${\rm Spaces}/{\rm rational\; homotopy \; equivalance}$}
\;\; \overset{\sim}{\longleftrightarrow} \;\;
\fbox{${\rm DGLAs}/{\rm quasi-}{\rm isomorphism}$}
$$
such that if a space $X$ corresponds to a DGLA $L$, then 
$H^*(L)=\pi_*(X; \mathbb{Q})$ as vector spaces. 
Every reasonably nice space 
\footnote{That is, simply connected spaces with rational homology of finite type.}
admits a minimal free Lie model $(L(v), \partial v=0)$, unique up to isomorphism. 
\footnote{Note that in modern perspectives, one considers $L_\infty$ rather than free Lie.}

\bigskip 
The Sullivan minimal model $M(Z)$ (see \cref{smm}) 
of each of the spaces $Z$ we are
considering has quadratic differential. This model 
is actually the symmetric algebra on a space of
generators:
$$
M(Z) = (S(Q(Z)[-1]^*), d),
$$
where $Q(Z)$ is the \emph{Quillen minimal model 
of} $Z$ \cite{Quillen}, which in our quadratic-differential
 case is given by the \emph{graded Lie algebra of real homotopy groups}
\[
Q(Z) := \pi_\bullet(Z) \otimes \RR[1]
\]
of $Z$.  
 An equivalent, more standard description of the corresponding
Lie bracket on $\pi_\bullet (Z)[1] \otimes \RR \cong \pi_\bullet
(\Omega Z) \otimes \RR$, is known as \emph{Whitehead product}, which does
not appeal to based loop spaces; see, e.g.,
\cite{felix-halperin-thomas}. We will not need it here
(see \cite{FSS-WZW}\cite[\S 3.2]{FSS-Chern} for detailed discussion in this context). 
As indicated in the Introduction, the two models are dual 
$$
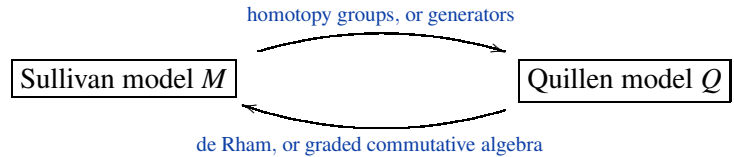
\xymatrix@R=.1em{
 \fbox{\text{Sullivan model $M$}}\qquad \qquad  \ar@/^1.5pc/[r]^-{\color{darkblue} \bf \rm homotopy\;groups, \; or\; generators} 
 & \quad  \qquad 
 \fbox{\text{Quillen model $Q$}}
 \ar@/^1.5pc/[l]^-{\color{darkblue} \bf \rm de \; Rham, \; or\; graded\; commutative\; algebra} 
}
$$

\medskip 
Now we start with the case
$Z = S^4$.
$k =0$.
The following is, in a  sense, dual (see \eqref{SulQuil}) to the description of the fields via the Sullivan 
minimal model of $S^4$ in \eqref{S^4}.

\paragraph{M-theory gauge algebra via the Quillen (free Lie) model of $S^4$.}
\label{Ex-Malgebra} 
The graded Lie algebra $\pi_*(\Omega S^4) \otimes \mathbb{Q}$ is a free graded 
Lie algebra on a generator $v_3$ of degree 3. 
In the tensor algebra $T(v_3)$ on a single degree 3 generator $v_3$, this is the free 
Lie subalgebra $\mathbb{L}(v_3)=\mathbb{R} v_3 \oplus \mathbb{R} [v_3, v_3]$. 
The Sullivan minimal model $M(S^4)$ can be extracted in a direct fashion from this model 
as what is called the cochain algebra $C^*(\mathbb{L}(v_3))$.
This algebra is free in that the differential is zero. 
Alternatively, one can use a priori independent generators, so that the Quillen model of $S^4$ is just the graded Lie algebra on two
generators
\vspace{-2mm} 
\begin{gather}
\label{M-Quil}
  Q(S^4) = \RR e_3 \oplus \RR e_6,\\
  \nonumber 
  \abs{e_3} = 3, \quad \abs{e_6} =
  6,\\ \nonumber 
  [e_3,e_3] = e_6, \quad [e_3,e_6] = 0, \quad [e_6,e_6] = 0,
\end{gather}
which is actually the free graded Lie algebra over $\RR$ generated by $e_3$. 
Remarkably, as highlighted in \cite{SV1}, this captures the algebra of 
gauge transformation of the C-field and its
dual and, hence, also captures the Dirac quantization of the M-branes
(see \cite{CJLP2}\cite{LLPS99}\cite{KS03}\cite{tcu}).

\medskip 
We will later consider the reduction of this algebra,
corresponding to cyclic loop spaces $\mc{L}_c^k S^4$ for $k > 0$
 (see \cref{Sec-IIAsym} and  \cref{symm-cycl-concrete}). 

\subsection{Algebraic topology vs.\ algebraic geometry: real vs.\ complex}
\label{Subsec-RvsC}

Adopting the perspective that $\CC \PP^2$ and del Pezzo surfaces 
$\mathbb{B}_k$ are about the complex world 
while $S^4$ and its cyclifications $\mc{L}_c^k S^4$ are  about the real world,  several aspects fall naturally and compatibly into place. These include

\vspace{-2mm} 
\begin{enumerate}[{\bf (i)}]  
\setlength\itemsep{-2pt}
\item the fact that, generally, M-theory is naturally a real theory; 

\item 
the appearance of real groups (the noncompact forms) as U-duality groups in toroidal reductions of M-theory; 

\item 
the passage from complex algebraic geometry to real homotopy theory as a transition from $\CC \PP^2$ to $S^4$ by modding out by complex conjugation, see below. Similarly, we can transition from  $\CC \PP^1 \times \CC \PP^1$ corresponding to type IIB
to $\CC \PP^2$ corresponding to M-theory (see \Cref{Sec-IIBS}).

\end{enumerate}

\subsubsection{From $\CC \PP^2$ to $S^4$ via complex conjugation} 

We consider a natural way of transitioning from $\CC \PP^2$ to $S^4$ as a possible way of directly relating algebraic geometry and algebraic topology in
Mysterious Triality, see \Cref{MT}.

\paragraph{Complex conjugation automorphism.} 
We intuitively describe how  to utilize the quotient by complex conjugation on the complex manifold 
$\mathbb{C}\PP^2$
being a smooth manifold diffeomorphic to $S^4$, via 
the Arnold-Kuipers-Massey theorem; 
see \cite{Arnold88}\cite{Massey}\cite{Ku74}\cite{Lawson}\linebreak[0]\cite{Letizia}\cite{Marin}\linebreak[0]\cite{AB03}\linebreak[0]\cite{HH11} for various approaches and proofs.

\paragraph{Effect of complex conjugation on middle (co)homology.} 
In transitioning from $\CC \PP^2$ to $S^4$, we would like to ask what happens to cohomology classes.
The group
of homotopy classes of self-homotopy 
equivalences of $\CC \PP^2$ is $\mathbb{Z}_2$ (see e.g. \cite{Baues}), 
generated by complex conjugation $\sigma:  
\mathbb{C}\PP^2 \to \mathbb{C}\PP^2$ which,
in homogeneous coordinates $[z_0:z_1:z_2]$ on $\mathbb{C}\PP^2$, 
takes the form
$$
    \sigma: [z_0:z_1:z_2]  \longmapsto [\overline{z}_0: \overline{z}_1: \overline{z}_2] \;.
$$
This is a diffeomorphism of $\mathbb{C}\PP^2$ which is orientation-preserving, while it is orientation-reversing on $\mathbb{C}\PP^1 \subset \CC \PP^2$ and
therefore reverses the signs of the second (co)homology classes. 
In fact, the action of $\sigma$ on the homology of $\mathbb{C}\PP^2$ is trivial on 
degree 4 classes.\footnote{Generally, trivial on $4k$-dimensional classes and multiplication 
by $-1$ on $(4k+2)$-dimensional classes \cite{Mostovoy}.} 
This is consistent with the discussion we will see in \cref{Sec-cell}.

\paragraph{Identifying the quotient via gluing disks.}
As in 
\cite{Mostovoy},
the conjugation action on $\mathbb{C}\PP^2={\rm Sp}^2(\mathbb{C}\PP^1)$, viewed as the symmetric
square of $\CC \PP^1$,
acts by reflecting pairs of points in $\mathbb{C}\PP^1$ with respect to the real line. Denote by $D_+$ ($D_-$) the subspace of $\mathbb{C}\PP^2$ formed by pairs of points 
$(z_1, z_2)$ such that $z_1$ and $z_2$ lie in the same (different) open hemisphere of $\CC \PP^1$. 
Both $D_{\pm}$  as well as their closures $\overline{D}_{\pm}$ are $\sigma$-invariant.
Let $\mathbb{C}\PP^1\cong \mathbb{D}^2 \cup_{S^1} \mathbb{D}^2$ be the presentation of $\CC \PP^1$ as two closed hemispheres
(or 2-disks $\mathbb{D}^2$)
glued at their intersection circle. 
Notice that $\mathbb{C}\PP^2/\sigma \cong (\overline{D}_+/\sigma) \cup (\overline{D}_-/\sigma)$.
The orbit spaces $\overline{D}_{\pm}/\sigma$ can be identified with the symmetric 
square of $\mathbb{D}^2$:
$$
\overline{D}_+/\sigma = \overline{D}_-/\sigma= {\rm Sp}^2(\mathbb{D}^2) \simeq \mathbb{D}^4,
$$
where the intersection $(\overline{D}_+/\sigma) \cap (\overline{D}_-/\sigma)$ is exactly the 
boundary of $\overline{D}_+/\sigma$ and $\overline{D}_-/\sigma$. Hence $\mathbb{C}\PP^2/\sigma$ 
is homeomorphic to a union of two 4-disks glued together along their boundary, i.e., to 
a 4-sphere
$$
\mathbb{D}^4 \cup_{S^3} \mathbb{D}^4 \simeq S^4.
$$

\subsubsection{Real vs.\ complex Lie algebras/groups} 
\label{Sec-RvsC}

In the spirit of the previous section, here we clarify the role 
of real vs.\ complex forms of Lie algebras and groups appearing 
in our context.

\paragraph{Real forms of Lie groups.}
Real forms of complex reductive algebraic groups arise in two equivalent ways. 
First, since $\CC/\RR$ is a Galois extension, for a complex affine algebraic group
$G$ (see  \cite[\S 11-14]{borel}\cite[\S 11]{Springer}),
one can 
define a real form
using an action of the Galois group on 
$G$. Since the Galois group has only one nontrivial element, a real
form of $G$ is given by a single map,
called \emph{conjugation}, 
$
\sigma: G \to  G
$.
Such a map comes from an antilinear
automorphism of order two of the (complex) Hopf algebra of regular functions on $G$. 
The corresponding \emph{real form} $G_0$ of $G$ will be the fixed-point subgroup of $\sigma$:
\[
G_0 := G^\sigma.
\]
Second, a \emph{real form} of the complex Lie group 
$G(\CC)$ of complex points of $G$ is
the subgroup 
$$
G_0(\RR) := G(\RR, \sigma) := G(\CC)^\sigma 
$$
of fixed points of
a real Lie group automorphism 
$\sigma$ on $G(\CC)$ of order two:
$\sigma: G(\CC) \to G(\CC)$,   $\sigma^2 = \id$,
whose differential at the identity of
$G(\CC)$ is
an 
antilinear Lie-algebra automorphism.
The real form $G_0(\RR)$ is a real Lie group of real dimension equal to the complex dimension of 
$G(\CC)$. 
This repeats the discussion of real forms in the algebraic-group language from above, but now in the
category of analytic manifolds rather than of varieties. 
For reductive algebraic groups, the two categories give rise to exactly the same real forms.

\paragraph{Real forms of Lie algebras.}
A real Lie algebra $\g_0$ is a \emph{real form} of a complex Lie algebra $\g$, if 
 $$
 \mathfrak{g} \cong \g_0 \otimes_\RR \CC = \mathfrak{g}_0 \oplus i \mathfrak{g}_0 \;,
$$
the \emph{complexification} of $\g_0$, in which case 
$
\mathfrak{g}_0 =  (\mathfrak{g})^{\sigma}\;,
$
where $\sigma$ is the complex conjugation on $\mathfrak{g}$.
Every complex reductive Lie algebra has two special real forms: the compact real form and the split one. In this work, split real forms play a prominent role. 
Split real Lie algebras follow their complex counterparts in many 
respects: one is (semi)simple if and only if the complex Lie algebra is, the real rank equals the complex rank, etc. (see \cite{OV}).

\vspace{-2mm} 
\paragraph{Real maximal tori.} 
A complex maximal torus $T \subseteq G$ of a complex linear algebraic group $G$ is defined over 
$\RR$ (with respect to a conjugation $\sigma$ on $G$) if it is preserved by $\sigma$. 
In that case, the subgroup
$$
T_0 :=
T^\sigma = G_0 \cap T \subseteq G_0
$$
is 
called a \emph{$($real$)$ maximal torus} of the real form $G_0$. 
All complex maximal tori of $G$ are conjugate by elements of $G(\CC)$ and isomorphic to $\GG_m^N$ for some $N \ge 0$, 
where $\GG_m := \GL(1)$ is the \emph{multiplicative group}, whose group of complex points $\GG_m(\CC)$ is $\CC^\times$, 
the multiplicative group of nonzero complex numbers. There can be more than one conjugacy class of real maximal tori 
in the real form $G_0$, and in that
case they are not all isomorphic over $\RR$ to each other or to $\GG_m^n$ for any $n$.
Real subgroups of $G_0$ which are isomorphic to $\GG_m^n$ for some $n \ge 0$ over $\RR$ are called \emph{split $($real$)$ tori} 
of $G_0$. A \emph{maximal split $($real$)$ torus} is one which is not contained in a bigger split torus. Maximal split tori 
are always conjugate by elements of $G_0(\RR)$, and the groups of their real points are isomorphic to $\GG_m(\RR)^n = (\RR^\times)^n$. When $G_0$ is a split real form of a reductive linear algebraic group, a maximal split torus will also 
be a maximal torus of $G_0$.
Of course, for us $\mathfrak{g}=\mathfrak{e}_k$, $0 \leq k \leq 8$, as in 
\Cref{table1}.

\paragraph{Example: Maximal tori in SL(2).}
\label{Ex-maxtorSL2}
In the split real form ${\rm SL}(2, \RR)$ of ${\rm SL}(2, \CC)$, defined by the conjugation automorphism $\sigma^s (g) = \overline{g}$ and
appearing in type IIB and in 9 dimensions,
there are two real forms of the maximal torus $\GG_m \hookrightarrow {\rm SL}(2, \CC)$: the split one is the standard diagonal torus $ {\rm diag}(t, t^{-1})   \subset {\rm SL}(2, \RR)$
with real points $\RR^\times$, and the compact one is $\on{SO}(2)\subset {\rm SL}(2, \RR)$ with
real points the (compact) circle $S^1$. The latter in this case coincides with 
the maximal compact subgroup of $\on{SL}(2, \RR)$; however, we are interested in 
the former. See also \cref{Sec-IIBS}.

\subsubsection{Algebraic topology vs. algebraic geometry: Hopf fibrations} 
\label{Subsec-Hopf}

The complex Hopf fibration $h_{\mathbb{C}}$ and quaternionic 
Hopf fibration $h_{\mathbb{H}}$ play  prominent roles here. 
We will see the former  arising from the blowup process (cf. \cite{LS}), hence
as a source for transitioning from the complex side, with 
$\mathbb{C}\PP^2$ appearing as the cone for $h_{\mathbb{C}}$. 
The latter already 
 plays a central role in describing the field
strengths  on   the M-theory real side in  
\cite{Sati13}\cite{FSS-WZW}\cite{FSS19b}\cite{MF-theory}. 

\paragraph{Rank 2 bundles over the point.} 
Over the point $*$, the normal bundle $\nu: \mathbb{C}^2 \to E \xrightarrow{\; \pi\;} \ast$ 
 has a complex structure. 
  Set $E_0=E \mathbin{\vcenter{\hbox{$\scriptscriptstyle\mathrlap{\setminus}{\hspace{.2pt}\setminus}$}}} \{\rm zero\, section\}$ 
    and consider the bundle 
 $\nu_0: \mathbb{C}^2 \mathbin{\vcenter{\hbox{$\scriptscriptstyle\mathrlap{\setminus}{\hspace{.2pt}\setminus}$}}} \{0\} \to E_0 \xrightarrow{\; \pi_0\;} \ast$. 
 Then $\mathbb{C}^*=\mathbb{C} \mathbin{\vcenter{\hbox{$\scriptscriptstyle\mathrlap{\setminus}{\hspace{.2pt}\setminus}$}}} \{0\}$ acts on each fiber in $E_0$ by
 complex multiplication, with orbit space $P_\nu= E_0/\mathbb{C}^*$,
 the projective bundle 
 $\mathbb{C}\PP^1 \overset{i}{\longhookrightarrow} P_\nu \xrightarrow{\pi'} *$.
 These maps fit into a commutative diagram 
 \vspace{-2mm} 
 $$
 \xymatrix@R=.5em{
 E_0 \ar[rr]^-q \ar[dr]_-{\pi_0} && P_\nu \ar[dl]^{\pi'}
 \\
 & \ast 
 }
 $$
 where $q$ is the quotient map. 
  Let $S^3 \to S_\nu \to *$ be the sphere bundle
  and 
  $D_\nu$ be the disk bundle $\mathbb{D}^4 \to D_\nu \to \ast$ 
 as the tubular neighborhood of the point $\ast$. 
  Since $\partial(\mathbb{D}^4)\cong S^3$, then $D_\nu \simeq S_\nu \to \ast$ is a homotopy 
 equivalence. Overall, we have the diagram 
 $$
 \xymatrix@R=1em{
 \mathbb{C}^2 \supset \hspace{-5mm} &    \mathbb{D}^4  & \; S^3 \ar[d] \ar@{_{(}->}[l] && \mathbb{C}\PP^1 \ar[d]^-i  
 \\
 & D_\nu \ar[drr] & \; S_\nu \ar[dr] \ar@{_{(}->}[l] \ar[rr]^-q && P_\nu \ar[dl]
 \\
 &&& \ast &
 }
 $$

 \vspace{-2mm} 
\noindent  We take the complex Hopf fibration as the map of fibers $h_{\mathbb{C}}: S^3 \to \mathbb{C}\PP^1$
 obtained as the restriction of the map $q: S_\nu \to P_\nu$ to the fibers over the point $\ast$.

   \paragraph{The complex Hopf fibration and orientations.}
 Consider the class $a\in H^2(P_\nu ; \mathbb{Z})$ which has fiber restriction 
 \(
 \label{class-a}
 a\in H^2(\mathbb{C}\PP^1 ; \mathbb{Z})
 \)
 given by the negative first Chern class 
 $-c_1(\gamma^1)$, where $\gamma^1$ is the restriction to 
 $\mathbb{C} \PP^1$ of the 
 tautological line bundle over the classifying space $\mathbb{C}\PP^\infty$.
  The fibers $\mathbb{D}^4$ of the disk bundle $D_\nu$ over the point $\ast$ has a canonical 
 orientation given by its complex structure $\mathbb{D}^4 \subset \mathbb{C}^2$ 
 and determines a relative cohomology class
 $
 u_{\mathbb{D}^4} \in H^4(\mathbb{D}^4, S^3; \mathbb{Z})
 $. Through the connecting homomorphism
 $\delta: H^3(S^4; \mathbb{Z}) \xrightarrow{\; \cong\;}H^4(\mathbb{D}^4, S^3; \mathbb{Z})$, this 
 determines an obstruction class 
 \(
 \label{class-u}
 u_{S^3} \in H^3(S^3; \mathbb{Z})\;.
 \)

 \paragraph{Extension to $\mathbb{C}\PP^2$.} 
 The Hopf fibration can be extended to a map $\widetilde{h}_{\mathbb{C}}: 
 \mathbb{D}^4 \to \mathbb{C}\PP^2$. For $z=(z_1, z_2) \in \mathbb{C}^2$, 
 set $Z=\big(|z_1|^2 + |z_2|^2 \big)^{1/2}$, so that the 4-disk and the 
 corresponding sphere are defined, respectively, as
 $$
 \mathbb{D}^4=\big\{z\in \mathbb{C}^2\;|\; Z \leq 1 \big\}
 \qquad 
 \text{and}
 \qquad 
 S^3=\big\{z\in \mathbb{C}^2\;|\; Z = 1 \big\}.
 $$
Now define the map
$   \widetilde{h}_{\mathbb{C}}: \mathbb{D}^4  \to \mathbb{C}\PP^2
   $ by $
    z \mapsto [z_1: z_2: 1-Z]$.
Then for the restriction we have $\widetilde{h}_{\mathbb{C}}\vert_{S^3} =h_{\mathbb{C}}$, where $\mathbb{C}\PP^1$
is considered as the hyperplane with last coordinate 0, and the two maps are related via the diagram 
$$
\xymatrix@R=1em{
 \mathbb{D}^4 \ar[rr]^{\widetilde{h}_{\mathbb{C}}} && \mathbb{C}\PP^2
\\
 S^3 \ar@{^{(}->}[u] \ar[rr]^{h_{\mathbb{C}}} && \mathbb{C}\PP^1 \ar@{^{(}->}[u]
}
$$
By a cofibrant replacement, as in \cite{LS}, the restriction is a cofibration 
$S^3 \to \mathbb{C}\PP^1$. We will use this 
further below in \eqref{cofib-seq}.

\paragraph{The Sullivan model and the generators.} The Hopf fibration 
$h_{\mathbb{C}}$ gives rise to the Sullivan models and projection maps
between them 
\vspace{-2mm}
$$
\xymatrix{
 \big(  \bigwedge (x_2, y_3); dx_2=0, dy_3=x_2^2  \big) \ar[d]^g \ar[r]^-{h_{\mathbb{C}}^*} &
\big( \bigwedge (y_3); dy_3=0 \big) \ar[d]^f
 \\
     M(\mathbb{C}\PP^1)  \ar[r] & M(S^3)
}
$$    
such that the corresponding classes coincide with the classes \eqref{class-a} and 
\eqref{class-u} above, namely 
$$[g(x_2)]=a
\qquad 
\text{and}
\qquad [f(y_3)]=-u_{S^3}\;.
$$

\subsubsection*{The transition via cellular structure}
\label{Sec-cell} 

We describe the transition from $\CC \PP^2$ to $S^4$ via three further related approaches.

\paragraph{(i) The transition via CW-structure.} 
Consider $(\mathbb{C}\PP^2, \mathbb{C}\PP^1)$ as a CW-pair by viewing 
$\mathbb{C}\PP^1 \subset \mathbb{C}\PP^2$ as a CW-subcomplex. 
The cell structure of $\mathbb{C}\PP^2$ is $e^0 \cup e^2 \cup e^4$ 
while that of $\mathbb{C}\PP^1$ is $e^0 \cup e^2$. 
The quotient space $\mathbb{C}\PP^2/\mathbb{C}\PP^1$, obtained by collapsing 
the subspace $\mathbb{C}\PP^1$, inherits a natural cell 
complex structure from $\mathbb{C}\PP^2$
and  is a CW-complex 
whose cells are the cells of the 
difference $\mathbb{C}\PP^2 \mathbin{\vcenter{\hbox{$\scriptscriptstyle\mathrlap{\setminus}{\hspace{.2pt}\setminus}$}}} \mathbb{C}\PP^1$ plus one new 0-cell,
the image of $\mathbb{C}\PP^1$ in $\mathbb{C}\PP^2/\mathbb{C}\PP^1$. 
With a single 0-cell 
and the original 4-cell from $\mathbb{C}\PP^2$, i.e., $e^0 \cup e^4$,
 the quotient map 
 \(
 \label{map-q}
 q: \mathbb{C}\PP^2 \longrightarrow S^4
 \)
 gives a 
homeomorphism $\mathbb{C}\PP^2/\mathbb{C}\PP^1 \simeq S^4$.

\paragraph{(ii) The transition via cofibrations.} The above can also 
be described starting from the complex Hopf fibration 
$S^1 \to S^3 \xrightarrow{h_\mathbb{C}} \mathbb{C}\PP^1$. Associated to this is 
a cofibration sequence 
\footnote{
This is the Barrat-Puppe sequence associated to a map $f: A \to Z$
$$
A \xrightarrow{f} X \xrightarrow{g} C_f \xrightarrow{\delta}
\Sigma A \xrightarrow{\Sigma f} \Sigma X \xrightarrow{\Sigma g \;}
\Sigma C_f \longrightarrow \cdots
$$
where $C_\phi$ is the mapping cone of the map $\phi$ and $\delta$ is the 
connecting homomorphism. 
}
\(
\label{cofib-seq}
\xymatrix{
S^3 \ar[r]^-{h_\mathbb{C}} & \mathbb{C}\PP^1 
\ar[r]^-{i_\mathbb{C}} & \mathbb{C}\PP^2
\ar[r]^-q & \Sigma S^3 \ar[r]^-{\Sigma h_\mathbb{C}\;} & 
\Sigma \mathbb{C}\PP^1 \ar[r]^-{\Sigma i_\mathbb{C}\;} & \Sigma \mathbb{C}\PP^2
\ar[r] & \cdots 
}.
\)
Any 3-term sequence in \eqref{cofib-seq} is a cofibration so, using 
the suspension $\Sigma S^3 \simeq S^4$, we extract 
\begin{equation}
\label{Eq-cofib}
\xymatrix{
 \mathbb{C}\PP^1 
\ar[r]^-{i_\mathbb{C}} & \mathbb{C}\PP^2
\ar[r]^-q &  S^4 ,
}
\end{equation} 
which exhibits $S^4$ the cofiber of $i_\mathbb{C}$.
This indicates that the origin of the transition  indeed has to do with the complex
Hopf fibration  $h_\mathbb{C}$.

\paragraph{ (iii) The transition via homogeneous coordinates.} 
Consider the usual embedding   $\mathbb{C}\PP^1 \hookrightarrow \mathbb{C}\PP^2$
given by the mapping of homogeneous coordinates 
$[z_0:z_1] \mapsto [z_0:z_1:z_2]$. Let $Z_0=\frac{z_0}{z_2}$ 
and $Z_1=\frac{z_1}{z_2}$, assuming $z_0 \neq 0$, and let 
$Z^2=|z_0|^2 + |z_1|^2$. Then the map $f: \mathbb{C}\PP^2 \to S^4$ is given by 
\vspace{-2mm} 
$$
[z_0:z_1:z_2] \mapsto 
\left\{
\begin{array}{cc}
\left( \frac{2Z_0}{1+Z^2}, \frac{2Z_1}{1+Z^2}, \frac{-1 +Z^2}{1+Z^2}  \right)    
&  z_2 \neq 0,
\\
(0, 0, 1)     & z_0=0.
\end{array}
\right.
$$
The case when $z_0=0$ corresponds exactly to a point being in 
$\mathbb{C}\PP^2$. Hence $f$ is constant on  $\mathbb{C}\PP^1$, 
and so gives
rise to a quotient map $\overline{f}:\mathbb{C}\PP^2/\mathbb{C}\PP^1 \to S^4$,
which is a homeomorphism.

\paragraph{Cohomotopy.}
From the map $q:\mathbb{C}\PP^2 \to S^4$ arising from collapsing 
$\mathbb{C}\PP^1$ to a point, the homotopy classes of maps are given 
by degree 4 Cohomotopy 
$
\pi^4(\mathbb{C}\PP^2) \cong \mathbb{Z}\;,
$
 with generator the class of the map $[q]$. 
  Here $\pi^4(X)=[X, S^4]$ is the base-point preserving homotopy classes of maps from $X$ to $S^4$;  see \cite{We71}\cite{GS21}. Since the two spaces $\mathbb{C}\PP^2$ and $S^4$ are of the same 
  (real) dimension, this is of course given by the usual degree.
 
\subsection*{(Co)homology vs. (Co)homotopy} 

We consider the transition from (co)homology on the del Pezzo side to 
(co)homotopy on the $S^4$ side. Additionally, and to further 
highlight the extent to which  the approach is distinguished, 
we explain why neither homotopy for $\CC \PP^2$ nor
 (co)homology for $S^4$ would work.

\paragraph{(Co)homology route for $S^4$?} Since the only nonzero (co)homology groups 
of $S^4$ are in degree 0 and 4, there are no higher (co)homology groups 
for $S^4$ and, in particular, none in degree 7 needed 
to account for the dual field $G_7$. Hence this is clearly not a viable 
route and, in fact, was among several considerations for proposing cohomotopy in 
\cite{Sati13}.

\paragraph{Homotopy route for $\CC \PP^2$?} 
From the long exact sequence on homotopy groups arising from the 
fibration $S^1 \to S^5 \to \mathbb{C}\PP^2$, 
$$
\cdots \longrightarrow \pi_k(S^1)  \longrightarrow \pi_k(S^5) 
\longrightarrow \pi_k(\mathbb{C}\PP^2) \longrightarrow \pi_{k-1}(S^1) 
\longrightarrow \cdots \;,
$$
we get 
$\pi_4(\mathbb{C}\PP^2)=0$, which does not lead to a degree four
as $\pi_4(S^4)$ would. 
We use the rational equivalence $\Omega S^5 \simeq_{\mathbb{Q}} S^4$ (or better the 
suspension $S^5\simeq \Sigma S^4$, nonrationally) to get 
$$
\pi_5(\mathbb{C}\PP^2) \cong_{\mathbb{Q}} \pi_4(S^4) \cong_{\mathbb{Q}} \mathbb{Q}\;.
$$
One could say that we perhaps continue with higher homotopy groups of 
$\mathbb{C}\PP^2$. For $k>5$ it is still true that $\pi_k(\mathbb{C}\PP^2) \cong \pi_k(S^5)$,
but all these groups are going to be pure torsion with no clear interpretation 
and with no sign of the dual field or dynamics arising. 
Hence this shows that the homotopy route for $\mathbb{C}\PP^2$ is not the way to go. 
An analogous statement holds, and even more so, for del Pezzo surfaces, which have quite
complicated homotopy groups with large unwieldy growth.

\paragraph{Relative homotopy.} However, 
from the complex Hopf fibration $h_{\CC}$, we get an isomorphism 
$$
\pi_n(\mathbb{C}\PP^2, \mathbb{C}\PP^1) \cong \pi_{n-1}(S^3)\;.
$$
The relative homotopy group on the left is defined so that
corresponding homotopy classes of maps are trivial on the 
$\mathbb{C}\PP^1$ factor which are are eventually contracting. 
For $n=4$, we get $\pi_4$ to appear naturally via $\pi_4(\mathbb{C}\PP^2, \mathbb{C}\PP^1)$.
We also highlight that in the process the group $\pi_2(\mathbb{C}\PP^2, \mathbb{C}\PP^1)$
is killed (note that $\pi_2(\mathbb{C}\PP^2)\cong \mathbb{Z})$) since
$S^3$ is simply connected.

\medskip
This now allows us to transition to $S^4$.

\paragraph{Common ground via Hurewicz.}
We have seen indications that the  map on homology $H_4(\mathbb{C}\PP^2) \to H_4(S^4)$ is nontrivial. 
Indeed, the map $q$ from \eqref{map-q} induces an isomorphism on $H_4$, so it is homotopically 
nontrivial. 
Aside from degree 0, homology only makes sense in top degree, which matches the lowest 
degree occurring in homotopy, where they match by the Hurewicz isomorphism 
$$
\pi_4(S^4) \cong H_4(S^4) \cong \mathbb{Z}\;,
$$
signalling the appearance of the generator $g_4$ (universally) and the  
 field $G_4$ (in spcetime). Hence, overall we have the following picture:
\[
\label{eq-Hur}
\xymatrix{ 
\protect\underbrace{\mathbb{C}\PP^2}_{\rm \bf \color{darkblue} (co)homology} 
\ar@{~>}[rrrr]^{{\pi_4-, \, H_4-{\rm isomorphism}}}_{{\rm `common\; ground'}}
&&&&
\protect\underbrace{S^4}_{\rm \bf \color{darkblue} (co)homotopy}
}
\]

\paragraph{Dynamics and the dual via higher homotopy.} 
In order to see the dual and the dynamics, we need higher homotopy 
groups; in particular, (rationally) $\pi_7(S^4)\cong \mathbb{Q}$. 
So this should be a new effect in homotopy and a feature that 
does not arise in (co)homology. Indeed, this confirms the viability of 
the proposal for using 
$S^4$ which achieves exactly that.

\medskip
The above also indicates that it is quite natural to use $S^4$ to extend, 
and to some extent explain, Mysterious Duality to include fields via the Triality
(Figure \ref{MT}).

\subsection{Trombone symmetry of M-theory in 11 dimensions via $S^4$}
\label{Sec-TorSymm} 

Here we describe the toroidal symmetry of $S^4$ and provide 
a topological interpretation.

\paragraph{Trombone symmetries in M-theory.} 
\label{trombone}
We will employ scaling symmetries of the fields in the theory, 
which we interpret topologically below
in terms of winding numbers (self-maps) of spheres in appropriate dimensions.
This novel perspective should be interesting in its own right, but it also allows us to  
study Mysterious Duality from  a topological viewpoint.

\medskip 
There is a rigid scaling symmetry, called \emph{trombone symmetry}, of 11-dimensional supergravity 
which preserves the ratio of the coefficients 
of the Einstein-Hilbert and gauge-field kinetic terms and also keep the coefficient of the 
Chern-Simons term in the same ratio.  This is the $\R^+$ symmetry 
\cite{DNP}\cite{CDF}\cite{CLPS}
on the metric $g$ and the C-field $C_3$ 
\(
\label{scale-metric} 
g \mapsto \lambda^2 g\;, \qquad C_3 \mapsto \lambda^3 C_3\;. 
\) 

\vspace{-3mm} 
\begin{itemize} 
 \setlength\itemsep{-2pt}
\item  The Lagrangian scales homogeneously as $L \mapsto \lambda^9 L$. As this
$\RR^+$ symmetry leaves 
invariant the EOMs, it is appropriate that we use them in describing the dynamics.
 
\item 
This can alternatively be viewed as an engineering scale invariance of the classical equations, 
as a consequence of the fact that there is just one overall dimensionful coupling constant, 
which sits in front of the entire eleven-dimensional action.

\item 
While this symmetry is not a symmetry of
the Cremmer-Julia type, it plays a crucial role in understanding the
occurrence of the Cremmer-Julia $E_{k(+k)}$ symmetries in the lower dimensional theories \cite{CJLP}.

\item
Generally, it acts on any $p$-form as multiplication by $\lambda^p$. 
 So for us, considering the duality-symmetric formulation, where we
 treat the C-field and its dual as independent variables (giving up
 metric and Hodge duality relations), 
$$
C_3 \mapsto \lambda^3\, C_3\;, 
\qquad 
C_6 \mapsto \lambda^6 \, C_6 \;.
$$
Instead of using this action, which is associated with the anticanonical class $-K_k$ in our context,
we will, however, use a simpler scaling arising from the action of tori in \cref{Sec-TorSymm} :
\(
\label{scale-C3C6}
C_3 \mapsto t \, C_3\;, 
\qquad 
C_6 \mapsto t^2 \, C_6 \;.
\)
The corresponding scalings for field strengths $G_4$ and $G_7$ are
of the same form.

\end{itemize} 

\paragraph{Trombone symmetries via weights associated to the 4-sphere.}
In \cite{SV1}, we studied real toroidal symmetries of minimal Sullivan
algebras $M$, i.e., diagonalizable actions $T \to \Aut M$ of a
\emph{real split torus}, an affine algebraic group $T$ isomorphic over
$\RR$ to the group $\GG_m= \GL(1)$, the multiplicative group (see \cref{Sec-RvsC}).
Here $\Aut M$ is the group of automorphisms of $M$ as a DGCA, 
and by an \emph{action} $T \to \Aut M$ we mean
a morphism of algebraic groups defined over $\RR$.
Toroidal symmetries of $M$ can be captured by considering a \emph{maximal $\RR$-split torus} $T
\subseteq \Aut M$. 
M-theory in 11 dimensions will correspond to rank 1,
while later (in \cref{Sec-IIAsym} and  \cref{symm-cycl-concrete}) 
we will see that higher ranks are obtained via dimensional reduction.

We start with the automorphism group $\Aut M(S^4)$ of the Sullivan minimal model $M(S^4)$
of $S^4$; see \eqref{S^4}. By the degree argument,
an automorphism of $M(S^4)$ must take $g_4$ to a scalar multiple of itself:

\vspace{-3mm} 
\begin{equation}
  \label{rg_4}
g_4 \mapsto tg_4
\end{equation}

\vspace{-1mm} 
\noindent for some $t \in \GG_m(\RR)$, and this determines the action of the
automorphism on $g_7$:
\vspace{-1mm} 
\begin{equation*}
  g_7 \mapsto t^2 g_7,
\end{equation*}

\vspace{-1mm} 
\noindent
and thereby on the whole DGCA $M(S^4)$. This gives an identification
$\Aut M(S^4) \cong \GG_m$ over $\RR$, and, therefore, $\Aut M(S^4)$
automatically coincides with its maximal split torus. Note that this
identification is unique up to automorphism of $\GG_m$, which could
only be $t \mapsto t^{-1}$ if not trivial. Thus, we get a {\it weight
decomposition} determined by
\begin{equation}
  \label{S4weights}
g_4 \in M(S^4)_{\epsilon_0}, \qquad g_7 \in M(S^4)_{2 \epsilon_0},
\end{equation}
with the weights defined up to common sign, that is to say,
\vspace{-2mm} 
\begin{equation}
  \label{S4weight-eps}
t^{\epsilon_0} = t
\end{equation}

\vspace{-2mm} 
\noindent
or $t^{\epsilon_0} = t^{-1}$.
Again, the weights are just determined by the weight of $g_4$, as
that is the only $d$-closed generator.  We can always normalize this
ambiguity so as to have positive weights and assume \eqref{S4weight-eps} is valid. This choice also has topological motivation, as we will see below. We will combine weights in 
\cref{subsec-EMdual}.

\medskip 
Now we will present an interpretation of the toroidal symmetries
as rational (real) homotopy equivalences, see detailed construction
in \cite{SV1}. These symmetries have physical
interpretation of trombone symmetries above.

\paragraph{Topological interpretation of toroidal symmetries coming from $S^4$.}
 \label{natural0}
Let us start with the symmetry \eqref{rg_4}. The $n$-folding map 
$$
\varphi_0(n): S^4 \longrightarrow S^4
$$ 
(actually, any continuous map of degree $n$) acts on the homotopy group $\pi_4(S^4)$ as multiplication by $n$. Over the rationals and reals, 
the map $\varphi_0(n)$ has a rational homotopy inverse, which acts on $\pi_4(S^4)$ as multiplication by $1/n$, provided $n \ge 1$. 
Combining this with an $m$-folding self-map of the sphere, we will get a map (morphism, to be precise) $S^4 \to S^4$ in rational homotopy theory, 
which acts on $\pi_4(S^4)$ as multiplication by $m/n$. This map gives an action of the multiplicative group of the rationals $\QQ$ on $S^4$ in 
rational homotopy theory, which is equivalent to an action on the Sullivan minimal model $M(S^4)$. This action extends by continuity to 
an action of the multiplicative group $\RR^\times$ of the reals. A bit more generally, we can view this as an action of the real algebraic 
group $\GG_m$ on $M(S^4)$:
\(
\label{GmonS4}
\GG_m \times M(S^4) \longrightarrow   M(S^4),
\)
expressed explicitly by Equations \eqref{rg_4}--\eqref{S4weight-eps}.

\paragraph{The Lie algebra.} 
As in \cite{SV1}, we can use the toroidal symmetries of  $S^4$ from above
and consider it at the Lie algebra level, by taking derivations. 
The idea is to use the Lie algebra $\h_0 = \Lie
(T^{1})$ of the maximal real split torus $T^{1}$ of $\Aut M(
 S^4)$ with its canonical factorization $T
\xrightarrow{\sim} \GG_m$. This Lie algebra constitutes the
infinitesimal symmetries corresponding to the toroidal symmetries of
$S^4$ and acts on the
Sullivan minimal model $ M(S^4)$ by derivations.
That is, we have a Lie algebra homomorphism:
\vspace{-2mm} 
\[
\h_0 \longrightarrow \Der M(S^4),
\]
which comes from taking the differential of the action
\[
T^{1}\simeq \GG_m \longrightarrow \Aut M(S^4).
\]

\medskip 
$$
\xymatrix@C=8em{
{\color{darkblue} \footnotesize \rm   Group}  \;\;  T^{1} \ \rotatebox[origin=c]{-90}{$\circlearrowright$}~~
\;\;
\fbox{$\Aut M(
 S^4)$}
\ar[r]^-{\rm differential \; of \; map}_-{\rm linearization}
& 
\fbox{$\Der M( S^4)$ }
\;\;\; 
\ \rotatebox[origin=c]{-270}{$\circlearrowright$}~~
\h_0 = \Lie
(T^{1}) \;\;
{\color{darkblue} \rm Algebra} 
}
$$ 

\smallskip 
Now in order to proceed by extracting root data from topology, we will recall the 
analogous situation of the corresponding del Pezzo surface. 
 
\paragraph{Complex Projective space.}
Consider the \emph{del Pezzo surface} $\mathbb{B}_0$ which is just  
the complex projective plane $\mathbb{C P}^2$. 
The Picard group ${\rm Pic}(\mathbb{C P}^2)$ of line bundles is isomorphic to 
the divisor class group and the 2nd cohomology group: 
$\Pic (\mathbb{C P}^2) \cong H^2(\mathbb{C P}^2;\ZZ)$. This is a rank-$1$ lattice with generator $\cH$:
\vspace{-1mm} 
$$
H^2(\mathbb{C P}^2; \Z) \cong \Z \cH 
$$
where $\cH$ is the class of the proper transform of the line
(here also a hyperplane) $\CC \PP^1$ in $\mathbb{C P}^2$. 
The 2nd integral cohomology has a natural inner product
given by the intersection form:
\(
\cH \cdot \cH=1.
\)
$\mathbb{C P}^2$ has the \emph{canonical class} 
$
\cK_0 := \Omega^2_{\mathbb{C P}^2} = - 3 \cH 
$
as well as the \emph{anticanonical class}:
$-\cK_0 = 3 \cH$.
The \emph{degree of the surface} $\mathbb{C P}^2$ is the self-intersection number $(-\cK_0) \cdot (-\cK_0)= 9$. 
The \emph{degree of a divisor $\cD \in \Pic (\mathbb{C P}^2)$} is measured with respect to the anticanonical map.
The
anticanonical class $-\cK_k$  ``acts'' on the
Picard group (degree-two cohomology) $\Pic (\mathbb{C P}^2) \cong H^2 (\mathbb{C P}^2;\Z)$ by degree:
\begin{equation}
  \label{degree-k=0}
\deg \cD := - \cK_0 \cdot \cD, \qquad \cD \in \Pic (\mathbb{C P}^2).
\end{equation}
Here the ``action'' is understood as the intersection product $\Pic
(\mathbb{C P}^2) \otimes \Pic (\mathbb{C P}^2) \to \Z$. The \emph{degree of a divisor}
$\cD \in \Pic (\mathbb{C P}^2)$ is defined using the anticanonical morphism
$f: \BB_0 \to \CC \PP^d$, and 
$$
d = h^0 (\mathbb{C P}^2, -\cK_0) - 1 = (-\cK_0) \linebreak[0] \cdot (-\cK_0) =
9 
$$ 
is known as the \emph{degree of the surface $\BB_0=\mathbb{C P}^2$},
whence $-\cK_0$ is the pullback $f^* \mc{H}$ of the hyperplane class $\mc{H} \in  \Pic ( \CC \PP^d)$ and formula \eqref{degree-k=0} 
makes sense; see, e.g., \cite{Dolgachev}
and also \cite{SV1}.

\paragraph{Degree in Quillen model of $S^4$.}
In the case of M-theory we have a 
notion of degree, natural for the
Quillen minimal model. The significance of this is coming from the fact that it corresponds to the degree of
the C-fields, i.e., the potentials $C_3$ and $C_6$ of the basic
fields, such as $G_4$ and $G_7$, see \eqref{C-fields} and
\eqref{scale-C3C6}, and thereby to the dimension of the corresponding
branes, the M2- and M5-branes, respectively. This other notion of degree
just differs from the degree we have been using on the generators of
the Sullivan minimal model by one, but has a homotopy-theoretic
origin, as we  now explain.

\medskip 
We have seen the Quillen model in \cref{Sec-Quillen}. The maximal split torus $T^{1}$ and its Lie algebra $\h_0$ act on the
Quillen minimal model $Q(S^4)$ with the same weights as on
the generators of the Sullivan minimal model $M = M(S^4)$. 
A remarkable fact is that the \emph{degree operator}
\begin{equation*}
x \mapsto \abs{x} \cdot x, \qquad x \in Q(S^4),
\end{equation*}
in the Quillen minimal model singles out a distinguished element of the Lie
algebra $\h_0$. 
Analyzing the degrees shows \cite{SV1} 
that there is a unique element of the Lie algebra $ \h_k$, namely,
\begin{equation}
  \label{elt0}
-K_0 := 3h_0 
\end{equation}
which acts on the Quillen minimal model $ Q(S^4)$ as the
degree operator:
\[
-K_0  x = \abs{x} \cdot x.
\]

Since the anticanonical class $- \cK_0$ of  $\mathbb{C P}^2$
acts on the Picard group by degree, it makes
all sense to use the element
$
-K_0 = 3h_0 
$
as a distinguished element, the analogue of the anticanonical class. 
Indeed, using \eqref{M-Quil}, we have 
$$
-K_0  v_3=3 h_0  v_3=|v_3| \cdot v_3
$$

\subsection{M-theory and del Pezzo cohomology classes} 
\label{subsec-EMdual} 

Here we provide an extension of \cite{INV} to include field 
strengths $G_4$ and $G_7$ and show how (combinations of) these correspond to  classes on the del Pezzo side.

\paragraph{Electric-magnetic/Hodge duality.}
This duality in M-theory was interpreted in \cite{INV} as corresponding to classes 
adding up to the anticanonical class
$
\mathcal{C}_{\rm Electric} + \mathcal{C}_{\rm Magnetic} = - \cK = 3 \cH$on $\CC \PP^2$.
We interpret this from the perspective of field strengths. 
We naturally take $\mc{C}_{E} := \mathcal{C}_{\rm Electric} = \cH$ to correspond to the field 
$G_4$ and  $\mc{C}_M := \mathcal{C}_{\rm Magnetic} = 2\cH$ to its dual $G_7$.
This also works universally for the elements 
$g_4$ and $g_7$ in the minimal model $M(S^4)$.

\begin{center}
\begin{tabular}{ccccc}
\hline 
\bf Degree & \bf del Pezzo class & \bf BPS state & \bf Form field & \textbf{Generator of $M(S^4)$}\\
\hline 
\hline 
\rowcolor{lightgray}  4 & $\cH$ & M2 & $G_4$ & $g_4$\\
7 & $2\cH$ & M5 & $G_7$ & $g_7$\\
\hline 
\end{tabular}
\end{center}
There are two interpretations, one via the Sullivan model and another via
the Quillen model. We will focus on the former, leaving the latter, which involves  
Whitehead products, for a separate treatment elsewhere
(as this requires a longer discussion). 

\paragraph{The anticanonical class.} 
In the Sullivan perspective, we take $-\cK$ to correspond to the Chern-Simons form 
$CS_{11} := C_3 \wedge G_4 \wedge G_4$ when using only $G_4$, or the composite 
$G_4 \wedge G_7$ when using both fields, $G_4$ and its dual $G_7$. 

\smallskip 
\begin{center} 
\renewcommand{\arraystretch}{1.2} 
\begin{tabular}{ccc}
\hline 
{\bf Entity} & {\bf Field interpretation} & \textbf{RHT interpretation}
\\
\hline 
\hline 
\rowcolor{lightgray}
$-\cK$ & Chern-Simons form $C_3 \wedge G_4 \wedge G_4$ & products $g_4^3$ and $g_4 g_7 \in M(S^4)$
\\
$+$ & wedge product $\wedge$ & product in $M(S^4)$ 
\\
\rowcolor{lightgray}
$-$
& 
division/contraction $\iota$ & division in $M(S^4)$
\\
\hline 
\end{tabular} 
\end{center}

\paragraph{Division/Contraction}
Let us explain the last row in the table above. This feature works in the simplest way on our universal model, the minimal Sullivan algebra $M(S^4) = (\RR[g_4, g_7], d)$. 
With the weight space decomposition \eqref{S4weights}, the weight space $M(S^4)_{3 \epsilon_0}$ of the distinguished weight $3 \epsilon_0$, which corresponds to the anticanonical class $-\cK = 3 \cH$, contains the element $g_4 g_7$:
\[
g_4 g_7 \in M(S^4)_{3 \epsilon_0}
\; .
\]
The weight $2 \epsilon_0$
corresponding to the class $-2 \cH = -\cK - \cH$,
contains the element $g_7$:
\[
g_7 \in M(S^4)_{2 \epsilon_0}
\; .
\]
The generator $g_4$ is a non-zero-divisor in the algebra $M(S^4)$, and we can interpret the subtraction $-\cK - \cH$ as division by $g_4$:
\[
g_4 g_7  \longmapsto g_4 g_7/ g_4 = g_7 \; .
\]
At the level of spacetime fields $G_4$ and $G_7$, which are actually not non-zero-divisors in the de Rham algebra $\Omega^\bullet (Y)$ of the spacetime $Y$, this correspondence could be reincarnated
as the contraction/interior product 
\[
`` -\iota_{\cH^\vee} \cK \,
" = \iota_{G_4^\vee} G_4 \wedge G_7
 = G_7 \; .
 \]
Here
we admit some frivolity in taking the 
interior product $\iota$ with a multi-vector $G_4^\vee$, some sort of dual of $G_4$.

\medskip 
Note a relation of $G_7$ to the degree-eight \emph{electric current form} $j_E := -2 d G_7 = G_4 \wedge G_4$, which may be obtained by mimicking the Lie derivative of $-\cK$ in the direction $\cH^\vee$:
\begin{align*}
`` -\mathbb{L}_{\cH^\vee} \cK = -\mathbb{L}_{G_4^\vee}\cK \,
" & = (d \iota_{G_4^\vee} + \iota_{G_4^\vee} d) C_3 \wedge G_4 \wedge G_4
\\
& = d \iota_{G_4^\vee} C_3 \wedge G_4 \wedge G_4\\
& = -2d G_7\\
& = j_E\; .
\end{align*}

\paragraph{Serre duality.} 
In relation to Serre duality $- \mathcal{C} \mapsto \cK + \mathcal{C}$,
we have (cf.\ \cite{INV} without fields):
:
\begin{itemize}
\setlength\itemsep{-2pt}
\item {\it Electric-to-magnetic}: The transformation $\cH \mapsto - \cK - \cH$ is 
manifested universally by
\[
g_4 \longmapsto g_4 g_7 /g_4 = g_7\;,
\]
and at the level of field strengths by 
\vspace{-2mm} 
$$
 G_4 \longmapsto \iota_{G_4^\vee}(G_4 \wedge G_7)=G_7\;.
$$
\item {\it Magnetic-to-electric}: Likewise, the transformation $2\cH \mapsto -\cK - 2\cH$ is the adjoint 
transformation 
$$
G_7 \longmapsto \iota_{G_7^\vee}(G_4 \wedge G_7)=G_4\;.
$$
\end{itemize}

\section{Dimensionally reduced theories and `root data' via iterated cyclic loop spaces}
\label{Sec-CyclicSugra}

\paragraph{M-theory on $T^k , k \leq  8$ and higher cyclic loop spaces.}
The reduction of the M-theory system on tori $T^k = (S^1)^k$ leads to a low-dimensional
system corresponding to duality-symmetric supergravity actions in
these dimensions matching the EOMs of M-theory
compactified on a $k$-dimensional torus $T^k$ in an iterative way \cite{SV1}. 
This arises via the iterated cyclic loop spaces
$\mc{L}_c^k S^4$ of the 4-sphere $S^4$, both for $0 \le k \le 8$ and for $k \geq 9$.
For $k=1$ this is studied extensively in \cite{FSS-pbranes}\cite{FSS-L00}
in relation to T-duality and the twisted K-theory description of the fields
in type II string theory (cf. \cref{IIB}). The generalization
is possible thanks to the general construction in \cite{BSS}.

\subsection{Type IIA in 10 dimensions and the reduction on $S^1$}
\label{Sec-IIAsym} 

\paragraph{The reduction to type IIA.}
\label{Sullmm}

The reduction from M-theory in 11 dimensions to type IIA string theory in 10 dimensions is 
captured by looping. Such a process has been utilized topologically 
at the level of bundles  
leading to loop bundles in 10 dimensions starting from an $E_8$ gauge bundle
(capturing the purely topological aspects of $G_4$) in 11 dimensions \cite{MaS}. 
In our current case, one is to do this at the level of universal spaces, and 
 taking into account the rotation of the circle. This leads to the concept of 
 a cyclic loop space or cyclification, advocated in \cite{FSS-L00}
$$
\mc{L}_c S^4: = \mc{L} S^4 \dslash S^1.
$$
The Sullivan minimal model of the cyclic loop space 
$M(\mc{L}_c Z)$ in terms of the Sullivan minimal model $M(Z)$
of a simply connected (or at least nilpotent) space $Z$ is identified in \cite{SV1}. For $Z=S^4$, 
this gives  a duality-symmetric reduction of fields 
in M-theory on tori, extending and organizing partial/local results in the 
supergravity and M-theory literature.

 \medskip 
Starting with $M(S^4)$ in \eqref{S^4}, the Sullivan minimal model $ M(\mc{L}_c S^4)$ of the cyclic loop
space of $S^4$ is presented in \cite{SV1} (cf. \cite[Ex. 3.3]{FSS17}\cite[Ex. 2.7]{FSS-L00})
\[
M(\mathcal{L}_cS^4)=\big(\mathbb{R}[f_2,f_4,f_6,h_3,h_7],\ df_2=0,\, dh_3=0,\, df_4=h_3f_2,\, df_6=h_3f_4,\,
dh_7= - \tfrac{1}{2} {f_4}^2 + f_2f_6\big).
\]
Being defined for $\mathcal{L}_cS^4$, these equations are universal, and we obtain the corresponding ones in
spacetime by pulling back, 
giving the datum of a closed 3-form $H_3$ and of 2-, 4- and 6-forms $F_2$, $F_4$ and $F_6$ on $X$ such that
\[
dF_2=0;\qquad dF_4=H_3\wedge F_2;\qquad dF_6= H_3\wedge F_4,
\]
together with a 7-form $H_7$, which is a potential for a certain
8-form:
\[
d H_7 = - \tfrac{1}{2} F_4\wedge F_4 + F_2 \wedge F_6;
\]
cf.\ \cite{Campbell}\cite{Huq}\linebreak[0]\cite{Giani} for the classical theory and \cite{BNS04}
for the duality-symmetric formulations of type IIA  supergravity. 

\paragraph{Scaling symmetries in type IIA.} 
In type IIA, one has two $\R^+$ symmetries: one, $\alpha$, is  inherited by dimensional reduction 
from M-theory (see \cref{Sec-TorSymm}) and similarly scales the Lagrangian, with weights 
$3$ to both $F_4$ and the once-reduced field  $s_1F_4$, 
the field strength of the NS B-field. 
The second, $\beta$, leaves the Lagrangian invariant.
We are interested in the first symmetry, which arises from  Kaluza-Klein dimensional reduction, where 
all $p$-index potentials scale with a factor $\lambda^p$, while all scalar
fields are left invariant. 
We will see the corresponding description for type IIB in \cref{Sec-torIIb}.

\paragraph{Type IIA and the cyclification $\mc{L}_c
S^4$ of the sphere $S^4$.}
A general framework for describing the automorphisms of Sullivan 
minimal models is given in \cite{SV1}. 
The maximal real split torus of $\Aut M(\mc{L}_c
S^4)$ compatible with the structure of $M(\mc{L}_c S^4)$ as the
Sullivan minimal model of the cyclic loop space of $S^4$
can be canonically identified 
with $\GG_m \times
\GG_m$, acting on $M(\mc{L}_c S^4) = \RR[g_4, g_7, sg_4, sg_7,
  w]$ as follows:
  $$
   t \cdot \phi_i = t^{\lambda_i} \phi_i , \qquad   t \in (\GG_m \times \GG_m) (\RR).
   $$

\(
\label{tg_4}
\begin{tabular}{ccc}
\hline 
{\bf Field} $\phi_i$ & {\bf  degree} & {\bf  weight} $\lambda_i$ 
\\ \hline 
\rowcolor{lightgray}
$g_7$ & $7$ & $2\epsilon_0$
\\
$sg_7$ & $6$ & $2\epsilon_0 - \epsilon_1$
\\
\rowcolor{lightgray}
$g_4$ & $4$ & $\epsilon_0$
\\
$sg_4$ & $3$ & $\epsilon_0 - \epsilon_1$
\\
\rowcolor{lightgray}
$w$  & $2$ & $\epsilon_1$ 
\\
\hline 
\end{tabular}
\)
\smallskip 
Here $w$ is the class corresponding to the circle (loop), and 
$S$ corresponds to reduction (i.e., contraction).
 This collection of fields $(g_4,  g_7,  sg_4, sg_7, w)$ corresponds to a weight
decomposition 
\begin{align*}
M(\mc{L}_c S^4)_{\epsilon_0}
\oplus M(\mc{L}_c S^4)_{2
  \epsilon_0}
  \oplus M(\mc{L}_c S^4)_{\epsilon_0 - \epsilon_1}
\oplus
 M(\mc{L}_c S^4)_{2 \epsilon_0 - \epsilon_1}
\oplus M(\mc{L}_c S^4)_{\epsilon_1}=M(\mc{L}_c S^4). 
  \end{align*}

\paragraph{Topological interpretation of toroidal symmetries coming from $S^1$.}
\label{Sec-toptor}
As in the case for $k=0$, here we present an interpretation of the toroidal symmetries
\eqref{tg_4}  as rational (real) homotopy equivalences; see details in \cite{SV1}. 
These symmetries have physical interpretation of torus rescaling symmetries. 
The situation with the action with weights $\epsilon_i$, $i > 0$, in \eqref{tg_4} is similar 
but a little subtler than the case of $S^4$.
The $n$-fold winding maps $S^1  \to S^1$, $n \in \ZZ \setminus \{0\}$, of the source $S^1$ for the loop space $\mc{L}_c S^4$ induce maps 
$$
\varphi_1(n): \mc{L}_c S^\longmapsto \mc{L}_c S^4
$$
in the rational homotopy category.
These morphisms are invertible, and the compositions of them with their inverses, as in \Cref{natural0},
give a group homomorphism 
$\Q^\times  \xrightarrow{\sim} \Aut
  M(\mc{L}_c S^4)(\Q)$, which defines naturally a morphism of $\RR$-algebraic groups
$$
\GG_m  \xrightarrow{\;\; \sim \;\;} \Aut M( \mc{L}_c S^4)\;,
$$
expressed through the contribution of $\epsilon_1$ in \eqref{tg_4}.

\medskip 
We now consider the Lie algebra level as in the case of 11d.

\medskip
\paragraph{The Lie algebra.} 
 Here we use the Lie algebra $\h_1= \Lie
(T^{2})$ of the maximal real split torus $T^{2}$ of $\Aut M(
\mc{L}_c^1 S^4)$ with its canonical factorization $T
\xrightarrow{\sim} \GG_m^{2}$. This Lie algebra constitutes the
infinitesimal symmetries corresponding to the toroidal symmetries above 
and acts on the
Sullivan minimal model $ M( \mc{L}_c S^4)$ by derivations.
That is, we have a Lie algebra homomorphism:
$\h_1 \to \Der M(\mc{L}_c^1 S^4)$
which comes from taking the differential of the action
$
T^{2} \to \Aut M(\mc{L}_c^1 S^4)$
$$
\xymatrix@C=8em{
{\color{darkblue} \footnotesize \rm   Group}  \;\;  T^{2} \ \rotatebox[origin=c]{-90}{$\circlearrowright$}~~
\;\;
\fbox{$\Aut M(
\mc{L}_c^{1} S^4)$}
\ar[r]^-{\rm differential \; of \; map}_-{\rm linearization}
& 
\fbox{$\Der M(\mc{L}_c^1 S^4)$ }
\;\;\; 
\ \rotatebox[origin=c]{-270}{$\circlearrowright$}~~
\h_1 = \Lie
(T^{2})\;\;
{\color{darkblue} \rm Algebra} 
}
$$ 
Under the action of the Lie algebra $\h_1$ on $M = M(\mc{L}_c^1 S^4)$,
the weight decomposition 
becomes
\[
M = \bigoplus_{\alpha \in P(\h_1)} M_{\alpha},
\]
where $P(\h_1) \subseteq \h_1^* = \Hom_\RR (\h_1, \RR)$ is the
\emph{weight lattice}, the image of the character group 
$$
\mathfrak{X}(T) =
\Hom_\RR (T, \GG_m) \linebreak[0] \cong \ZZ \oplus \ZZ
$$
under the differential map
  $\mathfrak{X}(T) \to  \h_1^*$
  taking $\beta$  to $d \beta$.

\paragraph{Bases for the Lie algebra of symmetries and its dual.}
$\,$

\vspace{-3mm} 
  \begin{enumerate}[{\bf (i)}]
   \setlength\itemsep{-1pt}
  \item
    The $2$-dimensional real abelian Lie algebra $\h_1 = \Lie
    (T^{2}) \subseteq \Der M(\mc{L}_c^1 S^4)$ of the maximal
    $\RR$-split torus $T = T^{2}$ of the algebraic group $\Aut
    M(\mc{L}_c^1 S^4)$ has a canonical basis $\{h_0, h_1\}$.

\item
  The weights $\epsilon_0$ of $g_4$ and $\epsilon_1$ of $w_1$ 
  give a canonical basis $\{\epsilon_0, \epsilon_1\}$ of the 
  vector space $\h_1^*$. This is also a basis of
  the weight lattice $P(\h_1) \subseteq \h_1^*$.
  \end{enumerate}

\paragraph{The dual lattices and inner product.}
$\,$

\vspace{-3mm} 
\begin{enumerate}[{\bf (i)}]
 \setlength\itemsep{-1pt}
\item
  The lattice $\h_1^\ZZ := \ZZ h_0 \oplus \ZZ h_1$ is the dual of the weight lattice
  $P(\h_1) = \ZZ \epsilon_0 \oplus  \ZZ \epsilon_1 \subset \h_1^*$.
\item
  The nondegenerate bilinear form $\h_1 \otimes \h_1 \to \RR$
  determined by the isomorphism
      $\h_1 \to \h_1^*$, which maps 
    $h_i$ to $\epsilon_i$,
   provides the vector space $\h_1$ with a canonical Minkowski inner
  product $(-,-)$, of signature $(+1, -1)$, i.e., satisfying 
  \vspace{-1mm}
  \[
  (h_0, h_0) = 1, \qquad (h_1, h_1) = - 1\;. 
  \]
 \item
The inner product induced on the dual space $\h_1^*$ is given by the formulas:
 \vspace{-1mm}
  \[
  (\epsilon_0, \epsilon_0) = 1, \qquad (\epsilon_1, \epsilon_1) = - 1\;.
  \]
  \end{enumerate}

Overall, we have the following picture 

$$
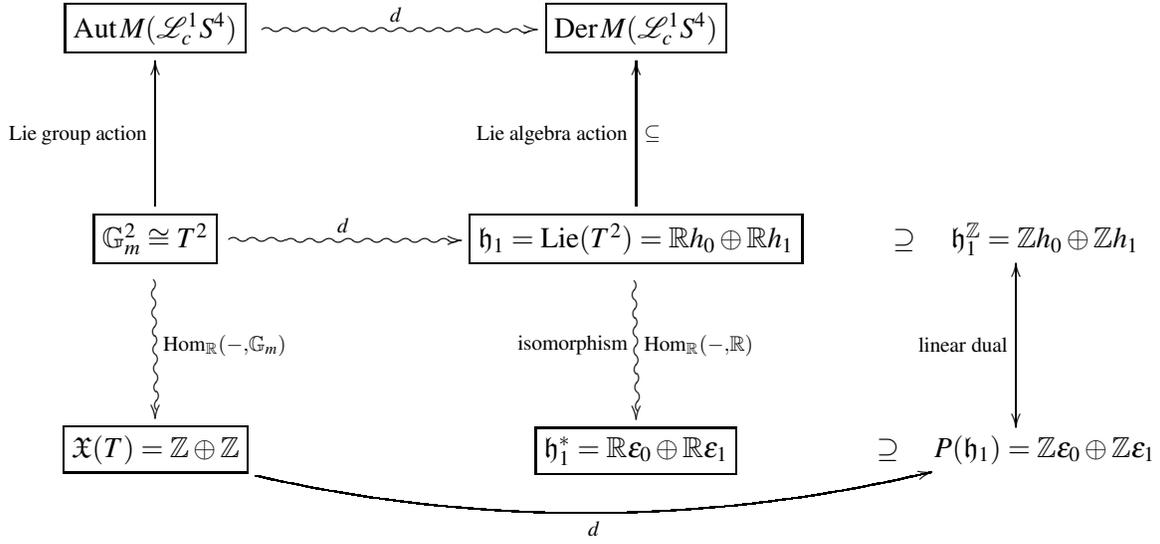
\xymatrix@C=2em@R=5em{
\fbox{$\Aut M(\mc{L}_c^{1} S^4)$}
\ar@{~>}[rrr]^-{d}
&&&
\fbox{$\Der M(\mc{L}_c^1 S^4)$}
&
\\
\fbox{$\GG_m^2\cong T^2$} \ar@{~>}[rrr]^-d 
\ar[u]^{\rm Lie \; group \; action}
\ar@{~>}[d]^{\Hom_\R(-, \GG_m)}
&&&
\fbox{$\h_1 = \Lie(T^{2})=\R h_0 \oplus \R h_1$}
\ar[u]^{\rm Lie \; algebra \; action}_-\subseteq 
\ar@{~>}[d]^{\Hom_\R(-, \R)}_{\rm isomorphism} 
&
\supseteq \quad \h_1^\Z = \Z h_0 \oplus \Z h_1
\ar@{<->}[d]_{\rm linear\; dual}
\\
\fbox{$\mathfrak{X}(T)=\Z \oplus \Z$}
\ar@/_2pc/[rrrr]_-d
&&& 
\fbox{$\h_1^*=\R \epsilon_0 \oplus \R \epsilon_1$}
& \supseteq \quad P(\h_1) = \ZZ \epsilon_0 \oplus  \ZZ \epsilon_1 
}
$$

We can compute the action of
the Lie algebra $\h_1$ on the other generators of the minimal DGCA
$M(\mc{L}_c^1 S^4)$.  

\paragraph{Action of the Lie algebra on the cyclic loop space in type IIA.}
The Lie algebra $\h_1$ acts on each weight space $M_\alpha$ with the
weight $\alpha$:
\[
M(\mc{L}_c S^4)_{\alpha} = \big\{\phi_i \in M(\mc{L}_c S^4) \; | \; h \cdot \phi_i = \alpha(h) \phi_i \quad
\text{for all $h \in \h_1$}\big\}.
\]
For the $\phi_i$ running over the universal fields in \eqref{tg_4}, and for $h=\{h_0, h_1\}$, we have 
\begin{gather*}
    h \cdot g_4 = \epsilon_0(h) g_4, \quad h \cdot g_7 = 2 \epsilon_0(h) g_7, \quad h \cdot
    sg_4 = (\epsilon_0(h) - \epsilon_1(h)) sg_4,\\
    h \cdot sg_7 = (2 \epsilon_0(h) - \epsilon_1(h)) sg_7, \quad h \cdot w =
  \epsilon_1(h) w\,.
\end{gather*}
With the Minkowksi metric and with the dual bases $\{\epsilon_0, \epsilon_1\}$ and $\{h_0, h_1 \}$, we have
 \vspace{-2mm}
\begin{align*}
  \epsilon_0(h_0)& = 1, \qquad \epsilon_0 (h_1)= 0,\\
  \epsilon_1(h_0)& = 0, \qquad \epsilon_1 (h_1) = -1\,,
\end{align*}

\vspace{-1mm} 
\noindent 
so that 
\begin{enumerate}[\bf (i)]

    \item For the action of $h_0$ we have 
    
\begin{tabular}{ll}
$\bullet$ $g_4$ and $sg_4$ have weight 1: & $h_0 \cdot g_4=g_4$, \; $h_0 \cdot sg_4=sg_4$
\\
$\bullet$ $g_7$ and $sg_7$ have weight 2: & $h_0 \cdot g_7=2g_7$, \; $h_0 \cdot sg_7=2 sg_7$
\end{tabular} 

        \item For the action of $h_1$ we have 
    
\begin{tabular}{ll}
$\bullet$ $sg_4$ and $sg_7$ have weight 1: & $h_1 \cdot sg_4=sg_4$, \; $h_1 \cdot sg_7=sg_7$
\\
$\bullet$ $w_1$ has weight $-1$: & $h_1 \cdot w_1=-w_1$,
\end{tabular} 
\end{enumerate}
   with all other actions having weight zero.

\medskip 
\paragraph{Del Pezzo surface  $\mathbb{B}_1$.}
This is obtained as the blowup of  $\mathbb{C P}^2$ at a point
$x_1$. 
The Picard group ${\rm Pic}(\mathbb{B}_1)$  in this case is a rank-$2$ lattice with 
two generators $\cH, \cE_1$:
$$
H^2(\mathbb{B}_1; \Z) \cong \Z \cH \oplus \Z \cE_1 
$$
where $\cE_1$ is the class of the exceptional divisor over $x_1$. 
See \cite{Man}\cite{Demazure}\cite{Beauville}\cite{KSC}. 
 This has a natural inner product
given by the intersection form:
\[
\cH \cdot \cH=1, \quad \cH\cdot \cE_1=0, \quad \cE_1 \cdot \cE_1 = - 1\, ,
\]
given by the Lorentzian form 
$
Q={\rm diag}(1, -1)
$.
The surface $\BB_1$ has the \emph{canonical class} 
$
\cK_1 = - 3 \cH + \cE_1 
$
as well as the \emph{anticanonical class}:
$
-\cK_1 = 3 \cH - \cE_1\;,
$
which  ``acts'' on the
Picard group $\Pic (\BB_1) \cong H^2 (\BB_1;
\Z)$ by degree:
\begin{equation*}
\deg \cD := - \cK_1 \cdot \cD, \qquad \cD \in \Pic (\BB_k).
\end{equation*}
Here the ``action'' is understood as the intersection product $\Pic
(\BB_1) \otimes \Pic (\BB_1) \to \Z$.
The \emph{degree} of  $\BB_1$ is the self-intersection number $(-\cK_1) \cdot (-\cK_1) = 8$. 
The \emph{degree of a divisor $\cD \in \Pic (\BB_1)$} is measured with respect to the anticanonical map as:
\(
\label{eqDegD}
\deg \cD := -\cK_1 \cdot \cD\;.
\)

\paragraph{The brane/reduced M-theory gauge algebra via the Quillen model of the cyclification of $S^4$.}
\label{Ex-reducedAlg} 
We have seen in \cref{Sec-Quillen} the description of the M-theory gauge algebra via Quillen model of $S^4$ 
is just the graded Lie algebra on two generators in \eqref{M-Quil}. 
The gauge algebra of the reduced fields in $11-k$ dimensions 
will correspond to the Quillen model of $\mc{L}_c^kS^4$.
We will not work out the details, as it is clear that matching the fields and EOMs 
of reduced M-theory with the generators and their differentials in the Sullivan minimal model of
the cyclification of $S^4$
implies similar matching between the gauge algebra and the Quillen model. 
Thus, the gauge algebra formulas in \cite{CJLP2}\cite{LLPS99} should be reproduced just by 
looking at the Quillen model of $\mc{L}_c^kS^4$. 

\paragraph{Degree in the Quillen model of $\mc{L}_cS^4$.}
 The maximal split torus $T^{2}$ and its Lie algebra $\h_1$ act on the
Quillen minimal model $Q(\mc{L}_c S^4)$ with the same weights as on
the generators of the Sullivan minimal model $M = M(\mc{L}_c S^4)$. 
Here again the degree operator
\begin{equation*}
x \mapsto \abs{x} \cdot x, \qquad x \in Q(\mc{L}_c S^4),
\end{equation*}
in the Quillen minimal model singles out a distinguished element of the Lie
algebra $\h_1$. 
Indeed \cite{SV1}, there is a unique element of the Lie algebra $ \h_1$, namely,
$
-K_1 := 3h_0 - h_1 
$,
which acts on the Quillen minimal model $ Q(\mc{L}_c S^4)$ as the
degree operator:
\[
-K_1 \cdot x = \abs{x} \cdot x\;.
\]
Here again, we use the element $-K_1$
as a distinguished element, the analogue of the anticanonical class. 
Extending the analogy with del Pezzo surfaces, as in \cite{SV1}, 
we define the \emph{degree of the cyclic loop space $\mc{L}_c  S^4$ of the four-sphere} as
\begin{equation}
\label{deg-L-k=1}
\deg \mc{L}_c  S^4 = (-K_1, -K_1) = 8.
\end{equation}

\medskip 
\paragraph{Type $IIA$ root system.} In contrast to the type IIB case, we will see later (see \eqref{IIB-root}), 
for type IIA we have 
(in the notation of \Cref{table1}):
\(
\label{IIA-root}
E_1 = R_1 = \big\{ x \in \h_1^\ZZ \; | \; (x, 3h_0 -h_1) = 0 , \; (x, x) = -2 \big\} = \varnothing = A_0\,.
\)
 While this might seem trivial, it is the zeroth step in an emerging pattern. 
This justifies the type IIA row of \Cref{table1}.

\subsection{Further reduction to $\leq 9$ dimensions and 
cyclifications of $S^4$}
\label{symm-cycl-concrete}

\subsubsection{Reduction to 9 dimensions}

\paragraph{Reduction to 9d  and the double cyclification $\mc{L}^2_c
  S^4$.}
The process can be iterated to give the same result for 
$\mc{L}_c^2Z$, so that we can now continue with further dimensional reduction.\footnote{
We will have multiple circle fiber directions and corresponding labels on 
the contractions $s_i$ and the classes of the circles $w_i$. As in \cite{SV1}, we realize that
the notation is not fully in parallel with the convention of using such labels to indicate the degree, but choosing another notation such as $s_{(i)}$ might overload the expressions when multiple such occur below. We hope the distinction 
will be clear from the context.}
Reduction on a 2-torus leads to the double cyclification
    $\mc{L}_c^2 S^4$
 whose Sullivan minimal model
$M(\mc{L}^2_c S^4)$ 
\cite{SV1} 
gives 
equations that are again universal, and we again obtain the corresponding ones in
spacetime by pullback. These are the EOMs and Bianchi identities
of type II string theory
at low energy, i.e., type II supergravity in 9 dimensions in the 
duality-symmetric formulation. 
The classical EOMs are given in \cite{BHO95}\cite{DR}, so the result
(established in \cite{SV1}) can be
viewed as a duality-symmetric extension.

 \paragraph{The group of automorphisms.} 
The maximal
torus of $\Aut M(\mc{L}^2_c S^4)$ is identified canonically with the
product $\GG_m^3 = (\GG_m \times \GG_m ) \times \GG_m$, where the
first factor $\GG_m \times \GG_m$ refers to the maximal torus of $\Aut
M(\mc{L}_c S^4)$ identified in the above 10d case. 
In this case, we have a weight
decomposition of $M(\mc{L}^2_c S^4)$, which is determined on its
generators as follows:
$$
\begin{tabular}{ccc}
\hline 
{\bf Field} $\phi_i$ & {\bf  degree} & {\bf  weight} $\lambda_i$ 
\\ \hline 
\rowcolor{lightgray}
$g_7$ & $7$ & $2\epsilon_0$
\\
$s_1g_7$ & $6$ & $2\epsilon_0 - \epsilon_1$
\\
\rowcolor{lightgray}
$s_2g_7$ & $6$ & $2\epsilon_0 - \epsilon_2$
\\
$s_2s_1g_7$ & $5$ & $2\epsilon_0 - \epsilon_1 - \epsilon_2$
\\
\rowcolor{lightgray}
$g_4$ & $4$ & $\epsilon_0$
\\
$s_1g_4$ & $3$ & $\epsilon_0 - \epsilon_1$
\\
\rowcolor{lightgray}
$s_2g_4$ & $3$ & $\epsilon_0 - \epsilon_2$
\\
$s_2s_1g_4$ & $2$ & $\epsilon_0 - \epsilon_1 - \epsilon_2$
\\
\rowcolor{lightgray}
$w_1$  & $2$ & $\epsilon_1$ 
\\
$w_2$  & $2$ & $\epsilon_2$
\\
\rowcolor{lightgray}
$s_2w_1$  & $1$ & $\epsilon_1-\epsilon_2$ 
\\
\hline 
\end{tabular}
$$

 \noindent This collection of fields 
$$
(g_4, \, g_7, \, s_1 g_4, \, s_1 g_7, \, w_1, s_2 g_4, \, s_2 g_7, \, s_2 s_1 g_4,
 \, s_2 s_1 g_7, \, s_2 w_1, \, w_2)
$$
 corresponds to a weight
decomposition
$$
M(\mc{L}^2_c S^4)=
   M_{\epsilon_0} 
  \oplus 
  M_{2\epsilon_0} 
 \oplus M_{\epsilon_0 - \epsilon_1}\oplus
 M_{2 \epsilon_0 - \epsilon_1}
\oplus 
M_{\epsilon_1}
\oplus 
M_{\epsilon_0 - \epsilon_2} \oplus 
M_{2 \epsilon_0 - \epsilon_2}
\oplus  M_{\epsilon_0 - \epsilon_1 - \epsilon_2}
\oplus M_{2 \epsilon_0 - \epsilon_1 - \epsilon_2}
\oplus M_{\epsilon_1 -\epsilon_2}
\oplus M_{\epsilon_2}\,,
 $$ 
where, for brevity, we have been writing $M$ for $M(\mc{L}^2_c S^4)$.

\smallskip 
The corresponding Lie algebra can be obtained from derivations, as in the case $k=1$. We will not spell this out explicitly but relegate a detailed discussion to the general case in 
\cref{Sec-8lower}.

\paragraph{Iterated $S^1$ vs.\ direct $T^2 = S^1 \times S^1$ reduction.}
\label{Rem-2S1vsT2}
We compare the two settings:

\noindent {\bf (i)} In the iterated case, $M(\mc{L}_c^2 S^4)$, notice the appearance of the \emph{axion} $s_2 w_1$, 
which would be absent in the direct reduction on $T^2$
 corresponding to the ``toroidification'' $M(\Map(T^2, S^4) \dslash T^2)$.
See \cref{eq-toroid} and the discussion around it for further details. 

\noindent {\bf (ii)} Note that $dw_1 = s_2 w_1 \cdot w_2$ above, whereas in the direct $T^2$-reduction, we would have $d w_1 = d w_2 = 0$. 
Likewise, note the two terms in the differential $ds_2g_4$ above, while in the direct reduction, it
  would be on an equal footing with $ds_1 g_4$, that is, $ds_2 g_4=s_1 s_2
  g_4 \cdot w_1$.

\subsubsection{Reduction to 8 or lower dimensions}
\label{Sec-8lower}

\paragraph{Reduction on a 3-torus and the triple cyclification
    $\mc{L}_c^3 S^4$.}
  The Sullivan minimal model
 $M(\mc{L}_c^3 S^4)$
worked out in \cite{SV1} leads to equations are again universal, and we obtain the corresponding ones in
spacetime by pullback, and capture the equations of motion and Bianchi identities 
of type II string theory
at low energy, i.e., type II supergravity in 8 dimensions in the 
duality-symmetric formulation, extending, for instance, \cite{AT85}. 

\paragraph{Weights of fields for  triple cyclification $\mc{L}^3_c
  S^4$.}
Again the maximal torus of $\Aut M(\mc{L}^3_c S^4)$ splits canonically to
become $\GG_m^4 = (\GG_m^3) \times \GG_m$, where the first factor
comes from the double cyclification. The resulting weight decomposition
of $M = M(\mc{L}^3_c S^4)$ repeats the formulas above verbatim
for the weights of those generators which are the generators of
$M(\mc{L}^2_c S^4)$. The weight of $w_3$ is $\epsilon_3$. For the
weights of generators of the type $s_3 g$, where $g$ is a generator on
the above list, the formulas are the same, except
that weight $\epsilon_3$ gets subtracted, e.g.,
\begin{gather*}
  s_3 g_4 \in M_{\epsilon_0 - \epsilon_3}, \qquad s_3 w_2 \in M_{\epsilon_2-\epsilon_3}.
  \end{gather*}
The weight $\epsilon_1 - \epsilon_2 -\epsilon_3$ will not be present, as
$s_3 s_2 w_1$ gets truncated to zero. See also \cref{Sec-toroid}.

\paragraph{Reduction on $T^k$ and 
$k$-fold cyclifications $\mc{L}_c^k S^4$ for $k \ge 3$.}
   The above pattern 
pertains in the reduction to 8 and lower dimensions \cite{SV1}.
The pullbacks to spacetime of the differentials  of the set of universal generators  
correspond, likewise, to the EOMs and Bianchi identities of 
duality-symmetric low energy string theory/supergravity in dimensions $11-k$.
Supplying these leads to a duality-symmetric extension of 
the non-duality symmetric versions studied, e.g., in \cite{LP96}\linebreak[0]\cite{LLP98},
and surveyed in \cite{CDF}\cite{Ta98}.

\paragraph{ Weights of fields for the $k$-fold cyclification $\mc{L}^k_c
  S^4$, $k\ge 3$.}
The general pattern continues for $k \ge
3$. All the weights of the generating space $V$ for $S(V) =
M(\mc{L}^k_c S^4)$ will be of the form
$$
\begin{tabular}{cc}
\rowcolor{lightgray}
$\epsilon_0 -  \sum_{j=1}^l \epsilon_{i_j}$, & \qquad \text{where $0 \le l \le
  3$ and $1 \le i_1 < \dots < i_l \le k$},
\\
 $2\epsilon_0 -  \sum_{j=1}^l \epsilon_{i_j}$, & \qquad \text{where $0 \le l \le
  6$ and $1 \le i_1 < \dots < i_l \le k $},
\\
\rowcolor{lightgray}
  $\epsilon_i$, & \; \qquad \qquad $1 \le i \le k$,
\\
 $\epsilon_i -  \epsilon_j$, & \qquad \quad  $1 \le i < j \le k$.
\end{tabular} 
$$
Each of the corresponding weight spaces in $V$ will be
one-dimensional.

\medskip 
Here we generally describe the toroidal symmetry of the iterated cyclic loop spaces
$\mc{L}^{k}_c S^4$, 
from \cref{Sec-CyclicSugra}, for which 
we provide topological interpretation, leading to root data.

\paragraph{Scaling symmetries in lower dimensions.}
Upon reduction on $T^k$, the above rescaling 
modifies the $\R$-part of ${\rm GL}(11- k,\R) = \R \times {\rm SL}(11- k, \R)$, which 
becomes as a result an internal symmetry.
It can be traced back to the eleven-dimensional action, as above,
to appropriate multiplicative redefinitions of the 3-form. 
The plain $\R$-symmetry rescales the volume of the compactifying space,
and must be combined with the trombone symmetry  to define a new internal scaling symmetry, 
which in 8 dimensions becomes ${\rm SL}(2, \R)$ \cite{CJLP}.
The symmetry is preserved under Kaluza-Klein dimensional reduction, after which 
all $p$-index potentials scale with a factor $\lambda^p$, while all scalar
fields are left invariant.

\paragraph{The $\R^q$ symmetry in the reduction on tori.} 
There is also a rescaling of the $k$-torus $T^k$
with coordinates 
$
y^i \mapsto \lambda y^i$, $i=1, \cdots, k$,
as part of the eleven-dimensional diffeomorphisms.

\vspace{-3mm} 
\begin{enumerate}[{\bf (i)}]
 \setlength\itemsep{-2pt}
     \item 
This defines  a scaling which leaves the scalars in 
$(11-k)$ dimensions
invariant and 
 is part of the Cremmer-Julia group
embedded as ${\rm GL}(1) \subset {\rm GL}(k) \subset E_{k(k)}$
\cite{DS}. 
Gauging this symmetry amounts to having ${\rm GL}(1)$ charge $\mathfrak{q}$ with 
$
C_3 \mapsto e^{-3\mathfrak{q}} C_3
$,
which is an alternative way of writing \eqref{scale-metric}.
See \cite{DS} for details. 

\item The $\R^q$ part of the symmetry comes from the local abelian gauge symmetry of
$G_4$. Specifically, it describes the global shift symmetries of the axionic scalars 
that are the potentials for 1-form field strengths coming from the dimensional reduction.
\end{enumerate} 

\vspace{-2mm} 
\noindent We will consider these matters from the point of view of topology below  and algebraic geometry in 
\cref{Sec-M}.

\paragraph{Toroidal symmetries of minimal algebras.}
\label{tor-symm}
Here, generalizing the discussion in \cref{Sec-TorSymm} (for $k=0$) and \cref{Sec-IIAsym}
(for $k=1$), 
our \emph{real split torus}, an affine algebraic group $T$ isomorphic over
$\RR$ to the group $\GG_m^k$ for some $k \ge 1$.
As we showed in \cite{SV1}, we can consolidate essentially all toroidal symmetries of $M$, which
could be done by considering a \emph{maximal $\RR$-split torus} $T
\subseteq \Aut M$.
A maximal split torus will play a role similar to that of a maximal torus in the theory of compact Lie
groups or that of a Cartan subalgebra in the theory of complex
semisimple Lie algebras. 
Indeed, a real split torus $T$ gives rise to a
\begin{itemize}
    \item 
\emph{weight decomposition}:
\[
M = \bigoplus_{\alpha \in \mathfrak{X}(T)} M_\alpha
\]
\item indexed by the \emph{character group} $\mathfrak{X}(T) = \Hom_\RR (T, \GG_m)$ of
real algebraic group homomorphisms from $T$ to the multiplicative group
$\GG_m$,
\item so that $T$ acts on each \emph{weight space} $M_\alpha$ by
the character $\alpha$:
\[
M_\alpha = \big\{m \in M \; | \; t \cdot m = \alpha(t) m \quad \text{for
  all $t \in T$}\big\}.
\]
\end{itemize}

 \vspace{-3mm} 
Note that a weight decomposition of a minimal Sullivan algebra $M = S(V)$
induces a weight decomposition 
\vspace{-2mm} 
$$
V = \bigoplus_{\alpha \in \mathfrak{X}(T)}
V_\alpha
$$ 

\vspace{-3mm} 
\noindent 
of the generating space $V$.

\smallskip 
We established the following result in \cite{SV1}:
\label{split-rank}
{\it The maximal $\RR$-split torus of the real algebraic group $\Aut M (\mc{L}_c^k S^4)$ for $k \ge 0$ is $T^{k+1}$, isomorphic to $\GG_m^{k+1}$ over $\RR$. 
The structure of $\mc{L}_c^k S^4$ as an iterated cyclification 
determines a canonical splitting $T^{k+1} \cong \GG_m^{k+1}$.
}

\paragraph{Topological interpretation of toroidal symmetries coming from iterated $S^1$'s.}
For the iterated cyclic loop space $\mc{L}_c^k S^4$, $k \ge 1$, the algebraic-group morphisms corresponding 
to different iterations commute and thereby define an algebraic-group morphism
$
  (\GG_m)^k  \xrightarrow{\sim} \Aut
  M(\mc{L}^k_c S^4),
 $ which provides an action of $(\GG_m)^k$ on the Sullivan minimal model of the $k$-fold cyclification $\mc{L}_c^k S^4$ 
 of $S^4$.
 Acting similarly to \eqref{GmonS4} through the target $S^4$ in the cyclic loop space $\mc{L}^k_c S^4$, we get an action 
$
\GG_m \times M(\mc{L}^k_c S^4) \rightarrow   M(\mc{L}^k_c  S^4)
$
by the above tables 
through the weight $\epsilon_0$.

\medskip 
We now unravel the general pattern $E_k$ symmetry of the iterated cyclic loop spaces
$\mc{L}^{k}_c S^4$.

\paragraph{The ``Cartan subalgebra'' and weight lattice.}
\label{Cartan}
Generalizing the cases $k=0$ from \cref{Sec-TorSymm} and $k=1$ from \cref{Sec-IIAsym},
we consider $k \geq 2$ and we use the toroidal symmetries of the cyclic loop
spaces $\mc{L}_c^k S^4$ to  build certain canonical data: ``a lattice
$N_k$ with an inner product $(-,-)$ and a distinguished element
$K \in N_k$'', similar to the triple $(N_k, (-,-), K)$
in the theory of del Pezzo surfaces; see \cref{dP}. This will
produce the $E_k$ root system in a similar way.

\medskip 
The idea \cite{SV1} is to use the weight lattice coming from the weight
decompositions 
and use the Lie algebra $\h_k = \Lie
(T^{k+1})$ of the maximal real split torus $T^{k+1}$ of $\Aut M(
\mc{L}_c^{k+1} S^4)$ with its canonical factorization $T
\xrightarrow{\sim} \GG_m^{k+1}$.
This Lie algebra constitutes the
infinitesimal symmetries corresponding to the toroidal symmetries 
and acts on the
Sullivan minimal model $ M( \mc{L}_c^{k+1} S^4)$ by derivations.
That is, we have a Lie algebra homomorphism:
$
\h_k \to  \Der M(\mc{L}_c^k S^4)
$,
which comes from taking the differential of the action
$
T^{k+1} \to \Aut M(\mc{L}_c^k S^4)$.
$$
\xymatrix@C=8em{
{\color{darkblue} \footnotesize \rm   Group}  \;\;  T^{k+1} \ \rotatebox[origin=c]{-90}{$\circlearrowright$}~~
\;\;
\fbox{$\Aut M(
\mc{L}_c^{k} S^4)$}
\ar[r]^-{\rm differential \; of \; map}_-{\rm linearization}
& 
\fbox{$\Der M(\mc{L}_c^k S^4)$ }
\;\;\; 
\ \rotatebox[origin=c]{-270}{$\circlearrowright$}~~
\h_k = \Lie
(T^{k+1})\;\;
{\color{darkblue} \rm Algebra} 
}
$$

As we explained in \cite{SV1}, the Lie algebra $\h_k$ is an avatar of the Cartan
subalgebra of the Lie algebra of type $E_k$ of ``hidden'' symmetries
of the cyclic loop spaces of the four-sphere.
The cases $k=0$ and $k=1$ are the beginnings of a pattern, the general structure of which is
as follows:

\paragraph{Bases for the Lie algebra of symmetries and its dual.}
$\,$

\vspace{-3mm} 
  \begin{enumerate}[{\bf (i)}]
   \setlength\itemsep{-1pt}
  \item
    The $(k+1)$-dimensional real abelian Lie algebra $\h_k = \Lie
    (T^{k+1}) \subseteq \Der M(\mc{L}_c^k S^4)$ of the maximal
    $\RR$-split torus $T = T^{k+1}$ of the algebraic group $\Aut
    M(\mc{L}_c^k S^4)$ has a canonical basis $\{h_0, h_1,
    \linebreak[0] \dots, \linebreak[1] h_k\}$.
\item
  The weights $\epsilon_0$ of $g_4$ and $\epsilon_i$ of $w_i$ for $1 \le i
  \le k$ give a canonical basis $\{\epsilon_0, \epsilon_1, \dots,
  \epsilon_k\}$ of the vector space $\h_k^*$. This is also a basis of
  the weight lattice $P(\h_k) \subseteq \h_k^*$.
  \end{enumerate}

As in the case for $k=0,1$ above, we can compute the action of
the Lie algebra $\h_k$ on the other generators of the minimal DGCA
$M(\mc{L}_c^k S^4)$. 

\paragraph{The dual lattices and inner product.} Generalizing the previous special cases, we have: 
$\,$

\vspace{-3mm} 
\begin{enumerate}[{\bf (i)}]
 \setlength\itemsep{-1pt}
\item
  The lattice $\h_k^\ZZ := \ZZ h_0 \oplus \ZZ h_1 \oplus \dots \oplus
  \ZZ h_k \subset \h_k$ is the dual of the weight lattice
  $P(\h_k) = \ZZ \epsilon_0 \oplus \dots \oplus \ZZ \epsilon_k \subset \h_k^*$.
\item
  The nondegenerate bilinear form $\h_k \otimes \h_k \to \RR$
  determined by the identification (isomorphism) 
    $\h_k \to \h_k^*$, sending 
    $h_i$ to $\epsilon_i$,
provides the vector space $\h_k$ with a canonical Minkowski inner
  product $(-,-)$. This inner product satisfies the formulas:
  \vspace{-1mm}
  \[
  (h_0, h_0) = 1, \qquad (h_i, h_j) = - \delta_{ij} \quad \text{ for \; 
    $i \ge 0$, $j \ge 1$.}
  \]
 \item
The inner product induced on the dual space $\h_k^*$ is given by the formulas:
 \vspace{-1mm}
  \[
  (\epsilon_0, \epsilon_0) = 1, \qquad (\epsilon_i, \epsilon_j) = - \delta_{ij} \quad \text{ for \; 
    $i \ge 0$, $j \ge 1$.}
  \]
  \end{enumerate}

\paragraph{The anticanonical class on del Pezzo.}
\label{can-section}
Consider the \emph{del Pezzo surface} $\mathbb{B}_k$ 
obtained as the blowup of 
$\mathbb{C P}^2$ at $k$ generic points 
$x_1, \dots, x_k$, $0\leq k \leq 8$. 
The Picard group 
$\Pic (\BB_k)$ is a rank-$(k+1)$ lattice with generators $\cH, \cE_1, \dots, \cE_k$:
$$
H^2(\mathbb{B}_k; \Z) \cong \Z \cH \oplus \Z \cE_1 \oplus \cdots \oplus \Z \cE_k,
$$
where now 
$\cE_i$ is the class of the exceptional divisor over $x_i$. 
See \cite{Man}\cite{Demazure}\cite{Beauville}\cite{KSC}. 
This has a natural inner product
given by the intersection form:
\(
\label{HE-inter}
\cH \cdot \cH=1, \quad \cH\cdot \cE_i=0, \quad \cE_i \cdot \cE_j = - \delta_{ij}, \qquad 1 \leq i, j \leq k\;,
\)
thus given by the Lorentzian form 
$
Q={\rm diag}(1, -1, -1, \dots, -1)
$.
Hence $H^2(\mathbb{B}_k;\ZZ) \cong \ZZ^{1,k}$
and $\mathbb{B}_k$ has Betti numbers $b_2^+=1$, $b_2^{-}=k$, with signature 
$\sigma=1-k$.

\medskip 
The \emph{anticanonical class}:
 \(
 \label{AntiK}
-\cK_k = 3 \cH - \cE_1 - \dots - \cE_k,
\)
in this general case ``acts'' on $\Pic (\BB_k) \cong H^2 (\BB_k;
\Z)$ by degree:
\begin{equation*}
\deg \cD := - \cK_k \cdot \cD, \qquad \cD \in \Pic (\BB_k).
\end{equation*}
Here the ``action'' is understood as the intersection product $\Pic
(\BB_k) \otimes \Pic (\BB_k) \to \Z$. The \emph{degree of a divisor}
$\cD \in \Pic (\BB_k)$ is defined using the anticanonical morphism
$f: \BB_k \to \CC \PP^d$, and 
$$
d = h^0 (\BB_k, -\cK_k) - 1 = (-\cK_k) \linebreak[0] \cdot (-\cK_k) =
9 - k
$$ 
is known as the \emph{degree of the del Pezzo surface $\BB_k$}.

\paragraph{The ``anticanonical class'' for $\mc{L}_c^k
S^4$ analogous to del Pezzo.}
The maximal split torus $T^{k+1}$ and its Lie algebra $\h_k$ act on the
Quillen minimal model $Q(\mc{L}_c^k S^4)$ with the same weights as on
the generators of the Sullivan minimal model $M = M(\mc{L}_c^k$.
Also in this general case, the degree operator
\begin{equation}
\label{degree-operator}
x \mapsto \abs{x} \cdot x, \qquad x \in Q(\mc{L}_c^k S^4),
\end{equation}
in the Quillen minimal model singles out a distinguished element of the Lie
algebra $\h_k$, namely \cite{SV1} 
\begin{equation}
  \label{elt}
-K_k := 3h_0 - h_1 - \dots - h_k,
\end{equation}
which acts on the Quillen minimal model $ Q(\mc{L}_c^k S^4)$ as the
degree operator:
\[
-K_k \cdot x = \abs{x} \cdot x.
\]
Extending the analogy with del Pezzo surfaces, 
We define the \emph{degree of the cyclic loop space $\mc{L}_c^k S^4$ of the four-sphere} as
\begin{equation}
\label{deg-L}
\deg \mc{L}_c^k S^4 = (-K_k, -K_k) = 9 - k.
\end{equation}

\paragraph{Exceptional root data from the Sullivan model of cyclic loop spaces of the 4-sphere.} 
The following result summarizes one of the main constructions in \cite{SV1} 
and establishes the entries in Table \ref{table1}: 

\label{rootdata}
\noindent {\bf (i)}  For each $k \ge 0$,
  the data 
   $$
  \big(\h_k^*, \linebreak[1] \{\epsilon_o, \epsilon_1, \linebreak[0] \dots,
  \epsilon_k\}, (-,-), K_k^*\big)
  $$ 
  associated to the cyclic loop space
  $\mc{L}^{k}_c S^4$ and its Sullivan minimal model $M(\mc{L}^{k}_c
  S^4)$ consists of
  \vspace{-1mm} 
  \begin{itemize}
   \setlength\itemsep{-1pt}
  \item[{\bf (a)}] a real vector space $\h_k^*$ with a basis $\{\epsilon_0, \epsilon_1, \dots,
    \epsilon_k\}$, which generates a lattice $P(\h_k) \subset \h_k^*$;
  \item[{\bf (b)}] a symmetric bilinear form $\h_k^* \otimes \h_k^* \to \R$
    given by
    \vspace{-1mm}
    \[(\epsilon_0, \epsilon_0) = 1, \qquad (\epsilon_i, \epsilon_j) = -\delta_{ij}, \qquad i > 0, j \ge 0;
    \]
    \vspace{-7mm} 
    \item[{\bf (c)}] a distinguished element $K_k^* = -3 \epsilon_0 + \epsilon_1 + \dots +
      \epsilon_k$.
        \end{itemize}
        
\vspace{-1mm}         
\noindent {\bf (ii)} 
This data replicates the data
\vspace{-3mm} 
$$\big(H^2(\BB_k; \R), \{\cH, \cE_1, \dots,
\cE_k\}, (-, -), \cK_k \big)$$ 

\vspace{-1mm} 
\noindent 
determined by the rational surface $\BB_k$,
considered as the blowup of $\CC \PP^2$ at $k$ points; see
\cref{dP} in the del Pezzo case, when $k \le 8$. For $k \le 8$, the data produces the 
{\it root system}
\vspace{-1mm}         
\begin{equation}
\label{roots}
R_k := \big\{ \gamma \in P(\h_k) \; | \; (\gamma, K_k^*) = 0 , (\gamma, \gamma) = -2 \big\}
\subset (K_k^*)^\perp \subset \h_k^*
\end{equation}

\vspace{-1mm}         
\noindent
of type \footnote{True/genuine $E_k$ for $k =6, 7$, and 8, and using the conventions of \Cref{table1}
for $0 \le k \le 5$.} $E_k$ and the Weyl group $W(E_k)$, generated by the
reflections in the hyperplanes orthogonal to the
roots $r \in R_k$, now in the context of cyclic loop spaces of $S^4$.

\subsubsection{The toroidification model}
\label{Sec-toroid} 

We have adopted the iterated cyclification model $\mc{L}_c^k S^4$ so far,
providing here and in \cite{SV1} a proper setting for matching of the fields and their dynamics. 
For comparison and completeness, 
we will now briefly consider a variant model, which has some clear virtues in 
capturing the description of M-theory configurations/fields, but which 
will also naturally come with its own limitations. 

\medskip 
Instead of reducing M-theory in 11 dimensions to 
a theory in $11-k$ dimensions on 
$k$ iterations of the circle $S^1$,
this model is obtained by viewing the reduction as directly over the $k$-torus $T^k = S^1 \times \dots \times S^1$.  
From a physics point of view, the former highlights how one goes to the low-dimensional
theory by going through all intermediate theories, 
while the former goes directly to 
(a variant of) that theory, skipping the intermediate stages. 
The toroidification of the sphere $S^4$ is given by 
\vspace{-2mm} 
\(
\mathbb{T}^k(S^4):={\rm Map}(T^k, S^4)/\!\!/T^k 
\label{eq-toroid} 
\)

\vspace{-2mm} 
\noindent and follows the general construction for any Lie group as source
  given in \cite[Theorem 2.44 and (28)]{BSS}
  (the Ext/Cyc-adjunction). Of course the toroidification model coincides with the cyclification model 
  when $k=1$, and so the difference arises when dimensionally reducing to nine dimensions or below.

\paragraph{Virtues.} The toroidification model puts the source circles on an equal footing and allows for the symmetric group $\Sigma_k$ action
on the minimal model $M(\mathbb{T}^k(S^4))$ (and related spacetime fields), rather than just on the weight lattice, cf.\ \eqref{wtrans}.
From the point of view of del Pezzo surface $\BB_k$, 
this corresponds to exchanging the points at which blowups take place. 
From our perspective, this corresponds to the exchange of source loops, arising from
the symmetry of the torus, in the toroidification model. 
This model will also be prominent in \cite{SV3}.

\paragraph{Limitations.} In the toroidification model, the generators 
$s_j w_i$ and the corresponding fields $s_j \omega_i$ are not detected, while all the 
$w_i$'s will be closed: $d w_i=0$. 
The fields $s_j \omega_i$ are {\it axions}, i.e., 1-form field strengths (corresponding to 0-form potentials). The missing axions $s_j \omega_i$
arise from the form-field
$F_2$, the latter corresponding to the universal element $w_2$. 
However, we will still find axions
arising from the reduction of the $C$-field, i.e., axions of type 
$s_3s_2s_1G_4$, in the toroidification model.
The fact that the $w_i$'s are closed reflects the fact that,
in the toroidification perspective,
there is no room for twisting which takes place in
the iterated cyclification.
Thereby, we will continue adopting our cyclification model.

\paragraph{Scalars.}
In deriving the above, we needed an assumption that amounts to setting the scalars to zero \cite{SV1}. Such scalars are 
treated from a physics perspective as being coordinates on the 
real moduli spaces $E_{k(+k)}/K$, where $K$ is the maximal compact subgroup.
Such spaces  are contractible hyperbolic spaces with no nontrivial 
homotopy, hence perhaps  not
that interesting from the rational homotopy theory point of view. 
From this angle, the scalars might seem orthogonal to the way 
we are probing the structure and 
symmetries. However, particularly the axionic scalars have made a prominent 
appearance here and will further appear in an interesting fashion in 
\cref{Sec-excvec}.

\subsection{Reduced electric-magnetic duality via topology and fields}

We have seen above,  in \cref{subsec-EMdual},
the interpretation of the anticanonical class from
the perspective of 11-dimensional M-theory for $k=0$. Now we consider the 
dimensional reduction of that perspective on electric-magnetic duality. 

\paragraph{Reduced M-theory fields and branes.}                 
In our perspective, we consider the fields that couple to the 
M-branes. Our interpretation of the classes is given in the fourth column of the following table (for $3\le k \le 8$): 
\begin{center}
\renewcommand{\arraystretch}{1.2}
\begin{tabular}{ccccc}
\hline 
\bf Degree & \bf del Pezzo class & \bf BPS state & \bf Form field & \textbf{Element of $M(\mc{L}_c^k S^4)$}\\
\hline
\hline 
\rowcolor{lightgray} 1 & $\cH-\cE_i-\cE_j-\cE_l$ & thrice-wrapped M2 & $s_ls_js_i G_4$ & $s_ls_js_i g_4$\\
2 & $\cH-\cE_i-\cE_j$ & twice-wrapped M2 & $s_j s_i G_4$ & $s_j s_i g_4$\\
\rowcolor{lightgray} 3 & $\cH-\cE_i$ & once-wrapped M2 & $s_i G_4$ & $s_i g_4$\\
$7-k$ & $2\cH-\sum_{p=1}^k \cE_p$ & $k$-wrapped M5 & $s_{k}\cdots s_1 G_7$ & $s_{k}\cdots s_1 g_7$\\
\hline 
\end{tabular}
\end{center}
 where, as before, for form fields, $s_i$ corresponds to reduction (i.e., contraction) 
 along the $i$th circle direction (in the notation of \cref{Sec-CyclicSugra}).

\paragraph{The reduced anticanonical class via the Sullivan model.} 
The canonical class (or divisor) $\cK_k$ of the 
del Pezzo surface $\mathbb{B}_k$  is given by 
the negative first Chern class $c_1(\mathbb{B}_k)$ of the tangent bundle:
\vspace{-1mm} 
\(
  \label{k-canclass}
\cK_k=- \Big( 3 \cH - \sum_{p=1}^k \cE_p\Big)\;.
\)

\vspace{-2mm} 
\noindent This has self-intersection number $\cK_k\cdot \cK_k=9- k$. Imposing positivity 
leads to $k < 9$. The same story happens with the ``canonical class'' $K_k$ of the $k$-fold cyclification of $S^4$, see \Cref{can-section}.
These generally correspond to the dimensional reduction of the description of M-theory above in  \cref{subsec-EMdual}.
For instance, for $\mathbb{B}_1$ and $\mc{L}_c S^4$, with
corresponding (see \cref{Sec-IIAsym})  $\cK_1=-3\cH + \cE_1$ and $K_1 = - 3 h_0 + h_1$, respectively, we have 
\begin{align*}
s_1(G_4 \wedge G_7) & = H_3 \wedge H_7 + F_4 \wedge F_6
\; ,\\
s_1(g_4 \wedge g_7) & = s_1 g_4 \cdot g_7 + g_4 \cdot s_1 g_7 \; .
\end{align*}
For form-fields, it seems common in the physics literature to leave only the first term $H_3 \wedge H_7$ and work modulo $F_4$ and $F_6$, 
which actually makes $H_3$ and $H_7$ closed.
This gives rise to electric-magnetic pairs, as expected in type IIA. 

\medskip 
The terms for higher $k$ are analogous and an obvious pattern emerges,
with the other terms being the various dimensional reductions of the 
cubic (Chern-Simons) term from eleven dimensions, along the lines of \cite{MaS},
and these terms are of the form 
$s_k \cdots s_1 (G_4 \wedge G_7)$.
These give topological terms familiar in the di\-men\-sion\-al\-ly-reduced theories. 
  
\medskip
Note that the analogous statements about electric-magnetic/Serre duality hold
for the reduced forms using $\cK_k$ in place of $\cK$ and the reduced classes
in the above table in place of the classes $\cH$ and $2\cH$. To a large extent, 
the results in  \cref{Sec-CyclicSugra} can be viewed as an illustration of this at the level of EOMs,  
and the reconstruction of the terms in the corresponding actions is straightforward.

\subsection{The $E_k$ root system and its Weyl group}
\label{Sec-Ek} 

We have seen the appearance of the Weyl group $W$ in obtaining exceptional root data from 
$\mc{L}_c^k S^4$ (see the end of \cref{Sec-8lower}). 
We now explain further the role of $W$, starting with a description that is close to our perspective
that involves tori and mapping degrees. 
 
\paragraph{The Weyl group via torus action.} 
 For a connected compact Lie group $G$ with $T \subset G$ its maximal torus, 
 a maximal connected abelian group, the normalizer $\mathcal{N}(T)$ of $T$ fits into a short
 exact sequence 
 \(
 \label{WfromN}
 1 \longrightarrow T \longrightarrow \mathcal{N}(T) \xrightarrow{\;\;\; p\;\;\;} W \longrightarrow 1\;.
 \)
 Since $T$ is abelian, $W$ acts on $T$ by conjugation, i.e., if $\sigma \in W$ and 
 $t \in T$ then $\sigma$ sends $t$ to $ntn^{-1}$, where $n \in p^{-1}(\sigma)$. 
Note that for such a $G$ is determined up to isomorphism by  $\mathcal{N}(T)$; see \cite{Osse}\cite{Not} for the general case, 
\cite{CWW} for semisimple case, and \cite{AH17} for reductive algebraic groups.
 From the short exact sequence \eqref{WfromN}, 
 $\mathcal{N}(T)$ is in turn determined up to isomorphism by three ingredients:
 \vspace{-2mm} 
\begin{enumerate}[{\bf (i)}]
 \setlength\itemsep{-2pt}

 \item the torus $T$,
 
 \item the finite group $W$, with its conjugation action on $T$, and 
 
 \item an extension class $\lambda \in H^2(W; T)$, the degree-2 group cohomology of $W$, i.e., of the classifying space $BW$.
 \end{enumerate} 
For (i) and (ii): $T=(S^1)^k$ for $k$ the rank of $G$. The natural map
 $\Aut (T) \to \Aut (\pi_1T) \cong {\rm GL}(k, \ZZ)$ is 
 an isomorphism and resulting conjugation homomorphism $W \to  {\rm GL}(k, \ZZ)$
 embeds $W$ as a finite  subgroup of ${\rm GL}(k, \ZZ)$
 generated by reflections. A \emph{reflection} here is a matrix which is conjugate in 
  $\Aut (\pi_1 T \otimes \QQ) \cong {\rm GL}(k, \QQ)$ to the diagonal matrix 
  ${\rm diag}(-1, 1, \cdots, 1)$. 
  For (iii) a characterization is given in \cite{DW05}. 
  Note that the above realizes the Weyl group itself as a quotient of a torus action, namely 
 $W=\mathcal{N}(T)/T$. 
 It is natural to identify the maximal torus with the spacetime torus $T^k$ in the 
 dimensional reduction. Note also that $E_k/T^k$ probes the nonabelian parts of $E_k$, as in 
 the discussion around \eqref{abelianization}.
 
 \paragraph{Meaning of reflections via covering.} 
 The relation between Weyl group actions and reflections in Euclidean spaces is as follows. 
 Let $\sigma \in W$ be a reflection. This means that if $\sigma: T \to T$ is lifted to the covering space
 $$
 \tilde{\sigma}: 
 \R^k \to \R^k\;, \qquad k={\rm rank}(G)
 $$
 then  $\tilde{\sigma}$ is a reflection in an $(k-1)$-hyperplane $H$ in $\R^k$. 
 Let $T^\prime=p^{-1}(\sigma)$ in \eqref{WfromN}. Since $\sigma^2=1=p(T)$, we see that 
 if $x \in T^\prime$ then $x^2 \in T$.

 \paragraph{Roots and Weyl via circle mapping degree.} 
 In connecting to the perspective in \cref{Sec-TorSymm}, \cref{Sec-toptor}, 
 and \cref{symm-cycl-concrete},
 it is interesting that in classical Lie theory one can interpret certain classical 
 notions via degrees. 
 Let $\phi: G \to G$ be the squaring map $\phi(x)=x^2$. Then 
 $$
 Q:=\{ t \in T ~|~ x\,t\,x^{-1}=t^{-1} \; \text{for} \; x \in T^\prime \} \subset T
 $$
 is topologically a disjoint finite set of circles. The identity component 
 $Q^0$ is a circle group. Then one can define \cite{CW75} the root and coroot 
 associated with a reflection $w$ to be a homomorphism 
 $\theta: T \to S^1$ and the homomorphic inclusion $\overline{\theta}: Q^0 \to T$, respectively. 
   Let $\sigma_1, \cdots, \sigma_r$ be 
   reflections generating $W$. 
  For each pair $\sigma_i, \sigma_j$, one has a map 
  $$
  S^1 \overset{\overline{\theta}_i}{\longrightarrow} 
  T \overset{\theta_j}{\longrightarrow} S^1
  $$
 and denote the degree of the composite 
 $\theta_j \circ \overline{\theta}_i$ by $d_{ij}$. 
 Furthermore, for each $\sigma_i$ associated is the degree $d_i$ of $[\overline{\theta}_i]$
 in $\pi_1(G)$ which is either 1 or 2. 
 Then  for each such pair of reflections, 
 the integers $C_{ij}=d_i d_{ij}$ are called the {\it Cartan integers}
 of $G$ and coincide with the usual definition \cite{CW75} 
 $$
 C_{ij}=2 \frac{\langle \theta_i, \theta_j \rangle}{\langle \theta_i, \theta_i \rangle} 
 $$ 
 where $\langle -, - \rangle$ is the ($W$-invariant) inner product on the dual Lie algebra $\mathfrak{g}^*$. 
 
 \medskip 
  Note  that, classically, the determinant of the Cartan matrix gives the index of the root lattice 
  inside the lattice of integral weights. We can relate to del Pezzo as follows: 
 For $G=E_k$,  $3 \leq k \leq 8$, the determinant of the Cartan matrix is  (see \cite{Wu82})
 $\det (C)=9-k$, which is the degree of $\BB_k$.

\paragraph{Weyl group as symmetry of symmetries.}
For each root $x$, an element of the set $R_k$ (see expression \eqref{roots})  
define a reflection
\vspace{-1mm} 
\(
\label{W-refl}
\sigma_x: y \;\; \longmapsto \;\; y - 2 \frac{(y,x)}{(x,x)} x = y + (y,x) x, \qquad y \in \h_k,
\)

\vspace{-1mm} 
\noindent 
of the vector space $\h_k$. As above, the reflections $\{\sigma_x \; | \; x \in R_k \}$ generate  the Weyl group $W(E_k)$ of the root system $E_k$. It is the group of all linear isometries of $\h_k^\ZZ$ which preserve the element $K_k$ \cite{Man}.
Thus, the Weyl group, being an isometry group of the abelian Lie algebra $\h_k$ of infinitesimal 
symmetries of $M(\mc{L}_c^k S^4)$, is a ``second derived'' object with respect to
$\mc{L}_c^k S^4$: $W(E_k)$ is the group of ``symmetries of symmetries'' of $\mc{L}_c^k S^4$.
This is typical for the role of Weyl groups in Lie theory: a Lie group is usually a group of 
symmetries of a certain mathematical object, the Cartan subalgebra is the maximal abelian Lie algebra of infinitesimal symmetries of that object. The Weyl group is a group of symmetries of the Cartan subalgebra.
In this sense, we can write \eqref{WfromN} as 
\vspace{-2mm} 
$$
T^k \longrightarrow \mathcal{N}(T^k) \longrightarrow \Aut (T^k)\,.
$$ 

\paragraph{Geometric aspects of the moduli space of $k$-fold cyclifications of $S^4$.}
As in \cite{SV1}, we can interpret an element $\omega$ of the linear dual space $\h_k^* \cong \R^{k+1}$ of the abelian Lie algebra $\h_k$ of infinitesimal symmetries of $M(\mc{L}_c^k S^4)$ as  an extra, geometric ingredient complementing the purely topological data carried by the real homotopy type of the $k$-fold cyclification $\mc{L}_c^k S^4$. Indeed, an element $\omega \in \h_k^*$ is a {\it weight}, and as such, it tells us which ``spectral parameters'' we might want to assign to the basic infinitesimal symmetries $h_0, h_1, \dots, h_k$.

\begin{itemize} 
\item For instance, recall from \Cref{natural0} that $h_0$ comes from the action of the real 1-torus $\R^\times$ by automorphisms of the real homotopy type of $S^4$ and the resulting action of $\R^\times$ on $\mc{L}_c^k S^4$. That action originates ultimately from the folding self-maps of $S^4$. The value $\omega(h_0) \in \R$ tells us how much we shall value the effect of the folding self-maps of $S^4$. In this sense, $\omega(h_0)$ is akin to the size of $S^4$, such as its radius $R_0$, or rather the logarithm $\log R_0$ thereof, since $\omega(h_0)$ is not necessarily positive. This value $\omega(h_0)$ is analogous to the logarithmic Planck scale $\log \ell_p$ in the 11-dimensional supergravity and the generalized K\"{a}hler volume $\omega(H) = \int_H \omega$ of the line $H = \CC \PP^1$ in $\CC P^2$ and its image in $\BB_k$ in the del Pezzo story; cf. \cite[\S 3.1]{INV}.

\item Similarly, from \cref{symm-cycl-concrete}, the element $h_i$ for each $i$, $1 \le i \le k$, comes from the action of $\R^\times$ by the folding self-maps of the $i$th source circle of the cyclic loop space $\mc{L}_c^k S^4$. Assigning $h_i$ a real value $\omega(h_i)$ tells us how much we shall value the effect of the folding self-maps of the $i$th source circle of $\mc{L}_c^k S^4$. In this way, $\omega(h_i)$ is analogous to the logarithm $\log R_i$ of the radius $R_i$ of the $i$th source circle and the $i$th compactification circle in M-theory on $T^k = (S^1)^k$, as well as the generalized K\"{a}hler volume $\omega(\cE_i)$ of the exceptional curve $\cE_i$ in $\BB_k$ in the del Pezzo story; cf. \cite[\S 3.1]{INV} again.
\end{itemize}

In this sense, the choice of a weight $\omega \in \h_k^*$ adds a certain ingredient of \emph{metric} flavor, missing in the real homotopy model $M(\mc{L}_c^k S^4)$ of $\mc{L}_c^k S^4$. For example, an arbitrary weight $\omega \in \h_k^*$ will not be a real homotopy invariant of $\mc{L}_c^k S^4$. The values $\omega(h_i)$, $ i = 0, 1, \dots, k$, that is to say, the logarithmic radii of the target sphere $S^4$ and the source circles $S^1$, may be thought of as the coordinates of the weight $\omega$ in the space of all weights.

\medskip 
As in \eqref{abelianization}, we will provide a concrete description of the picture:
$$
{\color{darkblue} \text{Nonabelian \; moduli \; space}} \quad 
\fbox{$K \backslash G / G(\ZZ)$} 
\xymatrix{\ar@{~>}[rr]^{\text{\bf \color{darkblue} abelianize}}&&}  
\fbox{$\mathcal{M}_k = A/W$} \quad  {\color{darkblue} \text{Abelian \; moduli \; space }}
$$

The Weyl group $W(E_k)$, being the group of symmetries of the $E_k$ data $(\h_k^\ZZ, (-,-), K_k)$, acts on the linear dual $\h_k^*$. 
Thus, 
it makes sense to identify the weights $\omega$ brought together by this action. 
We call the corresponding quotient orbifold the \emph{moduli space $\mc{M}_k$ of $k$-fold cyclifications of $S^4$}:
\vspace{-1mm}
\[
\mc{M}_k = \mc{M}_k (\mc{L}^k S^4)
:= [\h_k^* / W(E_k)].
\]

\vspace{-1mm}
\noindent This is a stacky quotient $[\h_k^* / W(E_k)]$, which is different from the naive, topological quotient
$\h_k^* / W(E_k)$. Another reincarnation of the quotient orbifold $[\h_k^* / W(E_k)]$ is the familiar homotopy 
quotient $\h_k^* \dslash W(E_k)$, which may be realized via the Borel construction, 
but 
in this context, the orbifold viewpoint would be more common. The $(k+1)$-dimensional topological quotient 
$\h_k^* / W(E_k)$ contains the $k$-dimensional quotient $K_k^\perp/ W(E_k)$, which for $k \le 8$ may be 
identified with the closure 
of a \emph{Weyl chamber}
for the Weyl group action in the Euclidean space $K_k^\perp = \{\omega \in \h_k^* \; | \; \omega (K_k) = 0\} 
\subset \h_k^*$, cf.\ \cite[Prop. 8.29]{Hall15}.

\medskip 
Another motivation for the form $\mathcal{M}_k = A/W$ of moduli space may come through the Chevalley Restriction Theorem $\g_\CC\dslash G_\CC \cong \h_\CC\dslash W$, where $\g_\CC\dslash G_\CC$ is, by definition, the spectrum
$\Spec (\CC[\g]^{G_\CC})$ of the algebra of polynomial functions on $\g_\CC$ which are invariant under the adjoint 
$G_\CC$-action and $\h_\CC\dslash W := \Spec (\CC[\h_\CC]^W)$. The spaces $\g_\CC\dslash G_\CC$ and $K \backslash G /G(\ZZ)$ 
cannot be isomorphic, as $\dim_\CC \g_\CC\dslash G_\CC = \dim_\CC \h_\CC\dslash W = \dim_\CC \h_\CC = \dim_\RR \h$ 
while $\dim_\RR K \backslash G /G(\ZZ) = \dim_\RR A + \dim_\RR N = \dim_\RR \h + \dim_\RR N$, but should nevertheless be related.

\paragraph{Further interpretation.}
\label{Interpretation}
We highlight the following:

\noindent {\bf (i)} Identifying the weights $\omega$ under a permutation of the radii of the circles entering
$\mc{L}_c^k S^4$ is similar to identifying punctured Riemann surfaces under a permutation of the punctures,
if we wish to consider the moduli space of Riemann surfaces with unlabeled punctures.
The identification of weights under the Weyl group action is also analogous to identifying the Planck scale 
and the sequence of radii of the circle components in compactified M-theory under U-duality. 
In the del Pezzo story, one also identifies generalized K\"{a}hler classes on a del Pezzo surface under
the action of the Weyl group, see \cite[\S 3.1]{INV}, and the fundamental Weyl chamber in the Picard 
group of a del Pezzo surface plays a prominent role in studying Cremona isometries \cite[\S 8.2.8]{Dolgachev}.

\vspace{1mm}
\noindent {\bf (ii)}  {\it The Weyl group as a symmetry:} 
There is an action by the group $W$ on the group ${\rm Pic}(\mathbb{B}_k)$ 
 of divisor classes of $\mathbb{B}_k$
  that fixes $\cK_k$ and  preserves
the intersection product of ${\rm Pic}(\mathbb{B}_k)$.
 For $\cE=\{\cH, \cE_0, \cdots, \cE_k\}$ an exceptional configuration on 
$\mathbb{B}_k$, any other configuration 
$\cE'$ is a $W$-translate of $\cE$ and, if $\mathbb{B}_k$ is a blowup of 
$\CC \PP^2$  at 
generic points, every $W$-translate of $\cE$ is an
exceptional configuration \cite{Nagata61}. 
This can be refined further \cite{Har85}:
For $\sigma \in W$ associated to $\cE$, we have that 
$\cE'=\{\cH'=\sigma \cH, \cdots, \cE_k'=\sigma \cE_k\}$
is an exceptional configuration on $\mathbb{B}_k$ if and only
if $\cH'$ is numerically effective and 
$\cE_i' - \cE_j'$  is not effective for $i>j>0$ \cite{Har85}, that is, if we keep the signs of the
intersection products intact. 

\vspace{1mm}
\noindent {\bf (iii)} 
 Ultimately, physical configurations have to be U-duality invariant, so 
it makes sense to mod out by that symmetry. 
The Weyl group $W= W(E_k)$ is traditionally taken as a subgroup 
of the discrete $\mathbb{Z}$-form of the U-duality group $E_k(\RR)$;
see the discussion at the end of \cref{dP}. 
And since $W$ already contains a substantial part of that symmetry, 
modding out by $W$ makes sense and is also `close' to the 
ultimate moduli space, in the sense of \eqref{abelianization}.

\subsection{Automorphisms of del Pezzo surfaces, roots systems, and Weyl groups}
\label{dP}

The connection between algebraic geometry and Lie theory comes from the fact that the Cartan matrices of the exceptional Lie algebras of type $E_k$ and their
root systems arise from the above data: a lattice with a distinguished element and inner product
(see \cite{Man}).
In general, even for a fixed $k$, the surfaces $\BB_k$ obtained from varying the blowup points are not isomorphic as complex manifolds. However, they are diffeomorphic, so these surfaces give rise to the same combinatorial data, and we will just speak of ``the'' del Pezzo surface $\BB_k$ for each $k$.
We have extracted analogous data from topology of the fields 
in \cref{Sec-TorSymm}, \cref{Sec-toptor}, 
 and \cref{symm-cycl-concrete}.


\paragraph{The data of the triple} 
For $3 \le k \le 8$, which is assumed throughout this subsection, there exists 
in the lattice 
$$
N_k := {\rm Pic}(\BB_k) = H^2(\BB_k; \ZZ)
$$ 
a free basis $(\cH, \cE_1, \cdots, \cE_k)$ such that \cite{Man} for
\vspace{-3mm} 
\(
\label{eqKk}
\cK_k := -3 \cH + \sum_{i=1}^k \cE_i,
\)

 \vspace{-2mm} 
\noindent
the scalar product on $\R \otimes_\Z N_k \cong \R^{k+1}$ defined by
\eqref{HE-inter}
induces on the orthogonal complement $\cK_k^\perp$ of $\cK_k$ the structure 
of a Euclidean vector space (with a negative-definite form) of dimension $k$.
Hence the triple 
$$
\big\{{\rm Pic}(\BB_k), \, \cK_k, \, \text{intersection form}\big\}\,
$$ 
up to isomorphism, depends only on $k$. The triple ($N_k$, $\cK_k$, $(-,-)$), consisting of a lattice with an integral inner product and a distinguished element of inner square $9-k$, carries rich enough combinatorics to give rise to a root system.

\paragraph{Special subsets.} 
The following three special subsets of $N_k$ are of interest
(for $3 \le k \le 8$):
\vspace{-2mm} 
\begin{enumerate}[{\bf (1)}]
  \setlength\itemsep{-2pt}

\item
{\it Root system:} 
The set 
\vspace{-2mm} 
\(
\label{RkIk}
R_k := \left\{ l \in N_k \; | \;  (l, \cK_k)=0, (l, l)=-2\right\}
\)
is a
rank $k$ root system in 
$\cK_k^\perp \cong \RR^{k}$ of type 
$
A_1 \times A_2,\, A_4, \, D_5, \, E_6, \, E_7, \, E_8,
$
respectively.

\item
{\it Root lattice:} The root lattice of type $E_k$ is the $\ZZ$-span $\mathbb{E}_k$ of $R_k$; here we also have $\mathbb{E}_k = \cK_k^\perp \cap N_k$.

\item {\it Exceptional vectors:} 
\(
\label{set-excep} 
I_k := \left\{ l \in N_k \; | \; (l, \cK_k)=(l, l)=-1\right\}\;.
\)
We will identify these groups later (see \cref{Sec-27}). 

\item {\it Simple roots:}
One can choose the simple roots 
of $E_k, 3\leq k\leq 8,$ as
\begin{eqnarray}
\label{simpleroots-E8}
\label{Eq-vectors0i}
\alpha_{0}
:= \cH- \cE_{1}- \cE_{2}- \cE_{3}  
\qquad 
\mbox{and}
\qquad 
\alpha_{i} := \cE_{i}-\cE_{i+1},~~i=1, \ldots, k-1.
\end{eqnarray} 

\vspace{-2mm} 
\noindent Their negative intersection numbers yield the $E_k$ 
Cartan matrix $C(E_k)$. See \cref{Sec-KM} for the case $k\geq 9$. 
\end{enumerate}

\paragraph{Special groups.}
With the subsets above, the following groups coincide:

\vspace{-2mm} 
\begin{enumerate}[{\bf (a)}]
  \setlength\itemsep{-1pt}
\item The group of automorphisms of the lattice $N_k$ preserving $\cK_k$
and the scalar product $(-,-)$;

\item The group of permutations of the vectors from $I_k$ preserving
their pairwise scalar products;

\item  As in \eqref{W-refl}, the Weyl group $W_k = W(R_k)$ of the system $R_k$ generated by the 
reflections with respect to the roots. 
Any vector $\alpha \in \ZZ^{1,k}$ with $\alpha^2=-2$ determines a reflection
$\sigma: \ZZ^{1,k}\to \ZZ^{1,k}$ by $y \mapsto y + (y, \alpha)\alpha$. 
The Weyl group $W_k \subset \operatorname{O}(\ZZ^{1,k})$
is the group generated by 
the reflections $\sigma_i$
through the
simple roots $\alpha_0, \alpha_1, \dots, \alpha_{k-1}$.
\end{enumerate} 

\vspace{-3mm} 
\noindent Note that this approach
differs from the usual treatment of considering 
$\mathbb{R}^k$. Using $\mathbb{R}^{k+1}$ as above is more convenient combinatorially 
and allows for the system of exceptional vectors $I_k$.

\paragraph{The Weyl group and the root lattice.} 
The set of roots $R_k=\bigcup_i W_k \alpha_i$ of $E_k$ is the union of $W_k$-orbits of the simple 
roots $\alpha_0, \cdots, \alpha_{k-1}$, where the latter form an integral basis 
for the root lattice 
$$
\mathbb{E}_k = \bigoplus_i \ZZ \alpha_i \subset  \ZZ^{1,k}\;.
$$

\vspace{-2mm} 
\noindent 
This root lattice is a lattice of rank $k$, invariant under the action of $W_k$, and is
\vspace{-1mm} 
\begin{itemize} 
\setlength\itemsep{-2pt}
\item negative-definite
for $k\leq 8$, 
\item negative semi-definite for $k=9$, 
\item and of signature 
$(1,k-1)$ for $k \geq 10$. 
\end{itemize}

\paragraph{Markings.} 
Let $\pi: \mathbb{B}_k \to \CC \PP^2$ be a rational surface presented as 
the blowup of $\CC \PP^2$ at $k$ points $(x_1, \dots, x_k)$ in general position.
Let $\ZZ^{1,k}$ denote the free abelian group $\ZZ^{k+1}$ with the standard basis 
$\{\epsilon_0, \epsilon_1, \dots, \epsilon_k\}$
and the Minkowski 
inner product 
\(
\label{Mink-inn}
(\vec{x}, \, \vec{y})= \vec{x}\cdot \vec{y}=x_0 y_0 - x_1 y_1 - \cdots - x_k y_k\;
\)
for $\vec{x} = (x_0, x_1, \dots, x_k)$ and $\vec{y} = (y_0, y_1, \dots, y_k)$.
There is a canonical {\it marking} isomorphism 
\begin{align}
\label{eq-mark} 
\phi: \ZZ^{1,k} & \longrightarrow H^2(\mathbb{B}_k; \ZZ) ={\rm Pic}(\mathbb{B}_k)
\\
\epsilon_0 & \longmapsto \cH
\nonumber \\
\epsilon_i & \longmapsto \cE_i, \;\; i=1, \dots, k. 
\nonumber 
\end{align} 
We call the pair $(\BB_k, \phi)$ a \emph{marked blowup}. Additionally, the canonical class \eqref{eqKk} of $\mathbb{B}_k$ is given by 
$$
\cK_{k}
=\phi ({\bf k}_k)\;
$$

\vspace{-2mm} 
\noindent the image of the special vector 
$
{\bf k}_k=(-3, 1, \dots, 1)=-3\epsilon_0 +\sum_{i=1}^k \epsilon_i
$.

\paragraph{Degree of del Pezzo surfaces.} 
A del Pezzo surface is (equivalently) 
a smooth birationally trivial surface $S$ on which the 
 anticanonical sheaf $-\cK$ 
is ample. The number $d=(\cK, \cK)=\cK.\,\cK$ is 
called the degree of $S$; cf. \eqref{eqDegD}. 
In general, $1 \leq d \leq 9$. For degree $d \ge 3$, it coincides with the 
degree of
the projective embedding
$i: S \hookrightarrow \mathbb{C} \PP^d$
defined by
$-\cK$, so as $-\cK = i^* \mathcal{O}_{\mathbb{CP}^d}(1)$.
One has the following characteristics
of a del Pezzo surface $S$ of degree $d$
(see \cite{Man}\cite{Dolgachev}):
\vspace{-1mm} 
\begin{enumerate}[{\bf (i)}]
  \setlength\itemsep{-2pt}
\item 
$S$ is isomorphic to the blowup $\BB_k$ of $\CC\PP^2$ at $k=9-d$ points as above;

\item Every irreducible curve with a negative self-intersection number on 
$S$ is exceptional;

\item If $S$ has no exceptional curves, then either $d=9$ and $S$ is isomorphic 
to $\mathbb{C}\PP^2$, or $d=8$ and $S$ is isomorphic to 
$\mathbb{C} \PP^1 \times  \mathbb{C} \PP^1$. 
\end{enumerate}

\vspace{-2.5mm} 
\paragraph{Automorphisms (of cohomology) of del Pezzo surfaces.} 
The marking \eqref{eq-mark} sends the Minkowski inner product
\eqref{Mink-inn} to the intersection 
pairing on $H^2(\mathbb{B}_k; \ZZ)$
\vspace{-3mm} 
$$
\xymatrix{
\ZZ^{1,k} \times \ZZ^{1,k} \ar[d]_{\phi \times \phi} 
\ar[rr]^{(-, -)} && \ZZ \ar[d]
\\
 H^2(\mathbb{B}_k; \ZZ)  \times H^2(\mathbb{B}_k; \ZZ) 
 \ar[rr]^{\cdot } && \ZZ 
}
$$
under which we get the relations \eqref{HE-inter}.
The marking determines the birational morphism $\pi: \mathbb{B}_k \to \CC \PP^2$ 
up to post-composition with an automorphism of $\CC \PP^2$.

\paragraph{Cremona transformation.} 
In slight generalization of \eqref{Eq-vectors0i},
the Cremona involution $\kappa_{ijl} \in \operatorname{O}(\mathbb{E}_k)$ is given by reflection in the vector 
$\alpha_{ijl}=\epsilon_0  - \epsilon_i - \epsilon_j - \epsilon_l$ for distinct indices $i,j, l \geq 1$. 
It acts by 
\begin{align*}
& \epsilon_0  \mapsto 2\epsilon_0 - \epsilon_i -\epsilon_j - \epsilon_l 
, \quad 
\epsilon_i  \mapsto \epsilon_0  -\epsilon_j - \epsilon_l 
, \quad 
\epsilon_j  \mapsto \epsilon_0 - \epsilon_i - \epsilon_l 
, 
\\
\;\;
& \epsilon_l  \mapsto \epsilon_0 - \epsilon_i -\epsilon_j 
, \qquad \quad \;
\epsilon_m  \mapsto  \epsilon_m \quad \text{if} \; m \not \in \{0, i, j, l\}. 
\end{align*} 
The standard quadratic Cremona transformation of $\CC \PP^2$ at the points
$[1,0,0]$, $[0,1,0]$, $[0,0,1]$
determines an
automorphism of $\ZZ^{1,k}$, which is the 
reflection $\sigma_0 = \kappa_{123}$ through the hyperplane orthogonal to 
$\alpha_0 = \cH - \cE_1 -\cE_2 - \cE_3$. 
Switching points $x_i$ and $x_{i+1}$ acts on  $\ZZ^{k+1}$
by the reflection $\sigma_i$ through the hyperplane orthogonal to $\alpha_i = \cE_i - \cE_{i+1}$ for $1\leq i \leq k-1$. 
The Cremona action on ${\rm Pic}(\BB_k)= \ZZ^{k+1}$ fixes the anticanonical class
$-\cK_k$.

\paragraph{$\Aut (\mathbb{B}_k)$ vs.\ $W(E_k)$.}  Now suppose there are two birational morphisms 
$\pi, \pi': \mathbb{B}_k \to \CC \PP^2$, exhibiting $\mathbb{B}_k$ as the blowup of 
$\CC \PP^2$ at the points $x_i$ and $x_i'$, respectively, $i=1, \cdots, k$. 
Then there is a birational map $f$ making the diagram 
$$
\xymatrix{
\mathbb{B}_k \ar[d]^{\pi} 
\ar@{=}[rr] && \mathbb{B}_k\ar[d]^{\pi'}
\\
\CC \PP^2
 \ar@{-->}[rr]^{f } &&  \CC \PP^2
}
$$
commute, and the corresponding markings are related by $\phi'=\phi \circ \sigma$ for a 
unique $\sigma \in \operatorname{O}(\ZZ^{1,k})$. The different blowups are related by the action of
the Weyl group \cite[p.\ 283]{Nagata60}\cite[Thm.\ 2, p.\ 90]{DOrt}: 
that is, 
$\sigma \in W_k$.
Hence, if $(\mathbb{B}_k, \phi)$ is a marked blowup, and $\sigma \in \operatorname{O}(\ZZ^{1,k})$
such that $(\mathbb{B}_k, \phi \circ \sigma)$ is  also a marked blowup, then 
$\sigma \in W_k$. With 
\(
\label{Eq-marked} 
W(\mathbb{B}_k, \phi)=\{ \sigma \in W_k ~|~ (\mathbb{B}_k, \phi \circ \sigma) \; \text{is a marked blowup}\} 
\)
denoting the subgroup of $W_k$ which is realized by isomorphisms of marked blowup structures on $\BB_k$
\cite[Thm 5.2]{McMullen} :

\noindent  {\bf (i)} The right action of the symmetric group $\Sigma_k$ simply reorders the basepoints
of a blowup, so $\Sigma_k \subset  W(\mathbb{B}_k, \phi)$; 

\noindent {\bf (ii)} The 
Cremona involution $\kappa_{123}$, see above,
is in $W(\mathbb{B}_k, \phi)$.

\paragraph{Weyl group action and the form fields.}
 In \cite{INV}, the dualities are restricted to compactifications with the
``rectangular, no $C$-field" condition.  The full duality group acting on 
such  compactifications is  isomorphic to the Weyl group
of 
$E_k$  \cite{BFM}\cite{OP} and is  generated by:

\vspace{-2mm} 
\begin{enumerate}[{\bf (i)}]  
\setlength\itemsep{-2pt}
\item The symmetric group $\Sigma_k$ permuting the $k$
radii; this group is generated by the $k-1$ exchanges  \cite[(48)]{INV}
\(
\sigma_{i} : R_i  \longleftrightarrow R_{i+1}, \;\;  i = 1, \dots , k-1.       
\label{wtrans}
\)
\item The action of T-duality (M2/M5 exchange, after $T^3$
compactification) which, when expressed in M-theory language, involves three radii, 
 requiring one extra generator when $k \geq 3$, namely \cite[(49]{INV} 
\(
\label{Eq-T3duality} 
\hspace{-5mm} 
\sigma_{0}: 2\pi R_1 \mapsto \frac{\ell_p^3}{ 2\pi R_2 2\pi R_3}\;,\quad 
2\pi R_2 \mapsto \frac{\ell^3_p}{ 2\pi R_1 2\pi R_3}
\;,\quad 
2\pi R_3 \mapsto \frac{\ell^3_p}{ 2\pi R_1 2\pi R_2}
\;,\quad 
\ell_p^3
\mapsto \frac{\ell^6_p}{ 2\pi R_1 2\pi R_2  2\pi R_3}\;.
\)
\end{enumerate} 
\noindent {\it At the level of roots}, we have, respectively, 
$$
\alpha_i \longleftrightarrow \alpha_{i+1} 
$$
and
\vspace{-2mm} 
$$
\alpha_1 \longmapsto \alpha_0 - \alpha_2 - \alpha_3\;, \quad 
\alpha_2 \longmapsto \alpha_0 - \alpha_1 - \alpha_3\;, \quad 
\alpha_3 \longmapsto \alpha_0 - \alpha_1 - \alpha_2\;, \quad 
\alpha_0 \longmapsto 2 \alpha_0 - \alpha_1 - \alpha_2 - \alpha_3\;.
$$


\medskip 
\noindent {\it At the level of fields}, these correspond, respectively, to the following: 

\vspace{-2mm} 
\begin{enumerate}[{\bf (i)}]  
\setlength\itemsep{-2pt}
\item The first, \eqref{wtrans}, 
is also related to the toroidification (see \cref{Sec-toroid}), 
in which the classes of the circle bundles are on an equal footing and may be exchanged 
$$
\omega_i  \longleftrightarrow \omega_{i+1} , \;\;  i = 1, \dots , k-1;
$$

\item The second, \eqref{Eq-T3duality}, is captured by
2-form and 4-form degree-preserving 
transformations
exchanging the Euler classes of the circle bundles with the 2-contractions of $G_4$
and exchanging $G_4$ with the 3-contraction of $G_7$:
\(
\label{eq-wssG}
\omega_1 \longmapsto s_2 s_3 G_4\;, \quad  
\omega_2 \longmapsto s_1 s_3 G_4\;, \quad 
\omega_3 \longmapsto s_1 s_2 G_4 \;, \quad 
G_4 \longmapsto s_1 s_2 s_3 G_7
\;.
\)
\end{enumerate} 

\vspace{-3mm} 
\noindent  Here $s_i$ corresponds to reduction (i.e., contraction) 
 along the $i$th circle direction (in the notation of 
 \cref{Sullmm} extended to \cref{Sec-toroid}).

\paragraph{Clarification of the relation between $W(G)$ and $G$.}
{\it Can we always realize $W$ as a subgroup of $G$?}
This is not possible in general: $W$ may rather be realized as a subquotient $W = \mc{N}(T)/T$, where $\mc{N}(T) \subset G$ is the normalizer of $T$ in $G$. 
However, in the physics literature, $W$ seems to be treated as
a subgroup of the discrete form of $G$; e.g. in \cite{INV}: 
The non-obvious part of the U-duality group is given by the 
Weyl group, hence it is natural to mod out by the action 
$W(E_k)$. In the following example, we consider the setting relevant for type IIB and the reduced theory in 9 dimensions.

\begin{example}[Weyl group of ${\rm SL}(2, \CC)$]
Note that already for rank 1, one has 
$W({\rm SL}(2, \CC)) \not\subset {\rm SL}(2, \ZZ)$.
The Weyl group has order 2, but the only element of order 2 in the normalizer $\mathcal{N}(T)$ is $\footnotesize\arraycolsep=0.5\arraycolsep\ensuremath
\begin{pmatrix}
-1 & \; 0\\ \;\; 0 & -1
\end{pmatrix} \in T$,
which is trivial in $W = \mc{N}(T)/T$;
any representative $\footnotesize\arraycolsep=0.7\arraycolsep\ensuremath\begin{pmatrix}
0 & a\\ -1/a & 0
\end{pmatrix}\in \mc{N}(T)$ of a nontrivial element of $W$ has order 4.
\end{example} 

In fact, for a simply connected group, this is essentially the only obstruction: 
The Weyl group can always be lifted (uniquely up to conjugation) to a group $\widetilde{W}$
which is an extension of $W$ by an elementary abelian 2-group of order $2k$,
where $k$ is the rank of $G$ \cite{Tits66}. 

\medskip
However, the right question to ask about the relation between $W(G)$ and $G(\ZZ)$ might not be   
 through the former being a subgroup of the latter, but rather through 
 {\it conjugation action},
which explains  the appearance of the moduli space $\mathcal{M}_k = A/W$ in \cite{INV} (see \eqref{abelianization} in the Introduction). The U-duality group of 11d supergravity reduced on a $k$-torus is a connected semisimple real form $G$ of type $E_k$.
With the Iwasawa decomposition $G=KAN$, the more standard moduli space $K \backslash G /G(\ZZ)$ of 11d supergravity reduced on a $k$-torus may be identified with $(AN)/ G(\ZZ)$. If we ignore the topologically trivial unipotent factor $N$, as explained in the Introduction, we turn the right action of $G(\ZZ)$ into a conjugation action of the normalizer $\mc{N}(A)$ (or its integral form) of $A$ on $A$. Since $A$ is abelian, its conjugation action on itself is trivial, so the normalizer action on $A$ reduces to an action of the Weyl group $W = \mc{N}(A)/A$.

\section{Type IIB: description and compatibility checks}
We have seen explicit connections to M-theory in 
\cref{Sec-S4cyc} and to its dimensional reductions in  \cref{Sec-CyclicSugra},
including type IIA in \cref{Sec-IIAsym}. Now we consider the type IIB case
and provide interrelations to the above.

\subsection{Type IIB and S-duality}
\label{Sec-IIBS}

We start by applying the discussion in \cref{Sec-RvsC} to type IIB string theory. 

\paragraph{${\rm SL}(2, \CC)$ and type IIB in nine dimensions.}
\label{Ex-SL2C}
${\rm SL}(2,\CC)$ has two real forms: $\theta^s$ and $\theta^c$
(see \cite[Ex. 12.20]{AdC}).
\vspace{-2mm} 
\begin{enumerate}[{\bf (i)}]  
\setlength\itemsep{-2.5pt}
\item
{\it Real forms via anti-holomorphic involutions: }
Up to conjugacy, there are two antiholomorphic involutions of 
${\rm SL}(2,\CC)$: $\sigma^s(g)=\overline{g}$ or $\sigma^c(g)={}^t\overline{g}^{-1}$.
These are the two real forms: the split $G(\RR)={\rm SL}(2, \RR)$ and 
the compact $G(\RR)={\rm SU}(2)$. 

\item
{\it Real forms via holomorphic involutions: } 
Equivalently, ${\rm SL}(2,\CC)$ has two equivalence classes of holomorphic 
involutions.

\vspace{-3mm} 
\begin{enumerate}[{\bf (a)}]  
\setlength\itemsep{-2pt}
\item
Let $\theta^s(g)=t gt^{-1}$ where $t={\rm diag} (i, -i)$. Then  
the maximal compact subgroup is
$K(\CC)=G^{\theta^s}= \CC^\times$, so $K(\RR)=S^1$, the maximal 
compact subgroup of the corresponding real form ${\rm SL}(2, \RR)={\rm SU}(1,1)$. 

\item Let  $\theta^c(g)=g$, i.e., $\theta^c=1$, so  
$K(\CC)=G^{\theta^c}= {\rm SL}(2, \CC)$, hence $K(\RR)={\rm SU}(2)$,
and the corresponding real form is ${\rm SU}(2)$. 
\end{enumerate} 

\vspace{-2mm} 
\item {\it Normalizer:} 
Let $H=\{{\rm diag} (z, 1/z) | z \in \CC^\times \}$, $W=+\{1, s\}$, 
$t={\rm diag} (i, -i)$, and $n$ be any element of $N_G(H)$ mapping to $s \in W$. 
Then $N_G(H)=H \cup Hn$. 

\item {\it Cartan subgroups:} As in \cref{Sec-RvsC},
${\rm SL}(2, \RR)$ has two conjugacy classes of Cartan subgroups. The compact
Cartan subgroup $H(\RR)=T\simeq S^1$ corresponds to $1\in W$, and the split 
Cartan subgroup $H(\RR)=A \simeq \RR^\times$ corresponds to $s \in W$. 

\end{enumerate}

We will describe S-duality in type IIB in a fashion that is compatible with the 
description using the 4-sphere for M-theory in \cref{smm} and with the perspective of modding out by complex conjugation in 
\cref{Subsec-RvsC}, building on  the constructions in \cite{Ku74}\cite{Lawson}\cite{HH11}.

\medskip 
The involution $(x, y) \mapsto (y, x)$ exchanging the two factors on $\CC \PP^1 \times \CC \PP^1 \cong S^2 \times S^2$ has as its fixed point set 
the diagonal $\Delta=\Delta(\mathbb{C} P^1)$ consisting of all pairs
$\{x, x\}$ and is homeomorphic to $\mathbb{C}\PP^1$.
It acts as the antipodal map on the anti-diagonal (see below), and the quotient is $\mathbb{C}\PP^2$.
We interpret this as corresponding to S-duality. 

\begin{center}
    \begin{tabular}{cc}
    \hline 
   \bf Type IIB with S-duality  &  \bf M-theory with no S-duality \\
   \hline
   \hline 
\rowcolor{lightgray} $\mathbb{Z}_2 
\;\; \rotatebox[origin=c]{-90}{$\circlearrowright$}~~ \; (\mathbb{C}\PP^1 \times \mathbb{C}\PP^1)$     &
$(\mathbb{C}\PP^1 \times \mathbb{C}\PP^1)/\mathbb{Z}_2$ 
\\
\hline
\end{tabular}
\end{center}
This is described by the symmetric square: ${\rm Sp}^2(\mathbb{C}\PP^1):=(\mathbb{C} \PP^1 \times  \mathbb{C} \PP^1)/\mathbb{Z}_2=\mathbb{C} \PP^2$. It is the orbit space under the action of the 
cyclic group $\mathbb{Z}_2$ generated by the involution $\iota$ that interchanges the two factors. 
Its elements are the unordered pairs $\{x, y\}$ for $x, y \in \mathbb{C}\PP^1$. The fixed point of 
$\iota$ determines the diagonal subspace $\Delta=\Delta(\mathbb{C} P^1)$.

\paragraph{The diagonal and self-dual points.} 
From a physics perspective, 
the diagonal $\Delta$  represents the elements fixed under the duality, i.e., the self-duality points. 
We interpret these as corresponding to the S-duality invariant combinations of field strengths, such as the 
combination 
$
G_3=\frac{1}{\sqrt{{\rm Im}\tau}} (F_3 - \tau H_3)
$
or the modified field strength 
$
\widetilde{F}_5=F_5 +\tfrac{1}{2} \epsilon_{ij}B^i \wedge H^j\;, \; i \in \{1, 2\}
$,
where $H^i=(H_3, F_3)$ is an ${\rm SL}(2, \RR)$ doublet of fields
and $\tau$ is the complex parameter formed of the axion and the dilaton (e.g., as in \cite{GSVY}).

\paragraph{The passage from type IIB to M-theory via $S^4$.} We have seen the effect of S-duality
given by interchanging the 
factors $(S^2 \times S^2)/(x, y) \sim (y, x)$. 
We now combine this with the other involution 
given  by complex conjugation $\sigma$, equivalent to $[(x, y)] \mapsto [(-x, -y)]$.
Consider the sequence 
$$
S^2 \times S^2 \longrightarrow 
(S^2 \times S^2)/J ={\rm Sp}^2(S^2)=\mathbb{C}\PP^2 \longrightarrow
(S^2 \times S^2)/L =S^4
$$
where  we have the pair of collections of involutions 
\vspace{-2mm} 
\begin{align*}
    J &=\{{\rm Id}, (x,y) \mapsto (y, x) \} \;,
    \\
    L &=\{{\rm Id}, (x,y) \mapsto (y, x), (x, y) \mapsto (-x, -y), (x, y) \mapsto (-y, -x) \} \;,
    \end{align*}
with the action of $L/J=\mathbb{Z}_2$ being equivalent to complex conjugation $\sigma$. 
We see that only products $xy$ survive, forming invariant combinations. Overall: 
$$
\xymatrix{
 S^2 \times S^2 \ar[rrrr]^-{\rm \bf  \color{darkblue} quotient\; by\; S-duality}_-{(x,y)\, \mapsto (y, x)} 
 &&&&
\mathbb{C}\PP^2 
\ar[rrrrr]^-{\rm \bf  \color{darkblue} quotient\; by\; complex \; conjugation}_-{(x,y)\, \mapsto (-x, -y), \;\; (x,y) \, \mapsto (-y, -x)} &&&&&
S^4
}.
$$
Alternatively, in terms of complex coordinates: 
Consider the $\mathbb{Z}_2$-action on  $\mathbb{C} \PP^1 \times  \mathbb{C} \PP^1$  
generated by  the involution $(z, w) \mapsto (w, z)$. It has the diagonal 
$\Delta=\{(z, z)| z\in \mathbb{C} \}$
as the fixed 
point set, and it acts as the antipodal map on the anti-diagonal 
$\overline{\Delta}=\{(z, -\frac{1}{\overline{z}})| z\in \mathbb{C} \}$.
Next, consider the combined $\mathbb{Z}_2 \times \mathbb{Z}_2$-action 
on $\mathbb{C} \PP^1 \times  \mathbb{C} \PP^1$ given by 
$$
f(z, w)= (w, z)
\qquad 
\text{and} 
\qquad 
 g(z, w)=\Big(-\frac{1}{\overline{z}},
-\frac{1}{\overline{w}} \Big),  
$$
where $g$ is the antipode $\times$ antipode on $\mathbb{C} \PP^1 \times  \mathbb{C} \PP^1$.
Modding out by this action gives the 4-sphere
$$
(\mathbb{C} \PP^1 \times  \mathbb{C} \PP^1)/\mathbb{Z}_2 \times \mathbb{Z}_2 \cong S^4.
$$
Therefore, we see that the S-duality symmetric description of type IIB 
is not only compatible with the 
4-sphere description of M-theory, but admits an explicit relation and a natural transition to it (cf. \cref{Subsec-RvsC}).

\subsection{Type IIB and T-duality} 
\label{IIB}

We have worked out the Sullivan minimal model for type IIB in \cite{SV1}. 
In terms of $S^4$, the expected equivalence in nine dimensions  
would amount to the equivalence of Sullivan models 
$$
M(\mathcal{L}_c IIB) \cong M(\mathcal{L}^2_c S^4)
$$
and a real homotopy equivalence at the level of spaces
$$
\mathcal{L}_c IIB \sim \mathcal{L}^2_c S^4.
$$
The relation between the various models is as appropriate from the physics picture 
$$
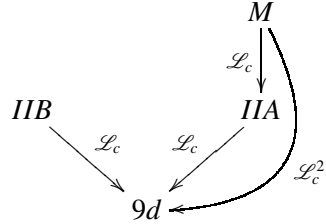
\xymatrix{
&& M \ar[d]_{\mathcal{L}_c}  \ar@/_-3pc/[ddl]^{\mathcal{L}_c^2}
\\
IIB \ar[dr]^{\mathcal{L}_c}  && IIA  \ar[dl]_{\mathcal{L}_c}
\\
& 9d
}
$$
and can be considered as a topological and physical analogue of the relations on the del Pezzo side
$$
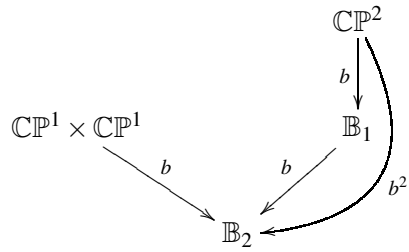
\xymatrix{
&& \CC \PP^2 \ar[d]_{b}  \ar@/_-3pc/[ddl]^{b^2}
\\
\CC \PP^1 \times \CC \PP^1 \ar[dr]^{b}  && \BB_1  \ar[dl]_{b}
\\
& \BB_2
}
$$
where $b$ denotes the process of blowing up (rather than the blowup map, which would go in the opposite direction).

\paragraph{The Sullivan minimal model $M(IIB)$ of the real homotopy type $IIB$.}
Consider the free graded commutative algebra $\RR[\omega_1, h_3, \omega_3, \omega_5, h_7, \omega_7]$, and define a differential by the following equations:
\begin{align}
\nonumber
d\omega_1&=0,\;\;\;\; \qquad \qquad \quad dh_3  =0,   
\\
\label{htypeIIB}
 d \omega_3&=h_3 \omega_1,  \qquad \qquad \;\; d\omega_5  =h_3 \omega_3,
\\
\nonumber
dh_7& =\omega_3 \omega_5 + \omega_7 \omega_1, \quad d\omega_7 =h_3 \omega_5.
\end{align}
Pulling back these universal differential forms to spacetime, 
we get the EOMs of $D=10$ type IIB supergravity
in the duality-symmetric formulation,
 and without imposing self-duality;
 cf.\  \cite{SW83}\cite{HW84}\cite{Schwarz83} for the classical formulation
 and \cite{DLS97}\cite{DLT98}
for the duality-symmetric formulation. 
Note that we do not include fields of degree greater than seven,
as in the dual picture of type IIA  (see \cref{Sec-IIAsym}),
this would require parametrized homotopy theory \cite{BSS}.
 Additionally, there are several subtleties in type IIB which makes a purely topological 
 perspective delicate due to the mixing between geometry and topology,
 in the sense that some of the fields arise from the metric, as we explain 
 further below.

\paragraph{Unstable T-duality: comparing $M(\mathcal{L}_c IIA)$ and $M(\mathcal{L}_c IIB)$.}
In \cite{SV1}, we computed $M(\mathcal{L}_c IIB)$, using the recipe of \cref{Sullmm}.
With 
\begin{gather*}
  c_2 := s h_3, \qquad \omega_2 := s \omega_3, \qquad \tilde{c}_2 := w, \quad 
    \omega_4 := s \omega_5, \qquad \omega_6 := s\omega_7,
  \end{gather*}
 we have the following for type IIB reduced to 9 dimensions:
    \begin{gather}
 \nonumber
 M(\mathcal{L}_c IIB) = \left(\RR[\omega_1, c_2,  \tilde{c}_2, \omega_2, h_3, \omega_3, \omega_4, \omega_5, sh_7, \omega_6, h_7, \omega_7], d \right)\\
  \nonumber
d\omega_1 = 0, \qquad d c_2 = 0, \qquad d \tilde{c}_2 = 0,\\
 \nonumber
   d \omega_2 = - c_2 \cdot \omega_1, \quad d h_3 = c_2 \cdot \tilde{c}_2, \quad d \omega_3 = h_3 \cdot \omega_1 + \omega_2 \cdot \tilde{c}_2,\\
\label{LIIB}
  d \omega_4 = - c_2 \cdot \omega_3 + h_3 \cdot \omega_2, \qquad d \omega_5 = h_3 \cdot \omega_3 + \omega_4 \cdot \tilde{c}_2,\\
  \nonumber
 d sh_7 = - \omega_2 \cdot \omega_5 + \omega_3 \cdot \omega_4 - \omega_6 \cdot \omega_1, \quad d \omega_6 = - c_2 \cdot \omega_5 + h_3 \cdot  \omega_4, \\
 \nonumber
  d h_7 = \omega_3 \cdot \omega_5 + \omega_7 \cdot \omega_1 + sh_7 \cdot \tilde{c}_2, \quad d\omega_7 =h_3 \cdot \omega_5 + \omega_6 \cdot \tilde{c}_2\,.
  \end{gather}
Similarly, for type IIA reduced to 9 dimensions we have the  model 
\begin{gather}
\nonumber
  M(\mc{L}_c IIA) = \left( \RR[\omega_1, c_2, \tilde{c}_2, \omega_2, h_3,  \omega_3, \omega_4, \omega_5, s h_7, \omega_6,  h_7], d \right),\\
 \nonumber
 d \omega_1 = 0, \qquad d c_2 = 0, \qquad d \tilde{c}_2 = 0, \\
 \nonumber
 d \omega_2 = \omega_1 \cdot \tilde{c}_2 , \quad dh_3 = c_2 \cdot \tilde{c}_2, \quad d \omega_3 = - c_2 \cdot \omega_2 + h_3 \cdot  \omega_1,\\
\label{LIIA}
d \omega_4 = h_3 \cdot \omega_2 + \omega_3 \cdot \tilde{c}_2, \qquad d \omega_5 = - c_2 \cdot \omega_4 + h_3 \cdot \omega_3,\\
  \nonumber
 d sh_7 = \omega_3 \cdot \omega_4 - \omega_5 \cdot \omega_2 -  \omega_6 \cdot \omega_1, \quad d \omega_6 = h_3 \cdot \omega_4 + \omega_5 \cdot \tilde{c}_2, \\
  \nonumber
 d h_7 = -\tfrac{1}{2} \omega_4^2 + \omega_6 \cdot \omega_2 + sh_7 \cdot \tilde{c}_2\,.
  \end{gather}

\paragraph{Matching in 9 dimensions.}
We compare $M(\mc{L}_c IIA)$ with $M(\mc{L}_c IIB)$. 
Up to replacement
\vspace{-2mm} 
\[
c_2 \leftrightarrow - \tilde{c}_2,
\]
all the equations for the differential of $M(\mc{L}_c IIA)$ in \eqref{LIIA} exactly match those of $M(\mc{L}_c IIB)$ in \eqref{LIIB}, except for the following mismatch for the generators of degree 7:
\begin{center}
\begin{tabular}{ccc} 
\hline 
\bf Model & \bf $M(\mc{L}_c IIA)$ & \bf $M(\mc{L}_c IIB)$ 
\\ 
\hline 
\hline 
\rowcolor{lightgray}
 Generators & $h_7$ & $h_7$, $\omega_7$\\
 \multirow{2}{2.2cm}{Differentials} & \multirow{2}{5.2cm}{$dh_7 = -\tfrac{1}{2} \omega_4^2 + \omega_6 \cdot \omega_2 + s h_7 \cdot \tilde{c}_2$} & $d h_7 = \omega_3 \cdot \omega_5 + \omega_7 \cdot \omega_1 - s h_7 \cdot c_2$,\\
 & & $d\omega_7 = h_3 \cdot \omega_5 - \omega_6 \cdot c_2$\\ 
 \hline 
\end{tabular}
\end{center}
 With relations imposed algebraically, this means
that
\begin{equation}
\label{T-duality}
M(\mc{L}_c IIA)/(h_7, dh_7) \cong M(\mc{L}_c IIB)/(h_7, \omega_7, d h_7, d \omega_7)\,,
\end{equation}
that is to say, the two DGCAs, $M(\mc{L}_c IIA)$ and $M(\mc{L}_c IIB)$, are isomorphic modulo the differential ideals generated by the generators of degree 7. Perhaps, topologically more interesting is an isomorphism between the dg-subalgebras $M_A \subset M(\mc{L}_c IIA)$ and $M_B \subset M(\mc{L}_c IIB)$ generated by all the generators except those of degree 7:
\begin{equation}
\label{T-duality-sub}
\xymatrix{
{}_{\phantom{A}_{\phantom{A}}}{M_A} \;\; \ar@{^{(}->}[d] \ar[r]^\sim &
\;\;\; M_B  {}_{\phantom{A}_{\phantom{A}}} \ar@{^{(}->}[d]\\
M(\mc{L}_c IIA) & M(\mc{L}_c IIB)\, .
}
\end{equation}
These dg-subalgebras are minimal Sullivan, and their isomorphism morally corresponds to the rational equivalence over $\RR$ of quotient spaces of $\mc{L}_c IIA$ and $\mc{L}_c IIB$.

\paragraph{Subtleties in comparing type IIB with type IIA.}
\label{Rem-notbad}
The situation above requires a discussion but it is not desperate, 
due to the several subtleties that
arise in type IIB, which once addressed properly will remove the mismatch. 
We will address these in detail below, but a summary is the following: 

\vspace{1mm} 
\noindent  {\bf (i)} Classical supergravity fields and their dynamics are 
accounted for, hence we certainly  have a clean picture at that level.

\vspace{1mm} 
\noindent   {\bf (ii)} The dual fields are subtle to include from the topological point of view. In M-theory the only such field is supplied by $S^4$, while in type IIA they are 
supplied by the cyclification 
$\mc{L}_c S^4$. The dual fields beyond rank 6 in type IIA 
require parametrized homotopy theory 
\cite{BSS}; something analogous would be needed in type IIB. 

\vspace{1mm} 
\noindent  {\bf (iii)}  In the quotients \eqref{T-duality}, the right-hand-sides 
of the expressions are zero,
as relations, i.e., 
the terms are zero as (linear combinations of) generators. 

\vspace{1mm} 
\noindent {\bf (iv)} The real reason for the mismatch is that, somewhat unexpectedly and contrary to the treatment of type IIA, one ultimately needs the metric $g$.

Physically (and modulo the above subtleties), the above reincarnation of T-duality reconfirms the validity of the basic fields and EOMs 
in $d=10$ type IIB supergravity, as dictated by the real homotopy type $IIB$; see \eqref{htypeIIB}. It also reconfirms the basic fields and equations in the 9d reduction of type IIB supergravity  for fields of degree 1 through 6, given by the real homotopy type $\mc{L}_c IIB$, see \eqref{LIIB}. At the same time, it calls for reassessment of the nature of higher-degree fields in $d=9$ type II supergravity.
In order to resolve this subtlety, we will dig into the origin and dynamics of the fields in nine dimensions.

\paragraph{The $d=9$ fields and their type II origin.} 
Maximal $d=9$ supergravity \cite{GNS86} arises either from the 
compactification of $d=11$ supergravity on a 2-torus or of type IIB supergravity 
on a circle, hence coincides with the dimensionally reduced version of both 
the IIA and the IIB supergravity theories in 10 dimensions \cite{BHO95}\cite{KK97},
by virtue of T-duality \cite{DLP89}\cite{DHS89}. 
The form-field content in $d=9$, i.e., the collection of
potentials $(A_1, A_1^m, B_2^m, \linebreak[0] C_3)$, 
is comprised of three 1-forms $A_1$ and $A_1^m$, two 2-forms $B_2^m$, 
and one 3-form $C_3$, where $m=1,2$:

\begin{center} 
\begin{tabular}{ccc} 
\hline 
{\bf d=9} & {\bf M-theory interpretation} & {\bf IIB interpretation} 
\\
\hline
\hline 
\rowcolor{lightgray} $A_1$ & $s_2 s_1 C_3$ &  $s_2' g_{{}_{\rm IIB}}$ 
\\
$A_1^m$ & $s_1g_{{}_M}$, $s_2g_{{}_M}$ & $s_1'B_2^m$
\\
\rowcolor{lightgray} $C_3$ & $C_3$ & $s_1' C_4$
\\
$B_2^m$ & $s_1C_3$, $s_2 C_3$ & $B_2^m$\\
\hline 
\end{tabular} 
\end{center} 

\vspace{2mm} 
\noindent Here $g_M$ and $g_{IIB}$ are the metrics in 11 dimensions and in type IIB, respectively. 
So there is a singlet, spanned by $A_1$, and a doublet, spanned by $A_1^m$, $m=1, 2$, with respect to ${\rm SL}(2, \R)$.
The former arises from the reduction of the {\it metric} in type IIB 
and hence couples to the massive Kaluza-Klein states, giving them 
charge. The reduction along the $i$th direction from the type IIB side is denoted by $s'_i$ in contrast with that on the type IIA side, denoted thus far by $s_i$.

\paragraph{Self-duality.}  
The compactification of  $F_5$
in type IIB supergravity is delicate. 
It is self-dual and  gives rise to a duality condition for the massive fields in $d=9$.
Hence $F_5$ and $F_4=s_1'F_5$ will be dual, so that using either
leads to equivalent formulations of the theory.

\paragraph{Reduction of cubic term.} Type IIA and type IIB are equivalent 
in $d=9$ by T-duality.
The $B$-field $B_2=s_1 C_3$ in type IIA reduces on the further circle 
to the 1-form $A_1=s_2B_2$, which measures the winding charge 
with respect to the circle. Under T-duality, this becomes the Kaluza-Klein 
1-form, which 
arises as the reduction of the 
10d metric in type IIB, i.e., as  $s_2' g_{\rm IIB}$. Then one gets a term 
$A_1 \wedge F_4 \wedge F_4$ in $d=9$ (see \cite{DM97}).
Hence this is geometric. 
Indeed, this is supported by the Lagrangian   \cite[(2.9)]{FOT11} 
as well as the EOMs \cite[(2.14)]{FOT11} in the 9-dimensional reduction of type IIB supergravity:
\footnote{The fields and corresponding EOMs involve shifts, as is common in supergravity -- see also footnote \ref{barefields}. We have freed the fields and separated into two types of fields, one primed and another unprimed.}
$$
dF_7=  - \tfrac{1}{2}F_4 \wedge F_4+ F_2 \wedge F_6 
+ F_2' \wedge F_6'.
$$
Hence the equation, which was missing in the real homotopy model of $\mc{L}_c IIB$, is not 
purely topological, but in fact arises from the metric. Therefore,
a description solely via rational homotopy theory would not 
encapsulate this (however, cf.\ the discussion in \cref{Sec-Ek} in regards to the moduli space of cyclifications).

\paragraph{Stable T-duality.}
If we agree to abandon the field $h_7$ and at the same time run the fields $\omega_n$ to unbounded degree $n$, we may wish to consider the following ``stable'' models of types IIA and IIB, which turn out to be in T-duality \eqref{st-T-duality}, in a way more perfect than \eqref{T-duality} and \eqref{T-duality-sub}:
\begin{align*}
M(IIA_{\on{st}}) 
  & := \big(
    \R[ \{\omega_{2p}, h_3\}_{p = 1, 2, \dots}] \;,\, d \omega_2 = 0, ~ dh_3=0, ~ d \omega_{2p+2} = h_3 \cdot \omega_{2p}  \text{ for $p \ge 1$}
  \big)
  \,,\\
M(IIB_{\on{st}}) 
  & := \big(
    \R[ \{\omega_{2p+1}, h_3\}_{p = 0, 1, \dots}] \;,\, d \omega_1 = 0, ~ dh_3=0, ~ d \omega_{2p+1} = h_3 \cdot \omega_{2p-1}  \text{ for $p \ge 1$}
  \big)
  \,.
\end{align*}
As shown in \cite{FSS-pbranes}, their cyclifications are naturally isomorphic
\begin{equation}
\label{st-T-duality}
M(\mc{L}_c IIA_{\on{st}}) \cong M(\mc{L}_c IIB_{\on{st}}),
\end{equation}
which may be viewed as a stable version (spectra-fication) of T-duality. 

\subsection{Toroidal symmetries of type IIB}
\label{Sec-torIIb}

\paragraph{Del Pezzo for type IIB.}
In addition to the sequence $\BB_k$, $k=0, \cdots, 8$, connecting to the dimensional reduction of M-theory 
on $T^k$, there is also an ``outlier'' del Pezzo surface $\BB_1' := \CC \PP^1 \times \CC \PP^1$ 
of degree 8 
with Picard group 
$$
\Pic (\CC \PP^1 \times \CC \PP^1) \cong H^2 (\CC \PP^1 \times \CC \PP^1; \ZZ)  
\cong \ZZ \mathfrak{l}_1 \oplus \ZZ \mathfrak{l}_2\;,
$$
where $\mathfrak{l}_1$ and $\mathfrak{l}_2$ are the classes of the two $\CC \PP^1$ factors. The intersection pairing works as follows (cf. \eqref{HE-inter}):
\vspace{-1mm}  
\(
 \label{IIB-inter}
\mathfrak{l}_1 \cdot \mathfrak{l}_1 = \mathfrak{l}_2 \cdot \mathfrak{l}_2 = 0, \quad \mathfrak{l}_1 \cdot \mathfrak{l}_2 = 1 \;. 
\)
The anticanonical class is given by $- \cK_{\BB_1'} = 2 \mathfrak{l}_1 + 2 \mathfrak{l}_2$. 
This surface is related to the del Pezzo surfaces $\BB_k$ by a single blowup: if we blow up a point in $\CC \PP^1 \times \CC \PP^1$, we will obtain a surface isomorphic to $\BB_2$. 

\paragraph{Scaling symmetries in type IIB.} 
The same rules highlighted in \cref{Sec-8lower} --  and already implemented for type IIA in \cref{Sullmm} --
apply to the trombone symmetry of the type IIB supergravity theory. 
In type IIB, there are also two scaling symmetries.  The first corresponds to 
one of three one-parameter subgroups of ${\rm SL}(2, \R)$, namely  ${\rm SO}(1,1)^+$ 
with hyperbolic conjugacy classes. The second is an $\R^+$ symmetry 
(which is different from the $\R^+$ subgroup of ${\rm SL}(2, \R)$ corresponding to 
parabolic conjugacy classes). 
 With respect to the first, the fields $(H_3, F_3, F_5)$ transform with weights 
 $(2,2,4)$, while with respect to the second they transform as $(1, -1, 0)$;
 see \cite{BdWGLR02}. It is manifest that these two symmetries leave the EOMs invariant. 
The two groups ${\rm SL}(2, \R)$ and $\R^+$ combine to form the type IIB classical 
global symmetry ${\rm GL}(2, \R)$. 
Discrete versions are further studied in \cite{CMS}.

\paragraph{Matching of symmetries in 9d:}
From a $d=11$ perspective, the diffeomorphism invariance of the 
2-torus $T^2$ gives rise to a global ${\rm SL}(2, \R) \times (\R^+)^2$ 
symmetry, where the first factor
corresponds to the modular parameters of the torus, while the second captures the 
rescalings of the torus, descending from $\alpha$ and $\beta$. 
Both $\R^+$ factors leave invariant the EOMs. 
From a type IIB perspective, 
the ${\rm SL}(2, \R)$ symmetry already exists 
as S-duality (strong-weak coupling self-duality, as in \cref{Sec-IIBS}), 
while there are two symmetries 
descending from type IIB: an $\R^+$ symmetry $\delta$ 
corresponds to the rescalings of  the circle
and an ${\rm SO}(1,1)^+\subset {\rm SL}(2, \R)$  symmetry $\gamma$. 
The four symmetries are linearly dependent. Under $\alpha$ 
and $\delta$, the potentials $\big(A_1, A_1^{(1)}, A_1^{(2)}, B_2^{(1)}, B_2^{(2)}, C_3\big)$
 transform as $(3, 0, 0, 3,3,3)$ and $(0, 2, 2, 2,2,4)$, respectively \cite[Table 4]{BdWGLR02}.
 For the dual magnetic potentials, one has that 
 $\big(A_6, A_6^{(1)}, A_6^{(2)}, B_5^{(1)}, B_5^{(2)}, C_4\big)$
 transform as $(6, 9, 9, 6,6,6)$ and $(8, 6, 6, 6,6,4)$, respectively 
  \cite[Table 3]{FOT11}. While we will not be using all these symmetries explicitly, we nevertheless 
  hope they are useful in clarifying the context and providing comparisons.

\paragraph{Toroidal symmetries in type IIB.}
Since the type IIB model (see \cref{IIB}) falls out of the previous sequence of cyclifications of $S^4$, 
we need to treat it separately. Indeed, in \cite{SV1} we worked with the ``unstable'' model
$M(IIB) = (S(V),d)$ of type IIB; 
see \eqref{htypeIIB}. 
To identify a maximal $\RR$-split torus $T^B$ of $\Aut M(IIB)$, we need to start with
 $\RR \omega_1 \oplus \RR h_3$ in the notation of the system \eqref{htypeIIB}. 
Thus, a maximal  $\RR$-split torus is 2-dimensional and could be identified as the group of diagonal matrices in the 
basis $\{ \omega_1, h_3\}$. Given that a homogeneous basis of 
$\RR \omega_1 \oplus \RR h_3$
is unique up to scalar
multiplication, a maximal $\RR$-split torus $T^B$ here is unique.

\medskip 
Denote by $\beta_0$ the weight of $h_3$ and by $\beta_1$ the weight of $\omega_1$, so that
we have the action 
\[
t \cdot h_3 = t^{\beta_0} h_3, \qquad t \cdot \omega_1 = t^{\beta_1} \omega_1, \qquad \qquad t \in T^B.
\]
Then, with \eqref{htypeIIB}, we  have the following \cite{SV1}:
\vspace{-2mm} 
\begin{gather}
\label{tIIB}
  t \cdot \omega_{\, 3} = t^{\beta_0 + \beta_1} \omega_{\, 3}, 
  \qquad   t \cdot \omega_{\, 5} = t^{2\beta_0 + \beta_1} \omega_{\, 5},\\
  \nonumber
  t \cdot h_7 = t^{3\beta_0 + 2 \beta_1} h_7,  \qquad
  t \cdot \omega_{\, 7} = t^{3\beta_0 + \beta_1} \omega_{\, 7}.
  \end{gather}

  \vspace{-2mm} 
\noindent The exponents on the right-hand sides of these equations are exactly the weights appearing in the weight 
decomposition of the generators of $M(IIB)$.

\medskip 
The maximal real split torus can be canonically identified 
with $\GG_m \times
\GG_m$, acting on $M(IIB)$ as follows:
  $$
   t \cdot \phi_i = t^{\lambda_i} \phi_i , \qquad   t \in (\GG_m \times \GG_m) (\RR).
   $$
Using the standard notation as in the case of the M-theory sequence, we have the 
following table 

\[
\begin{tabular}{ccc}
\hline 
{\bf Field} $\phi_i$ & {\bf  degree} & {\bf  weight} $\lambda_i$ 
\\ \hline 
\rowcolor{lightgray}
$h_7$ & $7$ & $3\epsilon_0 - 2 \epsilon_2 - \epsilon_1$
\\
$\omega_7$ & $7$ & $3\epsilon_0 - 2\epsilon_1 - \epsilon_2$
\\
\rowcolor{lightgray}
$\omega_5$ & $5$ & $2\epsilon_0 - \epsilon_1 - \epsilon_2$
\\
$h_3$ & $3$ & $\epsilon_0 - \epsilon_1$
\\
\rowcolor{lightgray}
$\omega_3$  & $3$ &  $\epsilon_0 - \epsilon_2$
\\
$\omega_1$ & $1$ & $\epsilon_1 - \epsilon_2$
\\
\hline 
\end{tabular}
\]

\medskip 
We notice from the table that the  doublet exchanges $h_7 \leftrightarrow \omega_7$
and $h_3 \leftrightarrow \omega_3$ as well as the self-duality $\omega_5 \leftrightarrow \omega_5$
are witnessed by the exchange of weights $\epsilon_1 \leftrightarrow \epsilon_2$, corresponding 
to the exchange $s_1  \leftrightarrow s_2$ of the torus directions. Once pulled back from the universal space 
to spacetime fields, this exhibits a manifestation of the duality relations 
\vspace{-2mm} 
$$
H_7 \leftrightarrow F_7\;, \qquad 
H_3 \leftrightarrow F_3\;, \qquad 
F_5 \leftrightarrow F_5
$$

\vspace{-1mm} 
\noindent for the doublets $(H_7, F_7)$, $(H_3, F_3)$, and $(F_5, F_5)$, appropriate for type IIB
(and its duality relation to M-theory). See \cref{Sec-IIBS}  and \cref{IIB} for S-duality and T-duality, respectively. 

\subsection{The root data of type IIB}

Given that the real homotopy type $IIB$ was worked out so as to match with type $IIA$ via T-duality
\eqref{T-duality} and \eqref{T-duality-sub},
we would like to choose a compatible trivialization $T^B \cong \GG_m \times \GG_m$ of the maximal $\RR$-split 
torus of $\Aut M(IIB)$ and a compatible basis of weights of the action of $T^B$ on $M(IIB)$.

\medskip 
Now, let us use these maps to create distinguished bases of the Lie algebra $\h_B$ of the maximal $\RR$-split 
torus $T^B$ and the dual space vector space $\h_B^*$ of weights, as we did in the case of the sequence of torus reductions 
of  M-theory (see \cref{Sec-IIAsym} and \cref{symm-cycl-concrete}). 
In order to emphasize the symmetries, we will use the weights of the universal elements $h_3$ and $\omega_3$ as fundamental, 
corresponding to spacetime fields $H_3$ and $F_3$ associated with the fundamental 
string and its S-dual, the D1-brane. With that, we write
\begin{equation*}
t \cdot h_3 = t^{\gamma_{\, 1}} h_3, \qquad t \cdot \omega_3 = t^{\gamma_{\, 2}} \omega_3,
\end{equation*}
where
\[
\gamma_{\, 1} := \epsilon_0-\epsilon_1, \quad \text{and} \quad \gamma_{\, 2} := \epsilon_0 -\epsilon_2,
\]
given that
$$
h_3  \mapsto s_1 g_4, \quad \text{and} \quad  
\omega_3  \mapsto s_2 g_4
$$
in the correspondence between $M(\mc{L}_c IIB)$ and $M(\mc{L}_c IIA) = M(\mc{L}_c^2 S^4)$. Here $\epsilon_0$, $\epsilon_1$, 
and $\epsilon_2$ 
are the generating weights of $\h_2^*$, the Lie algebra of the maximal $\RR$-split torus $T^A_c$ of 
$\Aut M(\mc{L}_c IIA) = \Aut M(\mc{L}_c^2 S^4)$. 

\medskip 
We will use the inner product induced on $\h_{B}^*$ from $\h_2^*$:
$$
(\gamma_{\, 1}, \gamma_{\, 1})  := (\epsilon_0 - \epsilon_1, \epsilon_0 - \epsilon_1) = 0,\quad 
(\gamma_{\, 2}, \gamma_{\, 2})  := (\epsilon_0 - \epsilon_2, \epsilon_0 - \epsilon_2) = 0, \quad 
(\gamma_{\, 1}, \gamma_{\, 2})  :=(\epsilon_0 - \epsilon_1, \epsilon_0 - \epsilon_2) = 1\;.
$$
The dual basis of $\h_{B}$ will be given as $\{\mathfrak{l}_1 := h_0 -h_1, \mathfrak{l}_2 := h_0 - h_2\}$.
Then one can check from Equations \eqref{tIIB} that there exists a unique element $-K \in \h_B$ which acts as the degree operator \eqref{degree-operator} 
on the Quillen model $Q(IIB)$, namely, the element
\vspace{-2mm} 
\[
-K = 2 \mathfrak{l}_1 + 2 \mathfrak{l}_2. 
\]
We can check that this gives the correct degree by forming the inner product 
\begin{align*}
(-K, -K) &= (2 \mathfrak{l}_1 + 2 \mathfrak{l}_2, 2 \mathfrak{l}_1 + 2 \mathfrak{l}_2)
\\
&= 8\,,
\end{align*} 
upon using the inner product relations \eqref{IIB-inter}. Indeed, this matches what is expected 
in comparison with the case of type IIA; see \eqref{deg-L-k=1}.

\medskip 
Similar to  the root system of type IIA in \eqref{IIB-root}, we can produce the 
root system associated with the corresponding del Pezzo surface, except that now the 
root system is nontrivial.

\paragraph{Exceptional root data from the rational model for type IIB.}
The data associated to the Sullivan minimal model $M(IIB)$
of type IIB  replicates the data  determined by the del Pezzo surface 
$\CC P^1 \times \CC P^1$
and produces the root system
\(
\label{IIB-root}
E_B = R_B := \big\{ x \in \h_B^\ZZ \; | \; (x, 2\mathfrak{l}_1 + 2 \mathfrak{l}_2) = 0 , \;  
(x, x) = -2 \big\} = \big\{\mathfrak{l}_1 - \mathfrak{l}_2, \; \mathfrak{l}_2 - \mathfrak{l}_1\big\}
\)
which is nonempty and may be identified as a root system of type $A_1$. 

 
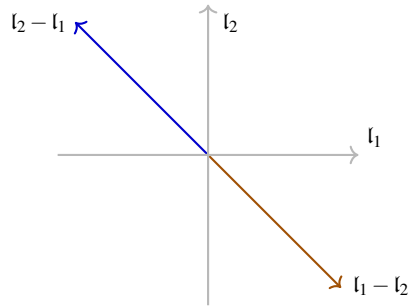
\begin{center}
\begin{tikzpicture}
    \foreach\ang in {135}{
     \draw[->,blue!80!black,thick] (0,0) -- (\ang:2.5cm);
    }
     \foreach\ang in {315}{
     \draw[->,orangeii!80!black,thick] (0,0) -- (\ang:2.5cm);
    }
    \foreach\ang in {0}{
     \draw[->,lightgray!80!black,thick] (-2,0) -- (\ang:2cm);
    }
    \foreach\ang in {90}{
     \draw[->,lightgray!80!black,thick] (0,-2) -- (\ang:2cm);
    }
    \node[anchor=south west,scale=0.8] at (2,0) {$\mathfrak{l}_1$};
    \node[anchor=north,scale=0.8] at (0.3,2) {$\mathfrak{l}_2$};
    \node[anchor=north,scale=0.8] at (-2.25,2) {$\mathfrak{l}_2-\mathfrak{l}_1$};
    \node[anchor=north,scale=0.8] at (2.3,-1.5) {$\mathfrak{l}_1-\mathfrak{l}_2$};
  \end{tikzpicture}
\end{center} 

This justifies the type IIB row of \Cref{table1}.

\newpage 

\section{
Algebraic geometric features within the Triality}
\label{Sec-M} 

We provide topological and physical interpretations of classical statements
in algebraic geometry that can be found, for instance, in \cite{Beauville}\cite{Dolgachev}.
We highlight that most statements hold at the universal level (i.e., for the generators $g_4$, $g_7$, 
etc., of the Sullivan minimal models of the cyclifications of $S^4$) and all hold in spacetime (i.e., for 
the pulled back physical fields $G_4$, $G_7$). 
The notation might use one or the other, with the understanding that the statements 
hold for both (unless otherwise specified; in particular, higher 
products
directly 
arise only in spacetime and not universally).

\subsection{Exceptional curves and exceptional vectors} 
\label{Sec-excvec}

\paragraph{Exceptional curves and Weyl groups.}
Recall
that the class $l = \cE_i \in H^2(\BB_k; \ZZ)$ of
the preimage of a blown-up point
satisfies the condition 
 \(
 \label{eq-lk} 
 (l, \cK_k)=(l, l)=-1\;,
 \)
and all such classes are called \emph{exceptional curves}, cf.\ \eqref{set-excep}. 
The number of exceptional curves on $\mathbb{B}_k$, for a given $k$, is presented in the following table: 
\(
\label{table-exvec}
\begin{tabular}{ccccccc}
\hline
\bf $k$ &  \bf 8 & \bf 7 & \bf 6 & \bf 5 & \bf 4 & \bf 3 
\\
\hline 
\hline 
\rowcolor{lightgray} \text{Exceptional curves} & 240 & 56 & 27 & 16 & 10 & 6
\\
\hline 
\end{tabular} 
\)
On the other hand, the orders of the Weyl groups are given as
$$
\begin{tabular}{c|ccccccc}
\hline 
\bf $k$ &  \bf 8 & \bf 7 & \bf 6 & \bf 5 & \bf 4 & \bf 3 & \bf 2
\\
\hline 
\hline 
\rowcolor{lightgray} $|W(R_k)|$ & $2^{14}\cdot3^5\cdot 5^2\cdot 7$ \;\; 
& $2^{10}\cdot3^4 \cdot 5\cdot 7$ \;\;\;
& $2^{7}\cdot3^4 \cdot 5$  \;\;\; 
& $2^{7}\cdot3 \cdot 5$ \;\;\;
&  $2^{3}\cdot3 \cdot 5$  \;\;\;
& 
 $2^{2}\cdot3$ & $2$ 
\\
\hline 
\end{tabular} 
$$

\vspace{1mm} 
\noindent Note that the ratio of the orders of the Weyl groups of
$E_i$ and $E_{i-1}$
is given by the number of exceptional curves corresponding to rank $i$.

\paragraph{Weyl groups and lattices.}
 As explained in \cite[\S 4.6]{OP}, the vectors
 $\vec{\varphi}=(\ln \ell_p^3, \ln R_1, \cdots, \ln R_k)$ can be represented as linear functionals
 on a vector space
 $V_{k+1}$ with basis $\{\epsilon_0, \epsilon_1, \cdots, \epsilon_k\}$ and used to assign to any weight vector 
 $\vec{\lambda}=y^0 \epsilon_0 + y^1 \epsilon_1 + \cdots + y^k \epsilon_k$ its `tension' 
 $$
\mathcal{T}=e^{\langle \vec{\varphi}, \vec{\lambda} \rangle} =\ell_p^{3y^0} R_1^{y^1} \cdots R_k^{y^k}.
 $$
 As we see in 
\eqref{wtrans} and \eqref{Eq-T3duality}, a minimal set of Weyl generators incorporates the exchange of the M-theory torus directions 
$S_i:  R_i  \leftrightarrow R_{i+1}$, where $i = 1 . . ., k-1$, 
as well as the T-duality $T_{123}$ on directions $1,2,3$ of the M-theory torus. 
These linear operators acting on $V_{k+1}$ are given explicitly in \cite[(4.20)]{OP}. 
When $k \leq 8$, the action of the Weyl group on $V_{k+1}$ is reducible. Indeed, the 
invariance of Newton's constant $\left(\prod_i R_i\right)/\ell_p^9$ implies that the roots are all 
orthogonal to the vector 
\(
\label{vec-delta} 
\vec{\delta}= \epsilon_1 + \cdots + \epsilon_k - 3\epsilon_0
\)
with length $\vec{\delta}\cdot \vec{\delta}= 9 - k$, so that the reflections actually restrict to the 
hyperplane $V_k$, given by $y^1 + \cdots + y^k + 3y^0=0$, normal to $\vec{\delta}$.
Note that $\vec{\delta}$ is
a vector invariant under the Weyl group, 
see  \cite[\S 4.6]{OP}. 

\subsubsection*{Roots and exceptional vectors}

\paragraph{The $E_k$ lattice.} 
Let $\ZZ^{1,k}$ be our rank-$(k+1)$ lattice with the standard Minkowski inner product and the special vector 
$
{\bf k}_k= -3\epsilon_0 +\sum_{i=1}^k \epsilon_i\;.
$
The $\mathbb{E}_k$ lattice is the sublattice of $\ZZ^{1,k}$ given by 
the orthogonal complement 
$
\mathbb{E}_k={\bf k}_k^\perp 
$
and satisfying ${\bf k}_k^2=9-k$. Hence $\mathbb{E}_k$ is a negative-definite lattice for $k \leq 8$. 
For $k \geq 3$, the vectors in \eqref{Eq-vectors0i}
form a basis of $\mathbb{E}_k$.
 The stabilizer subgroup $\operatorname{O}(\ZZ^{1,k})_{{\bf k}_k}$ of ${\bf k}_k$ 
 acts simply transitively on the set of canonical root bases in 
 $\mathbb{E}_k$ and coincides with the Weyl group $W(\mathbb{E}_k)$.

  \vspace{-1mm} 
 \begin{center} 
\begin{tabular}{ccc}
\hline 
\bf Lattice ${\rm Pic}(\mathbb{B}_k)$ & \bf Lattice $\mathbb{E}_k$
\\
\hline 
\hline 
\rowcolor{lightgray}  $\ZZ^{1, k}$ 
&  $\ZZ^{1, k-1}$ 
\\
\hline 
\end{tabular} 
\end{center}

\paragraph{Correspondence for roots.}
Roots correspond to 1-form field strengths or, equivalently, 
0-form potentials. Recall from \eqref{RkIk} that a vector $\alpha \in \mathbb{E}_k$ is called 
a root if $\alpha^2=-2$. For $k \leq 8$, the combinations  
(slightly generalizing \eqref{Eq-vectors0i})
\vspace{-1mm} 
\begin{align}
\label{pos-roots1}
\alpha_{ij}&=\epsilon_i - \epsilon_j, \qquad \qquad \quad 1 \leq i < j \leq k,
\\
\label{pos-roots2}
\alpha_{ijl}&=\epsilon_0 - \epsilon_i - \epsilon_j - \epsilon_l, \quad 1 \leq i < j< l \leq k,
\end{align} 
correspond, respectively,  to the spacetime fields
$$
s_j \omega_i, \;  \cdots, \; s_l s_j s_i G_4 \; 
$$
and the following (universal) generators of $M(\mc{L}_c^k S^4)$:
$$
s_j w_i, \;  \cdots, \; s_l s_j s_i g_4 \;.
$$
The roots in $\mathbb{E}_k$ are given by $\pm \alpha$,  the vectors
and their negatives, in the following table, where we also give a  correspondence with 
physical fields and elements of the Sullivan minimal model:

\begin{footnotesize}
\begin{center} 
\setlength{\tabcolsep}{3.8pt} 
\renewcommand{\arraystretch}{1.5} 
\begin{tabular}{ccccc} 
\hline 
{\bf Lattice} & {\bf Roots} & {\bf Associated fields} & \textbf{Elements of $M(\mc{L}_c^k S^4)$}  & {\bf Number} 
\\
\hline 
\hline 
\rowcolor{lightgray} $\mathbb{E}_3$ & $\alpha_{ij}$, $\alpha_{123}$ & $s_j \omega_i$, \; $s_3 s_2 s_1 G_4$ & $s_j w_i$, \; $s_3 s_2 s_1 g_4$ & 8 
\\
$\mathbb{E}_4$ & $\alpha_{ij}$, $\alpha_{ijl}$ & $s_j \omega_i$, \; $s_l s_j s_i G_4$
& $s_j w_i$, \; $s_l s_j s_i g_4$
& 20 
\\
\rowcolor{lightgray} $\mathbb{E}_5$ & $\alpha_{ij}$, $\alpha_{ijl}$ & $s_j \omega_i$, \; $s_l s_j s_i G_4$
& $s_j w_i$, \; $s_l s_j s_i g_4$
& 40 
\\
$\mathbb{E}_6$ & $\alpha_{ij}$, $\alpha_{ijl}$,
$2\epsilon_0-\epsilon_1- \cdots - \epsilon_6$ &
$s_j \omega_i$, \; $s_l s_j s_i G_4$, \; $s_6 \cdots s_1 G_7$
& $s_j w_i$, \; $s_l s_j s_i g_4$, \; $s_6 \cdots s_1 g_7$ & 72 
\\
\rowcolor{lightgray} $\mathbb{E}_7$ & $\alpha_{ij}$, $\alpha_{ijl}$,
$2\epsilon_0-\epsilon_1- \cdots - \epsilon_7 + \epsilon_i$ &
$s_j \omega_i$, \; $s_l s_j s_i G_4$, \; 
$s_7 \cdots \widehat{s}_i \cdots s_1 G_7$
& $s_j w_i$, \; $s_l s_j s_i g_4$, \; 
$s_7 \cdots \widehat{s}_i \cdots s_1 g_7$
& 126 
\\
$\mathbb{E}_8$ & $\alpha_{ij}$, $\alpha_{ijl}$,
$2\epsilon_0-\epsilon_1- \cdots - \epsilon_8 + \epsilon_i + \epsilon_j$, 
& $s_j \omega_i$,  $s_l s_j s_i G_4$, \; $s_8 \cdots \widehat{s}_j \cdots \widehat{s}_i \cdots s_1 G_7$, & $s_j w_i$,  $s_l s_j s_i g_4$, \; $s_8 \cdots \widehat{s}_j \cdots \widehat{s}_i \cdots s_1 g_7$,
\\
& 
$3\epsilon_0 - \epsilon_1 - \cdots - \epsilon_8 - \epsilon_i$
 &
$s_8 \cdots s_1  (s_i C_3 \wedge G_7)$ & $s_8 \cdots s_1  (s_i g_4 \cdot g_7)$
 & 240
\\
\rowcolor{lightgray} $\mathbb{E}_9$ & $\infty$ & $\infty$ & $\infty$ &
\\
\hline 
\end{tabular} 
\end{center} 
\end{footnotesize}

\vspace{1mm} 
\noindent Here, as before, for form-fields, $s_i$ means contraction (or integration over the fiber) in the direction of
the $i$th loop/circle and $\widehat{s}_i$ means omitting that. For 
form-fields as well as elements of $M(\mc{L}_c^k S^4)$, we also assume 
that $s_j s_i := - s_i s_j$ if $j < i$, and that each $s_j$ acts as a graded derivation of the wedge product. This allows us to understand compositions of the $s_j$'s via \emph{normal ordering}. For instance, formally speaking, $s_8 C_3 \wedge G_7$ would literally have to be zero, because it is supposed to take place in the spacetime reduced to 3 dimensions. However, after normal ordering, we get
\[
s_8 \cdots s_1  (s_8 C_3 \wedge G_7) := \sum \pm s_8 s_{i_7} \cdots s_{i_{p+1}}  C_3 \wedge s_8 s_{i_{p}} \dots s_{i_1} G_7,
\]
where the summation runs over all $(7-p,p)$ shuffles $(i_7, \dots, i_{p+1} \; | \;   i_{p}, \dots, i_1)$ of $(7, \dots, 1)$, over all $0 \le p \le 7$, with all but $p = 5$ and $6$ contributing nontrivially,
would make a more meaningful sense.
The case of $\mathbb{E}_8$ involves a field of degree 10, say $G_{10}$,
which can be interpreted as
an antiderivative of the composite
of $G_4$
and $G_7$, appearing as a spacetime field, 
but we have chosen to write a summand $C_3 \wedge G_7$ without loss of generality.
This also suggests a relation to 
a hypothetical
M8-brane; and the repetition of the $s_i$ operation applied to $G_{10}$ should have a  meaning different from the iterated contraction; we discuss this further in \cref{Sec-KM}.

\paragraph{Exceptional vectors.}  
Recall from \eqref{eq-lk} that a vector $v \in \ZZ^{1,k}$ is exceptional if and only if it satisfies $({\bf k}_k, v) =-1$
and $(v,v) \linebreak[1] = \linebreak[0] -1$. The
exceptional vectors correspond to the exceptional
curves on the del Pezzo
surface $\mathbb{B}_k$. 
We have indicated in the following table 
  the interpretation in terms of the fields. It is remarkable that
the number of fields in the Kaluza-Klein reduction is an exact match
(cf. Table \eqref{table-exvec}).
  Many of the fields in these two tables may be represented by elements, such as $s_j w_i$, $s_3 s_2 s_1 g_4$, etc., of the universal models $M(\mc{L}_c^k S^4)$.

\begin{footnotesize}
\begin{center} 
\setlength{\tabcolsep}{3.9pt} 
\renewcommand{\arraystretch}{1.5} 
\begin{tabular}{ccccc}
\hline 
{\bf Lattice} & {\bf Vectors} & {\bf Associated fields} & \textbf{Elements of $M(\mc{L}_c^k S^4)$} & {\bf Number} 
\\
\hline 
\hline 
\rowcolor{lightgray} $\mathbb{E}_3$ & $\epsilon_i$, $\epsilon_0-\epsilon_i-\epsilon_j$ & $\omega_i$, $s_j s_i G_4$ & $w_i$, $s_j s_i g_4$
& $6=\frac{|W(E_3)|}{|W(E_2)|}$ 
\\
$\mathbb{E}_4$ &  $\epsilon_i$, $\epsilon_0-\epsilon_i-\epsilon_j$ & $\omega_i$, $s_j s_i G_4$
& $w_i$, $s_j s_i g_4$
& $10=\frac{|W(E_4)|}{|W(E_3)|}$
\\
\rowcolor{lightgray} $\mathbb{E}_5$ &  $\epsilon_i$, $\epsilon_0-\epsilon_i-\epsilon_j$, $2\epsilon_0-\epsilon_1- \cdots -\epsilon_5$  &
$\omega_i$, $s_j s_i G_4$, $s_5  \cdots s_1 G_7$
& $w_i$, $s_j s_i g_4$, $s_5  \cdots s_1 g_7$
& $16=\frac{|W(E_5)|}{|W(E_4)|}$ 
\\
$\mathbb{E}_6$ 
 &  $\epsilon_i$, $\epsilon_0-\epsilon_i-\epsilon_j$, $2\epsilon_0-\epsilon_1- \cdots -\epsilon_6 +\epsilon_i$  &
 $\omega_i$, $s_j s_i G_4$, $s_6  \cdots \widehat{s}_i \cdots s_1 G_7$
 &  $w_i$, $s_j s_i g_4$, $s_6  \cdots \widehat{s}_i \cdots s_1 g_7$
 & $27=\frac{|W(E_6)|}{|W(E_5)|}$
\\
\rowcolor{lightgray} $\mathbb{E}_7$
 &  $\epsilon_i$, $\epsilon_0-\epsilon_i-\epsilon_j$, $2\epsilon_0-\epsilon_1- \cdots -\epsilon_7 +\epsilon_i + \epsilon_j$,
& $\omega_i$, $s_j s_i G_4$, $s_7 \cdots  \widehat{s}_j \cdots
 \widehat{s}_i\cdots s_1 G_7$, & $w_i$, $s_j s_i g_4$, $s_7 \cdots  \widehat{s}_j \cdots 
 \widehat{s}_i\cdots s_1 g_7$, &
 \\
 \rowcolor{lightgray}
& $-{\bf k}_7 - \epsilon_i$   &
 $s_7 \cdots s_1  (s_i C_3 \wedge G_7)$
& $s_7 \cdots s_1  (s_i g_4 \cdot g_7)$
&  $56=\frac{|W(E_7)|}{|W(E_6)|}$
\\ 

 \hline 
\end{tabular} 
\end{center} 
\end{footnotesize}

\vspace{4mm} 
\noindent 
We leave the case of $\mathbb{E}_8$
for the next
table, as several elements will involve $-{\bf k}_8$.
So these can also be described
using potentials $C_{i-1}$
and field strengths $G_i$, 
in a way, mixing the Sullivan and
Quillen perspectives. 
The interpretation in terms of 
2-form potentials yields:

\vspace{-2mm}  
\begin{footnotesize}
\begin{center} 
\setlength{\tabcolsep}{3.7pt} 
\renewcommand{\arraystretch}{1.3} 
\begin{tabular}{ccccc}
\hline 
 {\bf Lattice} \phantom{AAA}  & 
\phantom{AAA}  \quad {\bf Vector} \phantom{AAA}   & 
\phantom{AAA}  {\bf Associated potential field} \phantom{AA} & \textbf{Element of $M(\mc{L}_c^8 S^4)$} &
\phantom{AAA} {\bf Number} 
\\
\hline 
\hline 
 & 
  $\epsilon_i$  &
  $\omega_i$ & $w_i$
 &
  \\
 & $\epsilon_0-\epsilon_i-\epsilon_j$ 
 & $s_js_iG_4$ & $s_j s_i g_4$
 &
 \\
 & $2\epsilon_0-\epsilon_1- \cdots -\epsilon_7 +\epsilon_i +\epsilon_j$
 & 
 $s_7  \cdots \widehat{s}_j \cdots 
\widehat{s}_i\cdots s_1 G_7$
& $s_7  \cdots \widehat{s}_j \cdots 
\widehat{s}_i\cdots s_1 g_7$
&
\\
$\mathbb{E}_8$
& $-{\bf k}_8 + \epsilon_i -\epsilon_j$ & 
$s_8 \cdots \widehat{s}_i \cdots
s_1 (C_3 \wedge s_j G_7)$ & $s_8 \cdots \widehat{s}_i \cdots
s_1 (g_4 \cdot s_j g_7)$
& \quad $240=\frac{|W(E_8)|}{|W(E_7)|}$
\\
&$-{\bf k}_8 + \epsilon_0 - \epsilon_i  -\epsilon_j -\epsilon_l$ & $s_8 \cdots s_1 ( C_6 \wedge s_l s_j s_i G_7)$ & $s_8 \cdots s_1 ( g_7 \cdot s_l s_j s_i g_7)$
\\
& $-{\bf k}_8 + 2\epsilon_0 - \epsilon_{i_1}  - \cdots - \epsilon_{i_6} 
$ & 
$s_8 \cdots s_1 (C_3 \wedge C_6 \wedge s_{i_6} \cdots s_{i_1}G_7)$
& $s_8 \cdots s_1 (g_4 \cdot g_7 \cdot s_{i_6} \cdots s_{i_1}g_7)$
\\
& $-2{\bf k}_8 - \epsilon_i$ &
$s_8 \cdots s_1 ( C_3 \wedge s_i C_6) \cdot s_8 \cdots s_1 ( C_3 \wedge C_6) $
& $s_8 \cdots s_1 ( g_4 \cdot s_i g_7) \cdot s_8 \cdots s_1 ( g_4 \cdot g_7) $
\\
\hline 
\end{tabular} 
\end{center} 
\end{footnotesize}
\smallskip

\paragraph{Explanation of the approach in the tables.}
Note that the combinations of fields  in the above tables are not unique,
and other combinations of similar nature are possible. For instance,
the forms
$s_i \omega_j \wedge G_4$ and $\omega_i \wedge s_j G_4$ have the same weights and correspond to the
same `multiplet' in that they both generally 
appear in the same relevant step in the dimensional reduction process. We have chosen to pick one such 
representative for ease of recording the 
statements, but it should be understood that there is a clear general
pattern here. 

\subsection{27 Lines}
\label{Sec-27}

We work with spacetime field strengths. However, everything in this section also works universally, that is to say, for the generators of the Sullivan minimal models of the cyclifications $\mc{L}_c^6 S^4$ and $\mc{L}_c^7 S^4$.
We will map to constructions
in \cite[Ch.9]{Dolgachev}.

\subsection*{Systematics of dimensional reduction on $T^6$}

 We provide an exact mapping to the dimensional reduction
of M-theory on $T^6$, with a  
systemization of the latter and doing so within a rigorous mathematical perspective. 

\vspace{-2mm} 
\paragraph{Gravitational vs.\ nongravitational: The gravity line.}
Once we reach rank $k = 6$, the Dynkin diagram $E_6$ still contains the ``gravity line'' $A_5$,
corresponding to the Lie subalgebra $\mathfr{sl}_6$ of $\mathfr{e}_6$:
\begin{center}
  \begin{tikzpicture}[scale=.4]
    \draw (-1,0) node[anchor=east]  {$A_5$};
    \foreach \x in {0,...,4}
    \draw[xshift=\x cm,thick] (\x cm,0) circle (.3cm);
    \draw[
    thick] (0.25 cm,0) -- +(1.45 cm,0);
    \foreach \y in {1.15,...,3.15}
    \draw[xshift=\y cm,thick] (\y cm,0) -- +(1.4 cm,0);
  \end{tikzpicture}
\end{center}

\vspace{-2mm} 
\noindent 
and it also acquires 
a vertical node;
see \eqref{Dynkin}.
This corresponds to
when axions arise from the C-field.

\vspace{-2mm} 
\paragraph{Exceptional vectors and sixers.} The list of exceptional vectors 
in $\mathbb{E}_6$ in terms of the standard orthonormal basis in $\ZZ^{1,6}$, together with our identification 
with the corresponding spacetime and universal fields, is as follows:
\begin{center} 
\begin{tabular}{cccc}
\hline 
{\bf Exceptional vectors} && {\bf Fields} & \textbf{Generators of $M(\mc{L}_c^6 S^4)$}\\ 
\hline
\hline 
\rowcolor{lightgray} $
\epsilon_i, \quad i=1, \cdots, 6$ && $\omega_i$ & $w_i$ 
\\
$2\epsilon_0 - \epsilon_1 - \cdots - \epsilon_6 + \epsilon_i, \quad i=1, \cdots, 6$
&& $s_6 \cdots \widehat{s}_i \cdots s_1G_7$ & $s_6 \cdots \widehat{s}_i \cdots s_1g_7$ 
\\
\rowcolor{lightgray} $\epsilon_0 -\epsilon_i -\epsilon_j, \quad 1 \leq i < j \leq 6$ &&
$s_js_iG_4$ & $s_js_ig_4$ \\
\hline 
\end{tabular} 
\end{center}

\noindent
A \emph{sixer} in $\ZZ^{1,6}$ is a set of six mutually orthogonal exceptional 
vectors. An  example of this is $\{\epsilon_1, \dots, \epsilon_6\}$.
Associated to any sixer $\{v_1, \cdots, v_6\}$ is a unique root $\alpha$
such that $(v_i, \alpha)=1$ for $i=1, \cdots, 6$ (see \cite[§ 9.1.1]{Dolgachev}).
Two sixers with opposite associated roots form a ``double-six'' of exceptional 
vectors. Using the classification in \cite{Dolgachev} and the derivation 
and notation of 
the dilaton vectors in \cite{LP96}\cite{LPS96}, we interpret these 
via the following list of fields. The rows of the subsequent tables are the sixers along 
with the corresponding root in the last column:

\medskip 
\noindent {\bf (i)} 1 multiplet of 6 fields,
of type D (in the algebraic geometry notation of \cite{Dolgachev}),
corresponding to the dilaton vector $\vec{a}^*=-\vec{a}$
(in the physics notation of \cite{LP96}\cite{LPS96}):

\vspace{-1mm} 
\begin{center}
\begin{tabular}{cccccc|c}
\hline 
$\omega_1$ & $\omega_2$ &
$\omega_3$ & $\omega_4$ &
$\omega_5$ & $\omega_6$ &
$\alpha_{\rm max}$
\\
\hline
\hline 
\rowcolor{lightgray} 
$s_6 \cdots s_2 \widehat{s}_1G_7$ &
$s_6 \cdots \widehat{s}_2 s_1G_7$ &
$s_6 \cdots \widehat{s}_3 \cdots s_1G_7$& 
$s_6 \cdots \widehat{s}_4\cdots s_1G_7$ &
$s_6 \widehat{s}_5 \cdots s_1G_7$ &
$\widehat{s}_6 s_5 \cdots s_1 G_7$
& $-\alpha_{\rm max}$
\\
\hline 
\end{tabular} 
\end{center} 
Here $\alpha_{\rm max}$ is the maximal root of the root system $E_6$
equal to 
$2\epsilon_0-\epsilon_1-\cdots - \epsilon_6$. The reflection with respect to the associated root interchanges
the rows, preserving the order. 
See  \cite[Theorem 9.1.3]{Dolgachev}.

\vspace{1mm} 
\noindent {\bf (ii)} 15 multiplets of 6 fields,  of type $D_{ij}$,
corresponding to the dilaton vectors $\vec{b}_{ij}$:

\vspace{-2mm} 
\begin{center} 
\begin{tabular}{cccccc|c}
\hline 
$\omega_i$ & $s_6 \cdots \widehat{s}_i \cdots s_1G_7$ &
$s_ks_jG_4$
&
$s_ls_jG_4$
&
$s_ms_jG_4$
&
$s_ns_jG_4$
&
$\alpha_{ij}$
\\
\hline
\hline 
\rowcolor{lightgray} 
$\omega_j$ & $s_6 \cdots \widehat{s}_j \cdots s_1G_7$&
$s_ks_iG_4$
&
$s_ls_iG_4$
&
$s_ms_iG_4$
&
$s_ns_iG_4$
&
$-\alpha_{ij}$
\\
\hline 
\end{tabular} 
\end{center} 

\noindent {\bf (iii)} 20 multiplets of 6 fields,  of type $D_{ijl}$
corresponding to the dilaton vectors $\vec{a}_{ijl}$:

\begin{center} 
\begin{tabular}{cccccc|c}
\hline 
$\omega_i$ &
$\omega_j$ & $\omega_l$ &
$s_ls_mG_4$
&
$s_ms_nG_4$
&
$s_ls_nG_4$
&
$\alpha_{ijl}$
\\
\hline
\hline 
\rowcolor{lightgray} $s_l s_j G_4$
&
$s_l s_i G_4$
&
$s_j s_iG_4$
&
$s_6 \cdots \widehat{s}_n \cdots s_1G_7$& 
$s_6 \cdots \widehat{s}_l\cdots s_1G_7$ &
$s_6 \cdots \widehat{s}_m \cdots s_1G_7$ 
& $-\alpha_{ijl}$
\\
\hline 
\end{tabular} 
\end{center}

\paragraph{Weyl group action on sixers.} 
The number of sixers is equal to the number of roots. All the sixers form one orbit
under the Weyl group. 
The stabilizer of the sixer $\{\epsilon_1, \cdots, \epsilon_6\}$
(and hence of a root) is the symmetric group $\Sigma_{6}$ of order $6!$. 
There are 72 roots, and note that $|W(E_6)|=72\cdot 6!$. 
The stabilizer of the double-six $D$ is the subgroup 
$\langle \Sigma_6, \sigma_{\alpha_0} \rangle$ of order $2\cdot 6!$
(see \cite[Prop. 9.1.4]{Dolgachev}). 
This gives corresponding statements for the fields and 
 puts on firm ground (with the proper interpretation) 
discussions in the physics literature beyond simple 
combinatorics and numerology to 
proper setting and interpretation:
The number of 1-form field strengths $F_1$'s in the dimensional reduction is $20+15+1=36$,
together with their negatives form the 72-component multiplet of 
$W(E_6)$.

\paragraph{Interpretation of fields associated to roots of opposite sign.} 
Note that the lattice $\mathbb{E}_6$ is even, so the $\ZZ_2$-vector space $V=\mathbb{E}_6/2\mathbb{E}_6\cong \ZZ_2^6$ 
can be equipped with a quadratic form \cite{Dolgachev} 
\(
\label{eq-qE6}
q(x + 2\mathbb{E}_6)=\tfrac{1}{2}(x,x) \mod 2.
\)
Each pair of opposite roots $\pm \alpha$ defines a vector in $V$ with $q(v)=1$.
The quadratic form $q$ has Arf invariant 1, hence vanishes, on 28 vectors. 
The remaining 36 vectors correspond to 36 pairs of opposite roots or,
equivalently, double-sixes. 
The action of $W(E_6)$ on $V$ gives an isomorphism of groups $W(E_6)\cong \operatorname{O}^-(6; \ZZ_2)$, the orthogonal group of the odd quadratic form on $\ZZ_2^6$.
See \cite[p. 192]{Dolgachev}.

\subsection*{27 Lines in $\mc{L}_c^6 S^4$ and on a smooth cubic, and 27 fields}

Our interpretation of the statements in this part holds for the universal model $\mc{L}_c^k S^4$
(i.e., topologically)
as well as in spacetime (i.e., physically). We will first remind how the 27 lines arise on a smooth cubic, a del Pezzo surface of type $\BB_6$, along with presenting the list of 5d-spacetime 27 fields, extending the Mysterious Duality of \cite{INV}. Then we will detect the 27 lines on $\mc{L}_c^6 S^4$, confirming the prediction of Mysterious Triality.

\paragraph{Configurations of cubics.} 
Suppose that the points $x_1, \dots, x_k$ ($k \leq 6$) are in general position in $\CC \PP^2$. 
Then the linear system $\abs{-\cK_k} = \abs{3\cH - \cE_1 \ \dots - \cE_k}$ of cubics through these points defines the anticanonical  embedding  $j: \BB_k \hookrightarrow \CC \PP^{9-k}$. 
The image $j(\BB_k)$ contains a finite number of lines, which are the images 
under $j$ of the following curves in the del Pezzo surface $\BB_k$:

{\bf (i)} the exceptional curves $\cE_i$;

{\bf (ii)} the proper transforms of the lines $\overline{x_i\,x_j}$, $i \neq j$;

{\bf (iii)} the proper transform of the conics through 5 of the $x_i$.

\medskip 
\noindent Their number is given in the following table 
(see \cite[Prop. IV.12]{Beauville}):

\begin{center}
\begin{tabular}{lccccccc}
\hline 
\rowcolor{lightgray} $k=$ number of $\cE_i$ &  0 & 1 & 2 & 3 & 4 & 5 & 6
\\
Number of lines $\overline{x_i\,x_j}$ &  0 & 0 & 1 & 3 & 6 & 10 & 15
\\
\rowcolor{lightgray} Number of conics through 5 of the $x_i$ &  0 & 0 & 0 & 0 & 0 & 1 & 6
\\
Number of lines on $\BB_k$ &  0 & 1 & 3 & 6 & 10 & 16 & 27
\\
\hline 
\end{tabular} 
\end{center} 

 \noindent Observe that the last line corresponds to the 
 representations of 
$E_k$ identified in \cite[p. 26]{INV}.

\paragraph{Six points and the cubic.} 
Let $k = 6$ and consider the blowup $\BB_6$
at six points $x_1, \cdots, x_6$. 
Both $\BB_6$ and its image, the cubic surface
$S= \im  (\BB_6) \subset \CC \PP^3$,
 are isomorphic.
 There are 27 rational curves on $\BB_6$ of self-intersection $-1$; the image
 of each is a line in $S$. They correspond to 2-form field strengths built out of the 
 M-theory fields.
 
 \vspace{-3mm}
 \begin{enumerate}[{\bf (i)}]
 \setlength\itemsep{-2pt}
\item The exceptional divisors $\cE_1, \cdots \cE_6$, corresponding to the 
   classes of the circles/line bundles given by the (Chern) 
   2-forms $\omega_1, \cdots, \omega_6$. There are 6 of these.

\item A line passing through the points $x_i$, $x_j$ has class $\cH - \cE_i - \cE_j$.
The self-intersection is  $(\cH - \cE_i - \cE_j)^2=1-2=-1$. There are 
$\binom{6}{2}=15$ such lines through two of the blown-up points. 

These correspond to the 2-form fields  $s_j s_i G_4$. 

\item A conic passing through the five points $x_1, \cdots,  \widehat{x}_i, \cdots, x_6$ is a rational 
curve of class $2\cH - \cE_1 -   \cdots -  \widehat{\cE}_i \cdots -\cE_6$. The self-intersection is 
$(2\cH - \cE_1 -  \cdots -  \widehat{\cE}_i \cdots -\cE_6)^2=4-5=-1$.

These correspond to the 2-form fields $s_6 \cdots \widehat{s}_i \cdots s_1 G_7$. 

There are $6=\binom{6}{5}$ of those as the omitted index $i$ runs from 1 to 6
(conics through five of the blown-up points). 
\end{enumerate}

\vspace{-2mm} 
\noindent Overall, we have the precise matching: 

\vspace{-1mm} 
\begin{center} 
\begin{tabular}{cccc}
\hline 
{\bf Configuration} & {\bf Number} & {\bf Class} & {\bf Field}  
\\
\hline 
\hline 
\rowcolor{lightgray}  Exceptional curves & 6 & $\cE_i$ & $\omega_i$ 
\\
Lines & 15 & $\cH - \cE_i - \cE_j$ & $s_js_iG_4$
\\
\rowcolor{lightgray}  Conics & 6&  $2\cH - \cE_{i_1} -   \cdots  -\cE_{i_5}$  &
$s_{i_5} \cdots s_{i_1}G_7$ 
\\
\hline 
\end{tabular} 
 \end{center} 
Thus, we have a collection of distinguished 27 spacetime fields of degree 2 at our disposal. Depending on the geometry and topology of the reduced, 5d spacetime $Y/T^6$, these fields may not be necessarily independent, even linearly. However, at the universal level, that is to say, for the Sullivan minimal model $M(\mc{L}_c^6 S^4)$, we know that the space of generators of degree 2 is 27-dimensional and generated by the 27 elements $w_i$, $s_j s_i g_4$, and $s_{i_5} \dots s_{i_1} g_7$, as in the table above. 

\paragraph{Physical and topological interpretation of further statements.} 
We consider the following (the non-italicized text physically/topologically interprets the italicized text from algebraic geometry):

\vspace{-3mm} 
\begin{enumerate}[{\bf (i)}]
 \setlength\itemsep{-2pt}
\item 
{\it There are exactly 10 lines in $X$ meeting any given one, and exactly 5 lines in $X$ meeting any 
two disjoint given ones.} (see \cite[Rem. 11.6(b), 11.9]{Gathmann}).
We interpret ``meet" to mean ``has at most one component/tensor index in common with", 
in the sense of form components.

\item {\it Every exceptional curve
intersects the 5 lines and the 5 conics that pass through this blown-up point.}

\begin{sloppypar}
Fix $i$; then $s_i$ gives rise to five fields $s_j s_i G_4$ or $s_i s_j G_4$, as $j$ runs from 1 to 5. This 
also gives rise to five fields $s_1 \cdots \widehat{s}_j \cdots s_i  \cdots s_6 G_7$.
\end{sloppypar}

On the topological side, we also have similar the five generators $s_j s_i g_4$ or $s_i s_j g_4$ of $M(\mc{L}_c^6 S^4)$ for a fixed $i$, as well as the five generators $s_1 \cdots \widehat{s}_j \cdots s_i  \cdots s_6 g_7$. This pattern of universal fields repeating spacetime fields persists in (iii) and (iv) below.

\item  {\it Every line through two of the blown-up points meets}


{\bf  (a)} {\it the 2 exceptional curves of the blown-up points;} 
 
This means, say, taking 1 and 2 as indices for the blown-up points, the line through them corresponds to
the twice-reduced field $s_2 s_1 G_4$, while the exceptional curves
correspond to $\omega_1$ and $\omega_2$.

{\bf   (b)} {\it the $\binom{4}{2}=6$ lines through two of the four remaining points;} 
 
These correspond to the six fields: $s_4s_3 G_4$, \,
  $s_5s_3G_4$, \,
   $s_6s_3G_4$, \,
    $s_5 s_4G_4$, \,
     $s_6s_4 G_4$, \,
     and $s_6s_5 G_4$.
 
{\bf  (c)} {\it the 2 conics through the four remaining points and one of the blown-up points.} 
 
 These correspond to the two fields: 
 $s_6 s_5 s_4 s_3 s_1 G_7$ and 
 $s_6 s_5 s_4 s_3 s_2 G_7$.

\item {\it Every conic through five of the blown-up points meets the 5 exceptional 
 curves $\cE_1$, $\cE_3$, $\cE_4$, $\cE_5$, $\cE_6$ at these points, as well as the 5 lines through one of these five points and the remaining point.}

Take the first blowup point to be $x_1$. Then
$s_6 s_5 s_4 s_3 s_1 G_7$ has one index in common with each of 
$\omega_1$, $\omega_3$, $\omega_4$, $\omega_5$, $\omega_6$
and with each of
$s_2s_1G_4$, 
$s_3s_2G_4$, 
$s_4s_2G_4$, 
$s_5s_2G_4$, 
$s_6s_2G_4$. 
Similarly for the other cases. 
\end{enumerate} 

\paragraph{27 Lines in the cyclic loop space $\mc{L}_c^6 S^4$.}
Given that the data $\left(\h_6^*, \{\epsilon_0, \dots, \epsilon_6\}, (-,-), K_6^*\right)$ arising in \Cref{rootdata} is exactly the same as that for the del Pezzo surface $\BB_6$, we can identify the 27 ``lines'' in $\mc{L}_c^6 S^4$. These lines are generated by the $\RR$-homotopy classes of 27 maps $\CC\PP^1 \to \mc{L}_c^6 S^4$ (to be more precise, linear combinations of such, i.e., 27 elements of $\pi_2^\RR (\mc{L}_c^6 S^4) = \pi_2 (\mc{L}_c^6 S^4) \otimes \RR$) supplied by the following result \cite{SV1}:

\begin{quote}
\it 
The 27 exceptional vectors $\alpha \in (\h_6^\ZZ)^*$, $(\alpha, \alpha) = (\alpha, K_6) = -1$, give rise to 27 canonically defined lines in the $\RR$-vector space $\pi_2^\RR (\mc{L}_c^6 S^4)$. Moreover, these lines freely generate $\pi_2^\RR (\mc{L}_c^6 S^4)$ and thus $\dim \pi_2^\RR (\mc{L}_c^6 S^4) = 27$.
\end{quote} 

Since 27 is also remarkable as the dimension of the standard representation of the Lie algebra $\mathfr{e}_6$, this theorem is suggestive of an action (of a real form) of
it 
on $\pi_2^\RR (\mc{L}_c^6 S^4)$ and possibly on all $\pi_\bullet^\RR (\mc{L}_c^6 S^4)$. Does the corresponding Lie group act by rational homotopy self-equivalences over $\RR$ on $\mc{L}_c^6 S^4$? It would be interesting to find out. Of course, the question is equivalent to asking if there is a Lie-algebra homomorphism from a form of $\mathfr{e}_6$ to $\on{Lie}(\Aut M (\mc{L}_c^6 S^4))$.

\subsection{28 Bitangents}
\label{Sec-bitan}

A similar count of ``lines in $\mc{L}_c^6 S^4$'' works for $0 \le k < 6$, with the numbers of lines given by the ``Number of lines on $\BB_k$'' in a table 
earlier in the previous  subsection. This count starts to break for $k > 6$, because some of the exceptional weights will start having trivial weight spaces. 
This is reflected in the irregularity of the form of spacetime fields in the last two tables of \Cref{Sec-excvec}.

\subsection*{28 Bitangents on a quartic etc.}

\paragraph{Seven points and fields in four dimensions.} 
Let $\pi: \BB_7 \to \CC \PP^2$ be the blowup of seven points $x_1, \cdots, x_7$ in general position. 
There are 56 lines paired into 28 pairs corresponding to 
28 bitangents of the ramification quartic curve $C$. 
The 28 pairs of lines are the proper inverse transforms of the isolated pairs 
of curves, split into:

\vspace{-2mm} 
\begin{enumerate}[{\bf (i)}]  
\setlength\itemsep{-2pt}
\item {\it 21 pairs}: (a line through $x_i, x_j$;  the conic through the complementary five points);

\item {\it 7 pairs}: $\left( \text{a cubic with a double point at $x_i$ passing 
through the other points;
the exceptional curve $\pi^{-1}(x_i)$}\right)$. 
\end{enumerate}   


\vspace{-1mm}
\noindent 
Our interpretation of the above is as follows: 
The first 21 pairs of lines  correspond to the 21 pairs of spacetime 2-form field strengths 
$s_j s_i G_4$ and $s_7\cdots \widehat{s}_j \cdots \widehat{s}_i \linebreak[0]  \cdots s_1 G_7$, respectively. 
Similarly, we have 21 pairs of
universal fields, the elements $s_js_i g_4$ and $s_7\cdots \widehat{s}_j \cdots \widehat{s}_i \cdots s_1 g_7$ of $M(\mc{L}_c^7 S^4)$.
Recalling that cubic means having the class $3H$ with weight $3\epsilon_0$, and 
a double point means having $s_is_i$, for that direction $i$, 
the second group of 7 pairs of lines corresponds to the pairs of 2-form field strengths
$s_7 \dots \widehat{s}_i \dots s_1  (s_i C_3 \wedge s_i G_7)$ and $\omega_i$. 
However, only the second components of these pairs are 
represented by universal fields, $w_i$'s. 

\smallskip 
Since the degree of the field strengths is half of the spacetime dimension
$(2=4/2)$, there is a duality symmetry that interchanges the EOMs and the 
Bianchi identities and which give rise to an enlarged symmetry group 
under which the 2-forms and their duals are treated on an equal footing
(as classically in \cite{LPS96}).

\paragraph{Formulation in terms of dilaton vectors and weights.} 
The dilaton vectors are constants that characterize the couplings of the dilatonic 
scalars $\vec{\phi}$ to the various gauge fields; see \cite{LP96}\cite{LPS96}. 
The following are obtained by the dimensional reduction of the kinetic term. 
While we have mostly not adopted that perspective here, the end results are reassuring.  

\vspace{-2mm} 
\begin{enumerate}[{\bf (i)}]
 \setlength\itemsep{-6pt}
\item 
{\it $E_6$ and reduction on $T^6$}:
The 27 dilaton vectors are the weights of the 27-dimensional fundamental 
representation of $E_6$. The Weyl group $W(E_6)$ is isomorphic to the group of 27 lines on a cubic surface, 
i.e., to the subgroup of the permutation group $\Sigma_{27}$ which preserves the 
incidence relation between the lines (see \cite{Dol1}).
Our interpretation of the lines
is as follows:

\vspace{-2mm} 
\begin{enumerate}[{\bf (a)}]
 \setlength\itemsep{-2pt}
\item 
 6 exceptional divisors correspond to the dilaton vectors $\vec{b}_i$
associated to the vierbein;

\item 15 lines connecting two points corresponding to the dilaton vectors 
$\vec{a}_{ij}$;

\item 6 conics through 5 of the 6 points correspond to $\vec{a}_i^*=-\vec{a}_i$, the dilaton
vectors of the dualized 3-forms. 
\end{enumerate} 
 
\item {\it $E_7$ and reduction on $T^7$}: Similarly, the 28 2-form field strengths 
correspond to the dilaton vectors $\vec{a}_{ij}$ (of which there are 21) and 
$\vec{b}_i$ (of which there are 7). 
Together with their negatives, these correspond to the smallest representation of $E_7$,
which, under the decomposition ${\rm SL}(8, \RR) \hookrightarrow E_7$, is
${\bf 56}\simeq {\bf 28} \oplus {\bf 28}^*\simeq \Lambda^2 \RR^8 \oplus \Lambda^2 (\RR^8)^*$.

There are 63 1-form field strengths corresponding to 35 dilaton vectors 
$\vec{a}_{ijk}$, 21 dilaton vectors $\vec{b}_{ij}$, and
7 dilaton vector $\vec{a}_i^*=-\vec{a}_i$, $i=1, \cdots, 7$. Together with their
negatives, these give rise to 126 1-forms, corresponding  to the nonzero weights of 
the adjoint representation of $E_7$.
\end{enumerate}

\vspace{-1mm} 
\noindent We view this as a testament to the power of dualization of physical fields in spacetime. 
It would be interesting to see this aspect universally at the level of cyclifications, as for $E_6$ and the 27 lines above, but 
we leave this treatment for elsewhere.

\subsection{Higher and exotic states} 

Here we first provide interpretation via fields and hence via universal models 
(but we choose to write explicitly in terms of the former) of certain special, including 
the exotic, states 
appearing in Mysterious Duality. We will use 
the interpretation in \cref{subsec-EMdual} of sums in the Picard group of $\BB_k$ as
corresponding to (wedge) products of fields to provide an additive  
factorization of terms in the classes.

\paragraph{Fields corresponding to (exotic) states in $d=8$ spacetime.} 
The elements
\begin{equation}
\label{aHbiE}
\alpha=a\cH - \sum_{i=1}^k b_i \cE_i
\end{equation}
of the cohomology lattice $H^2( \BB_k; \ZZ)$
with relatively high values of $a$ and/or the $b_i$ do not easily lend themselves
to interpretation. Here we interpret these (exotic) states as corresponding 
to decomposable fields, demonstrating the utility of factorization. 
We illustrate this for several entries from the Table on p. 24 of \cite{INV}: 

\vspace{-2mm} 
\begin{enumerate}[{\bf (1)}]
\setlength\itemsep{-2pt}
\item 
The state $3 \cH-2\cE_1 -\cE_2 - \cE_3$ (or $3H-2E_M -E_i -E_j$ in the notation of \cite{INV}) 
is represented by the wedge product of fields 
 $s_2 s_1 G_7 \wedge s_3 s_1 G_4$,
which, in a more familiar notation, would be $F_5 \wedge F_2$. 

\item The state $3\cH-2\cE_1 -\cE_i$, $i = 2, 3$, is represented by the wedge product of fields
 $s_is_1G_7 \wedge s_1 G_4$,
which, in a more familiar notation, would be $F_5 \wedge H_3$. 

\item  The state $4\cH-2\cE_1 - 2\cE_2 - 2\cE_3$ is represented by the wedge product of fields 
 $s_3s_2s_1G_7 \wedge s_3 s_2 s_1 G_7$,
which in more familiar notation would be $F_4 \wedge F_4$ in 8 dimensions. 
This state was not identified in \cite{INV} in terms of branes. 
In our perspective, using factorization, interpretation via fields, and then 
field/brane correspondence, we see that 
it corresponds to a configuration of two M5-branes wrapped on $T^3$. 

\item  The state $3\cH-2\cE_1$ is represented by the wedge product of fields 
 $s_1 G_7 \wedge s_1 C_3$ or $s_1 C_6 \wedge s_1 G_4$,
which in more familiar spacetime notation would be $F_6 \wedge B_2$ or $C_5 \wedge H_3$, respectively. 
On-shell, these are locally equivalent to $F_8$ and indeed corresponds to the D6-brane. 
\end{enumerate}

\paragraph{Higher classes.} 
We consider a higher generalization of the anticanonical class $-\cK_k$ from 
\eqref{k-canclass} to more general coefficients
(specializing \eqref{aHbiE}):
\[
a\cH - b \sum_{i=1}^k \cE_i\;.
\]
Such a class 
is known to be ample for $a \gg b$ (see \cite{Kuchle}\cite{Xu95}). 
Nagata's conjecture asserts that 
essentially $a/b=\sqrt{k}$ \cite{Nagata59}.
It would be interesting to find a physical interpretation of these statements. For instance, 
recall that we have interpreted $a\cH$ to correspond to the $a$-fold 
wedge product of $G_4$ with itself (i.e., $a$ times) and $b\cE_i$ to correspond to 
contraction with respect to the $i$th circle direction $b$ times. 
This places a limit on $a$ and $b$ coming from the dimension of spacetime. 
For higher values of $a$ and/or $b$, this will take us outside the conventional
dimensions of M-theory. This is somewhat analogous to the situation 
in the $E_{11}$ approach, where the interpretation of higher degree fields
or representations is generally desirable, yet challenging.

\paragraph{Reduction and Chern classes with multiplicities.}
If $x_1, \dots, x_k$ are the points in $\CC \PP^2$ blown up to the exceptional divisors $\cE_1, \dots, \cE_k$, then, for $a$ large enough compared to the $b_i$'s, the class
\eqref{aHbiE} may be represented by the proper transform of a curve of degree $a$ in $\CC \PP^2$ passing through each point $x_i$ with multiplicity $b_i$.
Accordingly, we interpret classes in \eqref{aHbiE} with
an exceptional 
class $\cE_i$ entering with multiplicity as
corresponding to taking multiple
Kaluza-Klein reductions along the same circles.
From the point of view of M-theory
this means that we have identification of fibers in the reduction.
This explains the corresponding exotic states in the table on page 24 in
\cite{INV}, as we did in the discussion right after \eqref{aHbiE}.

\medskip 
Note that in \cref{Sec-bitan} above we have been able to allow for multiple occurrences of the contraction by 
the circle by utilizing the compositeness of the fields corresponding to the canonical classes.

\medskip 
Note also that as before, for $k \leq 8$, the only exceptional curves
without multiplicities in $\cE_i$'s
are $\cE_i$, $\cH - \cE_i -\cE_j$, and $2\cH - \cE_{i_1} - \cdots - \cE_{i_5}$.
Also, interestingly, these arise as the 27 lines in \cref{Sec-27} for $k = 6$, and this is the highest $k$ for which 
there are no multiplicities in $\cE_i$'s needed to describe exceptional
curves (see, for instance, the table on p.\ 50 in \cite{Roc}).

\section{Reduction to 2 and lower dimensions: extension 
to the Kac-Moody setting} 

\subsection{Blowing up $\geq 9$ points, cyclifying $\ge 9$ times, and Kac-Moody algebras}
\label{Sec-KM}

As in the case for $E_8$  in the reduction to 3 dimensions \cite{MarcuSchwarz}, 
the Kac-Moody algebras of type 
$E_9$ and $E_{10}$ have been suggested as symmetries of
supergravity when dimensionally reduced to 2 and 1 dimensions, respectively
\cite{Julia80}\cite{Julia82}\cite{Julia85}.
Indeed, 
reduction to 1 dimension witnesses the emergence of $E_{10}$ (see \cite{N92}\cite{Miz}), 
and further reduction to 0 dimension points to possible emergence of $E_{11}$ (see \cite{Ch04}\cite{FK12} for some speculations).

\medskip 
The Kac-Moody algebra $E_9$ is affine,
 $E_{10}$ is hyperbolic (all of whose proper regular subalgebras are either finite or
affine), while $E_{11}$ is Lorentzian. 
 The theory of real forms of complex Kac-Moody algebras 
(see \cite{KW} \cite{BBMR}\cite{BenM})
generalizes the theory of
real semi-simple Lie algebras
(see \cite{Hel}\cite{Knapp02}\cite{HPS}). 
Extending the case $E_{8(+8)}$, one obtains
the split real Kac-Moody algebras of type $E_{9(+9)}$, $E_{10(+10)}$ 
and $E_{11(+11)}$ 
(see \cite{HPS}); cf. the Introduction and \cref{Subsec-RvsC} for the corresponding 
forms in the Lie case.

\paragraph{Beyond del Pezzo surfaces.}
Recall from \eqref{k-canclass} that the canonical divisor of the blowup $\pi: \BB_k \to \CC \PP^2$ at $k$ ordinary points is 
$
\cK_k= -3 \pi^* \cH + \sum_{i=1}^k \cE_i
$, 
where $\cH \subset \CC \PP^2$ is a hyperplane and $\cE_i$ are the exceptional curves. 
The degree of $-\cK_k$, given by the self-intersection 
$$
(-\cK_k)^2 = \cK_k^2=9-k\,,
$$ 
shows that $-\cK_k$ can be ample only if $0 \leq k \leq 8$. Hence 
for $k \geq 9$, we do not have a del Pezzo, but rather a more general rational surface $\BB_k$.
Let $P=\{ x_1, \cdots, x_k\}$ be a finite set of 
$k$ points
in
the projective plane $\CC \PP^2$ whose blowing up results in a rational surface $\BB_k$. As for lower values of $k$, the geometry of the surface $\BB_k$ depends on the relative position of the points $P$: for example, $\Aut \BB_k$ is trivial if $P$ is generic \cite{Giz80}\cite{Hir88}\cite{Ko88}. For what we do in this paper, the structure of $P$ does not matter, but we would not be surprised if the intricacies of the algebraic geometry might be reflected in the subtler structure of physical fields or rational homotopy theory.

\paragraph{Why $k \le 8$ vs. $k \geq 9$ in the exceptional root data.}
\label{k-le-8}
  The construction of the data $\big(\h_k^*, \{\epsilon_0, \dots, \epsilon_k\}, (-,-), K_k^*\big)$ analogous to 
  \eqref{roots}
  extends beyond $k = 8$ verbatim. However, the identification of the root system for $k > 8$ needs to be treated with care. 
  For $0 \le k \le 8$, the degree $\deg (\mc{L}_c^k S^4) = (K_k, K_k) = (K_k^*, K_k ^*) = 9-k$ of the cyclic loop space is positive, 
  just like the degree of the del Pezzo surface $\BB_k$. From simple linear algebra of Minkowski inner products, we can see that
  $(K_k, K_k) > 0$ implies that the inner product induced on the subspace
  $K_k^\perp = \{ x \in \h_k \; | \; (x, K_k) = 0 \}$ of
  $\h_k$ by the Minkowski inner product $(-,-)$ is negative-definite. (If we switch the sign and use $-(-,-)$, it would be a more 
  familiar positive-definite inner product). This also implies that the root system $R_k$ is finite; see \cite{Man}. For $k \ge 9$, 
  the inner product loses its negative-definiteness, and the root
  system $R_k$ becomes infinite and can be identified as the set of \emph{real roots} of a
  more general Kac-Moody algebra; see the discussion of the $k=9$ case further below. In fact, for $k = 9$, the subspace $K_k^\perp$ 
  is negative semi-definite of nullity 1. For $k > 9$, the orthogonal complement $K_k^\perp$ gets a Minkowski inner product.

\paragraph{Beyond 8-fold  cyclifications of $S^4$.}
When we continue cyclifying the four-sphere $S^4$, that is to say, consider $\mc{L}_c^k S^4$, beyond $k = 8$, its degree in the sense of \eqref{deg-L} loses its positivity, the metric induced on $K_k^\perp$ loses its positive-definiteness, and the corresponding root system becomes infinite.
 
 \paragraph{Beyond Lie groups.}
One can
define the \emph{simple roots} to be, as in \eqref{Eq-vectors0i} with the markings 
\eqref{eq-mark},
\begin{equation}
\label{simple-roots}
\alpha_{0}=\epsilon_0 - \epsilon_1 - \epsilon_2 - \epsilon_3
\qquad \text{and} \qquad 
\alpha_{i}=\epsilon_i - \epsilon_{i+1}\;. 
\end{equation}
Note that $(K_k^*, \alpha_i) =0$ and $(\alpha_i, \alpha_i) =-2$, 
$0 \leq i \leq k-1$. 
 For $k\geq 9$, these elements
generate the root lattice $\mathbb{E}_k$ of the Kac-Moody root 
system, cf.\ \cite{Loo80}\cite{Loo81}\cite{Kac80}\cite{Moody79}.

\paragraph{The Weyl group.} 
The intersection matrix $({\bf \alpha}_i,  {\bf \alpha}_j)$  is equal to 
$\Gamma_k - 2I_k$, where $\Gamma_k$   is the incidence matrix
of the Dynkin graph $E_k$.
In particular, each class ${\bf \alpha}_i$  has self-intersection $-2$ and determines an involutive 
 isometry of $\ZZ^{1,k-1}$ by
$\sigma_i: y \mapsto y + (y \cdot {\bf \alpha}_i) {\bf \alpha}_i$.
Together, the reflections $\sigma_0, \sigma_ 1, \ldots,\sigma_{k-1}$
generate the Weyl group $W_k$ with Dynkin diagram $E_k$,
and the inequality $k \geq 9$ ensures that this Weyl group is infinite. In the algebraic geometric context, it is known that the same group $W_k$ is infinite if and only if 
$\rank \Pic (\BB_k) = k+1 \geq10$  \cite[Lemma 0.5]{Har85}.

\vspace{-1mm} 
\(
\label{Dynkin}
\hspace{-1cm} 
\mbox{\small Dynkin diagram $E_k$}
\qquad
  \scalebox{.9}{$
  \raisebox{-30pt}{\begin{tikzpicture}[scale=.6]
    \foreach \x in {0,...,5}
    \draw[thick,xshift=\x cm] (\x cm,0) circle (2.5 mm);
    \foreach \y in {0, 1,2, 4}
    \draw[thick,xshift=\y cm] (\y cm,0) ++(.25 cm, 0) -- +(14.5 mm,0);
    \foreach \y in {3,4}
    \draw[dotted, thick,xshift=\y cm] (\y cm,0) ++(.3 cm, 0) -- +(14 mm,0);
    \draw[thick] (4 cm, -2 cm) circle (2.5 mm);
    \draw[thick] (4 cm, -3mm) -- +(0, -1.45 cm);
    \node at (0,.8) {$\sigma_1$};
    \node at (2,.8) {$\sigma_2$};
    \node at (4,.8) {$\sigma_3$};
    \node at (6,.8) {$\sigma_4$};
    \node at (8,.8) {$\sigma_{k-2}$};
    \node at (10,.8) {$\sigma_{k-1}$};
    \node at (5,-2) {$\sigma_0$};
  \end{tikzpicture}
  }
  $}
\)

\noindent 
Note that the Dynkin diagram can also be labeled by the roots rather 
than by the Weyl reflections.
As in \eqref{RkIk}, an element of $\mathbb{E}_k$ of inner square $-2$ is called a \emph{root}. 
The lattice is spanned by the (simple) roots $\alpha_0, \dots \alpha_{k-1}$. 
A \emph{real root} is a root which belongs to the $W_k$-orbit of one (or any)
of these simple roots. 

\paragraph{Ten dimensions.} 
Let us 
consider for $k \ge 0$ the \emph{$\RR$-split rank} of $\Aut M(\mc{L}_c^k S^4)$, the dimension of the maximal $\RR$-split torus, described in \Cref{split-rank}:
$$
\rank_\RR \Aut M(\mc{L}_c^k S^4) = k+1\; .
$$
This is also the rank of the Picard group of the complex (del Pezzo for $k \le 8$) surface $\mathbb{B}_k$:
$$
\rank \Pic (\BB_k)=k+1\;.
$$ 
The degree of the cyclification $\mc{L}_c^k S^4$, see \eqref{deg-L}, is equal to the Minkowski inner square
of the ``anticanonical class''
\vspace{-2mm} 
$$
(-K_k , - K_k) =9-k\;.
$$
Thus, the following relation holds:
$$
\rank_\RR \Aut M(\mc{L}_c^k S^4) + \deg \mc{L}_c^k S^4 = 10\;.
$$
This is replicated for surfaces $\BB_k$:
$$
\rank \Pic (\mathbb{B}_k) + \deg \BB_k
=10\;.
$$
We find these as a curious possible reference to the critical dimension of (super)string theory. 

\paragraph{$(-1)$-curves in $\BB_k$ for $k \geq 9$.}
The blowup $\BB_k$ of $\CC \PP^2$ at
$k = 9$ points in (Cremona) general position contains infinitely many $(-1)$-curves
 \cite[Lemma 2.5 \& Thm. 4a]{Nagata61}. This is generalized to include $k > 9$ in \cite[Lemma 2.2]{Totaro}. Here, as before, \emph{$(-1)$-curves} are understood as smooth rational curves of degree 1. They will automatically have self-intersection index $-1$. They are particular case of \emph{$(-1)$-divisors} $D$, which by definition satisfy the relations $D \cdot \cK_k = D \cdot D = -1$, i.e., they represent the elements of $I_9 \subset \Pic (\BB_k)$. It is easy to see that the set $I_9$ is infinite, but finding infinitely many $(-1)$-curves on the surface is a subtler issue, addressed in the papers quoted above.

\paragraph{$(-1)$-lines in $\Aut M(\mc{L}_c^k S^4)$ for $k \ge 9$.}
Similarly, for $k \ge 9$ there are infinitely many $(-1)$-lines in $\Aut M(\mc{L}_c^k S^4)$, defined as the exceptional weights $\alpha \in P(\h_k) = \ZZ \epsilon_0 \oplus \dots \oplus \ZZ \epsilon_k$ such that $(\alpha, \alpha) = (\alpha, K_k) = -1$, as for $k = 6$ as in 
\cref{Sec-27}. Indeed, the Weyl group $W_k$ acts on the weight space $\h_k^*$ of the maximal $\RR$-split torus $T^{k+1}$ of $\Aut M(\mc{L}_c^k S^4)$ by isometries preserving $K_k$ and the weight lattice $P(\h_k)$. Therefore, the Weyl group acts on the exceptional lines. The examples show there are enough exceptional lines to generate the whole vector space $\h_k^*$. This implies that an element of $W_k$ is determined by its action on the exceptional lines.
Then the fact that the group $W_k$ is infinite for $k \ge 9$ implies that there are infinitely many exceptional lines.

\medskip
Recall from \cref{Sec-27} that in the $k=6$ case, the 27 exceptional weight spaces of the action of $\h_k$ on the Quillen model $Q(\mc{L}_c^6 S^4) = \pi_\bullet^\RR (\mc{L}_c^6 S^4)[1]$, i.e., on $\pi_2^\RR (\mc{L}_c^6 S^4)$ were all one-dimensional and presented 27 lines in $\mc{L}_c^6 S^4$.

\medskip 
On the contrary, for $k \ge 9$, the infinitely many exceptional weights have all but finitely many weight spaces equal to zero, both on $Q(\mc{L}_c^k S^4)$ and $M(\mc{L}_c^k S^4)$. This is because the underlying real vector space $\pi_\bullet^\RR (\mc{L}_c^k S^4)$ of the Quillen model and the degree-$2$ subspace of $M(\mc{L}_c^k S^4)$ of the Sullivan model are finite-dimensional for all $k \ge 0$, as follows from the identification of $M(\mc{L}_c^k S^4)$ in \Cref{Sullmm}.

\medskip
We now consider in detail the higher ranks, one case at a time.

\subsection{Blowing up 9 points and cyclifying 9 times} 
\label{sec-blow9}

Compactification of M-theory on a nine-torus $T^9$ to two spacetime dimensions 
gives rise to  a qualitative change in the U-duality group (see \cite[\S 4.12]{OP}):
the invariant vector $\vec{\delta}$ in \eqref{vec-delta} 
corresponding to the dimensionless Newton constant becomes light-like
(null) in the Lorentzian metric, so that the induced metric on the orthogonal complement to $\vec{\delta}$ becomes negative semi-definite and the U-duality group is no longer compact.
Accordingly, the generators $S_i$ and $T_{123}$ no longer span a finite group.
Instead, restricting to rectangular tori and vanishing C-field potential,
they
give rise to the Weyl group of the affine Lie algebra $E_9 = \widehat{E}_8=
E_8^{(1)}$
\cite{EGKR98}.

\paragraph{The  algebra $\mathfrak{e}_9$ of type $E_9$.} 
For $k=9$, the Dynkin diagram \eqref{Dynkin} corresponds to
the affine Lie
group of type $E_9 = \widehat{E}_8 = E_8^{(1)}$. 
 This is consistent with the occurrence of infinitely many conserved currents
 in two spacetime dimensions (see \cite{N87}\cite{NW89}). 
The group $E_9$ is the infinite-dimensional affine group that appears as the global
symmetry of maximal two-dimensional supergravity \cite{Julia81}\cite{Julia82}\cite{Julia85}\cite{Julia86}\cite{N87}\cite{NW89}.
The affine Lie algebra $\mathfrak{e}_9$ is
a centrally extended loop algebra 
$\widehat{\mathfrak{e}}_8$ over $\mathfrak{e}_8$.

\paragraph{The Picard and root lattice.}
The surface\footnote{As indicated earlier, for $k \geq 9$, the spaces $\mathbb{B}_k$ are
no longer del Pezzo, even though at times we continue to use the same notation to highlight 
the pattern, and we hope this does not cause confusion. The symbol is more meant to indicate the number of blowups rather than the nature of the surface. 
}
$\mathbb{B}_9$ obtained by blowing up 9 points in general position
$x_1, \cdots, x_9$ in $\CC \PP^2$ is an elliptic fibration.
 The exceptional curves $\cE_1, \cdots, \cE_9$ 
are sections of this fibration.
  The Picard group ${\rm Pic}(\mathbb{B}_9)\cong H^2(\mathbb{B}_9; \Z)$ 
is a rank 10 lattice with generators $\cH, \cE_1, \cdots, \cE_9$, where 
$\cH$, as before, is the class of the proper transform of the line $\CC \PP^1$ 
in $\mathbb{C} \PP^2$.
As in the $k \le 8$
case, the intersection relations \eqref{HE-inter} are satisfied,
so that 
$H^2(\mathbb{B}_9; \Z)\cong \Z^{1,9}$,
and from \eqref{k-canclass} the anticanonical class is 
$
-\cK_9=3\cH - \cE_1 - \cdots - \cE_9
$, which is fixed by the 
Weyl group
action on ${\rm Pic}(\BB_9)\cong \ZZ^{1,9}$.
The sublattice $\cK_9^\perp$ of the Picard lattice 
 is a root lattice for $E_9 = \widehat{E}_8$
  when $k=9$ (see  \cite{Loo80}\cite{Loo81}\cite{Kac80}\cite{Moody79}\cite{LXZ12}). 

\paragraph{The $9$-fold cyclification $\mc{L}_c^9 S^4$ and root lattice.} The same $E_9$ root system data arises as in the previous paragraph verbatim; 
see details in \Cref{rootdata}. The ``phase transition'' from $k \le 8$ to $k=9$ is summarized in the following table.

\begin{center} 
\setlength{\tabcolsep}{7pt} 
\renewcommand{\arraystretch}{1.2} 
\begin{tabular}{cccccc}
\hline 
\rowcolor{lightgray} \textbf{Case} & {\bf $\deg \mc{L}_c^k S^4$} & {\bf Inner product on $K_k^\perp$} & {\bf $K_k$ and $K_k^\perp$} & {\bf Weyl group $W_k$}
& {\bf Exceptnl vectors for $\mc{L}_c^k S^4$}
\\
\hline 
\hline 
 $k \le 8$ & $>0$ & negative definite & $K_k \not \in K_k^\perp$ & finite & finitely many
\\
\rowcolor{lightgray} $k=9$ & 0  & negative semi-definite & $K_k \in K_k^\perp$ & infinite & infinitely many\\
\hline 
\end{tabular} 
 \end{center} 
Here are some details on the dramatic difference which occurs for the Weyl group $W_9$ as compared to $W_8$.
The Weyl group $W_9$ is isomorphic to the affine Weyl group of type $\mathbb{E}_8$ and fits in the extension
\(
\label{W9fromW8}
0 \to \mathbb{E}_8 \xrightarrow{\;\iota \;} W_9 \to W_8 \to 1\;,
\)
so that we have 
$W_8=W_9/\mathbb{E}_8$.
The lattice $\mathbb{E}_8$ can be identified with 
$\mathbb{E}_9 / \ZZ K_9^* = (K_9^*)^\perp / \ZZ K_9^*$
and the map $\iota$ with
a homomorphism 
$\mathbb{E}_9 / \ZZ K_9^* \to W_9$,
This allows for an effective reduction from an infinite to a finite-dimensional setting.

\paragraph{The affine root system.}
We use the description of an explicit Kac-Moody root lattice of affine
 $E_9 = \widehat{E}_8$ from \cite{LXZ12}, such that all $(-2)$-classes form a real affine root system, 
 while the anticanonical class $-\cK_9$ (or $-K_9^*$, if we are talking about $\mc{L}_c^9 S^4$, rather than $\BB_9$) is the null root which generates the
imaginary roots. 

\vspace{-2mm} 
 \begin{enumerate}[{\bf (i)}]  
\setlength{\itemsep}{-2pt}
\item In contrast to the $k \le 8$ case in \cref{Sec-excvec}, 
 the set of exceptional divisors
$$
I_9=\left\{ l \in H^2(\mathbb{B}_9;\ZZ)\; \vert \; l \cdot \cK_9 =
l \cdot l = -1 \right\}
$$
is now infinite \cite{Demazure}\cite[Proposition 8]{LXZ12}, and so is the set of $(-1)$-curves \cite[Corollary 12]{LXZ12}.
Similarly, the set
\[
I_9=\left\{ \alpha \in P(\h_9) \; \vert \; (\alpha, K^*_9) =
(\alpha, \alpha) = -1 \right\}
\]
of exceptional weights for $\mc{L}_c^9 S^4$ is also infinite, just like for any $k \ge 9$; see the discussion just before this subsection.

\item The root system $R_9$ given by \eqref{RkIk} is no longer the standard Lie-theoretic root system of type $E_9$, but only part of it, called the {\it real roots}:
\begin{align*}
\Delta^{\rm Re}=R_9 & :=\big\{ l \in H^2(\mathbb{B}_9;\ZZ) \; \vert \; l \cdot \cK_9=0, \;  l^2=-2\big\} & & \text{\small in the alg. geometric context}\\
& = \big\{ \alpha \in P(\h_9) \; \vert \; (\alpha, K_9^*) =0, \;  (\alpha, \alpha) =-2 \big\} & & \text{\small in the topological context}.
\end{align*}
Explicitly, the system of simple roots of $R_9$ is given by extending the description of 
$R_8$ in  \eqref{simpleroots-E8} to (see, e.g., \cite[\S 3.2]{Chen15}) 
\vspace{-3mm} 
\begin{align}
\label{simpleroots-E9}
\hspace{-2mm}
\phantom{AA}
& \alpha_0= \cH- \cE_1 - \cE_2 - \cE_3, \;\;
    \alpha_1=
    \cE_1 - \cE_2, \;\;
\alpha_2=\cE_2 - \cE_3, \;\;
\cdots ,\;\; 
\alpha_8=\cE_8 - \cE_9 
\\
\nonumber 
\text{or} \quad 
&\alpha_0= \epsilon_0- \epsilon_1 - \epsilon_2 - \epsilon_3, \quad \;
    \alpha_1=
    \epsilon_1 - \epsilon_2, \;\;\;
\alpha_2=\epsilon_2 - \epsilon_3, \;\;
\cdots ,\;\;\; 
\alpha_8=\epsilon_8 - \epsilon_9\; ,
\end{align}
a system that extends the finite-dimensional case (rank $\leq 8$)
in \eqref{simpleroots-E8}.

\item The imaginary roots are given by nonzero integral multiples of the anticanonical class
\(
\label{Im-roots} 
\Delta^{\rm Im}= \big\{ m(-\cK_9)\,\vert \, m \in \Z 
\mathbin{\vcenter{\hbox{$\scriptscriptstyle\mathrlap{\setminus}{\hspace{.2pt}\setminus}$}}}
\{0\} \big\} = \big\{ m(-K_9^*)\,\vert \, m \in \Z 
\mathbin{\vcenter{\hbox{$\scriptscriptstyle\mathrlap{\setminus}{\hspace{.2pt}\setminus}$}}}
\{0\} \big\}\;,
\)
such that 
$
\Delta=\Delta^{\rm Re} \cup \Delta^{\rm Im} = 
R_9 \cup \big\{ m(-K_9^*)\,\vert \, m \in \Z 
\mathbin{\vcenter{\hbox{$\scriptscriptstyle\mathrlap{\setminus}{\hspace{.2pt}\setminus}$}}}
\{0\} \big\} 
$. See \cite{HL02}\cite{Kac}\cite{LXZ12}.
\end{enumerate}

\paragraph{Infinite number of dualizations and imaginary roots.}
We highlight the following observations:

\vspace{-2mm} 
\begin{enumerate}[{\bf (i)}]  
\setlength{\itemsep}{-2pt}

\item {\it Real roots:} 
To each positive real 
root of the real split affine algebra $E_9$ corresponds a BPS solution of 11d
supergravity  \cite{EHKNT}.

\item {\it Imaginary roots:} The infinitely many fields parametrizing the symmetric space  
$E_9/K(E_9)$, where $K(E_9)$ is the maximal compact subgroup,
are defined on-shell by virtue of an infinite set of duality relations typically formulated in terms 
of a linear system and its expansion through a spectral parameter
\cite{BCIKS}. Given our interpretation of the anti-canonical class in 
\cref{subsec-EMdual},
we observe that this is reminiscent of the infinite multiples 
of $-K_9^*$ in the set of imaginary roots \eqref{Im-roots}.

Hence from our perspective, what we have is an infinite number of dualizations 
associated with the occurrence of $-m K_9^*$ as spanning the imaginary roots. 
\end{enumerate}

\subsection{Cyclifying 10 times and blowing up 10 points} 
\label{sec-blow10}

We now consider the reduction of M-theory on a 10-torus $T^{10}$ down 
to 1 dimension. In \cite{BGH}, brief speculations were given about
the connection to $E_{10}$
and possibly extending Mysterious Duality to this case of $k=10$
via $\tfrac{1}{2}K3$ surface $S$, whose cohomology
lattice $H^2(S; \ZZ)$ is the root lattice $\mathbb{E}_{10}$.

\medskip 
Here we provide a precise and systematic approach. 
Given our perspective on Mysterious Duality via (universal) fields, 
we see that this readily extends, with admissible fields with at most of form degree 1 arising from the reduction of 
$G_4$ or $G_7$ or their combinations corresponding to $-\cK_{10}$ or $-K_{10}^*$
(see the table below). Furthermore, Triality also holds in this case, with 
cyclic loops spaces $\mathcal{L}^k_c S^4$ extending to $k=10$. 

\vspace{-1mm} 
\paragraph{The roots.} We recall the following:
  \vspace{-1mm} 
\begin{enumerate}[{\bf (i)}] 
 \setlength{\itemsep}{-2pt}
\item {\it Real roots:}
The real roots \cite{Kac80} of the root system are the $W_k$-orbits of the simple roots \eqref{simple-roots}. Real roots have an inner square of $-2$. If $k > 4$,
then the simple roots are $W_k$-translates of each 
other so that the real roots comprise a single orbit of $W_k$. 
The familiar instantons such as Kaluza-Klein particles, M2-branes, M5-branes, and
Kaluza-Klein monopoles all correspond to real
roots of $E_{10}$ \cite{BGH}. In fact, 
as we encountered earlier in the Lie algebra case, 
the Weyl group of $E_{10}$ formally acts as U-duality on instantons
\cite{OPR}\cite{OP}. 

   \item {\it Imaginary roots:} 
The physical interpretation of the fields corresponding to the \emph{imaginary roots}, which
are the roots (the weights of the corresponding Kac-Moody algebra) of nonnegative inner square, is explored in \cite{BGH} and proposed 
to be matched with some brane configurations.
Also in \cite{DN05}, the higher derivative quantum corrections to the action
of M-theory are associated to
certain
positive imaginary roots of $E_{10}$.

\end{enumerate} 
We will provide interpretation from our perspective; see after the table
below. 

\vspace{-1mm} 
\paragraph{Cyclifying 10 times and the lattice $\mathbb{E}_{10}$.}
 Let $\mc{L}_c^{10} S^4$ be the 10-fold cyclification of $S^4$. Then, by \Cref{rootdata}, we will have a maximal $\RR$-split torus $T^{11} = \GG_m^{11}$ in $\Aut M (\mc{L}_c^{10} S^4)$, and the dual space $\h_{10}^*$ of its Lie algebra $\h_{10}$ will acquire a basis $\{\epsilon_0, \epsilon_1, \dots, \epsilon_{10}\}$, an inner product $(-,-)$ of signature $(1,10)$, and a ``canonical class'' $K_{10}^*$. The root lattice $\mathbb{E}_{10}$ is generated by the simple roots of \eqref{simple-roots}. This time, when the number $k$ of cyclifications goes up to 10, 
 \begin{equation}
 \label{E10}
 \mathbb{E}_{10} \cong \ZZ^{1,9}
  \end{equation}
 becomes the unique even unimodular lattice of signature $(1,9)$.
 
\vspace{-1mm} 
\paragraph{Blowing up at 10 points}
Let $\BB_{10}$ be the blowup of 10 points on $\CC\PP^2$. Different configurations of the blowup points will result in different geometry, one famous example being the Coble surface \cite{DZ01}\cite{Do-Coble} (see more on it below), but the Picard group and the topological data, including the intersection pairing, will be the same:
\[
\Pic (\BB_{10}) \cong H^2 (\BB_{10}; \ZZ) \cong \ZZ^{1,10},
\]
and the root lattice $\mathbb{E}_{10}$ will be the same as above, \eqref{E10}.
A Coble surface $S_C$, 
a special type-$\BB_{10}$ surface obtained by blowing up $10$ 
 points $x_1, \dots, x_{10}$ in $\CC \PP^2$ with the
property that the 10 blowup points are the ordinary double points on an irreducible curve of degree 6 in $\CC\PP^2$. 
Let $W_{10}(2)$ denote the congruence subgroup of $W_{10}$ 
that acts trivially on the coset $\mathbb{E}_{10}/2\mathbb{E}_{10}$. 
Then every element of $W_{10}(2)$
is realized on $S_C$; cf.\ the case for $\mathbb{E}_6$ in \eqref{eq-qE6}.
Indeed, the image of $\Aut (S_C)$ in
 $W(E_{10})=W_{10}$ is the subgroup
\(
\label{WE102}
W_{10}(2) := \{ g  \in W_{10} \; 
| \; g(x) - x \in 2\mathbb{E}_{10} \; \text{ for all} \;
x \in \mathbb{E}_{10} \}\;.
\)
 This subgroup is normal, and the quotient group is
isomorphic to the finite orthogonal group $\operatorname{O}^+(10, \ZZ_2)$, with the $+$ superscript indicating quadratic form 
of even type (see  \cite[Thm. 2.9]{CD89}).
 This  makes connection to 
the study of orbifolds of  $T^{10}$ in \cite{BGH}, and we offer it as 
explaining and strengthening what was suggested there as a possible 
description via assigning a $\ZZ_2$-grading on the root lattice $\mathbb{E}_{10}$.

We have the following table. 
\begin{center} 
\setlength{\tabcolsep}{7pt} 
\renewcommand{\arraystretch}{1.2} 
\begin{tabular}{cccccc}
\hline 
\rowcolor{lightgray} \textbf{Case} & {\bf $\deg \mc{L}_c^k S^4$} & {\bf Inner product on $K_k^\perp$} & {\bf $K_k$ and $K_k^\perp$} 
& {\bf Weyl group $W_k$} & {\bf Exceptnl vectors for $\mc{L}_c^k S^4$}
\\
\hline 
\hline 
 $k \ge 10$ & $<0$ & Lorentzian & $K_k \not \in K_k^\perp$ & infinite & infinitely many\\
\hline 
\end{tabular} 
 \end{center}

\paragraph{Infinite number of dualizations and imaginary roots.}
We highlight the following:

\vspace{-2mm} 
\begin{enumerate}[{\bf (i)}]  
\setlength{\itemsep}{-2pt}

\item {\it Dual graviton:} Level decompositions of 
$E_{10}$ and $E_{11}$
under the gravity subalgebras $\mathfrak{gl}(k, \mathbb{R})$ (for $k=10,11$)
 contains  a 9-form potential $C_{8, 1}$
with a 10-form field strength $F_{10}$, which is traditionally associated with 
the dual graviton;

\item {\it M8-brane:} 
U-duality considerations point to the existence 
of an M8-brane in 11 dimensions \cite{EGKR98} which also appears
in Mysterious Duality \cite{INV}. 
Since this brane should  be associated with a field of dimension 10, it is tempting to 
ask whether it is related to the above dual graviton or to the field $G_{10}$ 
we have seen to make an appearance in our context (see \cref{Sec-excvec}). 
Recent results \cite{BKPPS} indicate that 
the statement about the dual graviton might not hold. 
However, it could be the analogue  of the MK6-brane, which 
reduces to D6-brane in type IIA string theory, and in this sense it is in two degrees higher. 
All these matters are also related to the parametrized homotopy theory 
perspective on the higher rank fields, as indicated towards the end of the Introduction 
(see \cite{BSS} for details);

\item {\it Infinite duality relations:} 
Physical interpretation
of the higher level generators in the decompositions of 
$E_{10}$ and $E_{11}$
under their respective gravity subalgebras $\mathfrak{gl}(k, \mathbb{R})$
has been pursued for a long time. 
The early suggestions stated that such generators correspond to higher excited
string modes or higher spin fields, while more recent 
evidence points to possible ‘deformations’ (such as gauged supergravities),
or that such generators should somehow be ‘dual’ to the low order generators/degrees of freedom  -- see \cite{RW}\cite{dWNS} (and references therein). 
 Indeed, in the covariant  approach of \cite{We03}, these representations would correspond to ‘dual fields’ 
generalizing the relation between p-form fields and $(D -p -2)$-form fields. 

Hence, as in the $k=9$ case (see towards the end of \cref{Sec-KM})
and with the perspective in \cref{subsec-EMdual}, 
we have the interpretation of an infinite number of dualizations 
associated with the occurrence of $-m K_{10}^*$
among the imaginary roots.

\end{enumerate}

\paragraph{Relating to $\mathbb{E}_8$ via quadratic forms.}
Let $\mathbb{L}_2 = \mathbb{E}_{10}/2\mathbb{E}_{10}
\cong \ZZ_2^{10}$.
Equip this $\ZZ_2$-vector space with
the quadratic form
 $$
 q(x) =\tfrac{1}{2} \, x \cdot x \mod 2\,.
 $$
There are exactly $496 = 2^4(2^5-1)$ 
vectors in the set $q^{-1}(1) = \{ x \in \mathbb{L}_2 \; | \; q (x) = 1\}$ of ``mod-2 roots'' \cite[\S 6.4]{CD12}, and all of them
are represented by actual roots in $\mathbb{E}_{10}$ \cite[Remark 4.7]{CD85}, as in the table below.

\paragraph{Physical/field/universal interpretation of classes.}
In this effectively finite perspective, we
 provide a form-field interpretation of a collection of roots of $\mathbb{E}_{10}$ which uniquely represent the 496 mod-2 roots:

\vspace{3mm} 
{\renewcommand{\arraystretch}{1.3}
\begin{tabular}{ccc}
\hline
\bf Root & \bf Associated Field & \bf Count 
\\
\hline 
\hline 
\rowcolor{lightgray}  $\epsilon_i - \epsilon_j$ 
&  $s_j \omega_i$ & $\binom{10}{2}=45$
\\
$\epsilon_0 - \epsilon_i - \epsilon_j - \epsilon_k$ 
& $s_k s_j s_i G_4$ & $\binom{10}{3}=120$
\\
\rowcolor{lightgray}  $2\epsilon_0 - \epsilon_i - \epsilon_j - \epsilon_k- \epsilon_m - \epsilon_n - \epsilon_r$ 
& $s_r s_n s_m s_k s_j s_i G_7$
& $\binom{10}{4}=210$ 
\\
 $3\epsilon_0 - 2\epsilon_i - \epsilon_j - \epsilon_k- \epsilon_m - \epsilon_n - \epsilon_r -\epsilon_s -\epsilon_t$ 
&
$s_t s_s s_r s_n s_m s_ks_j(s_i C_3 \wedge s_i G_7)$
& $\binom{10}{3}=120$
\\
\rowcolor{lightgray}  $4\epsilon_0 - 3\epsilon_1 - \epsilon_2- \cdots -\epsilon_{10}$ 
&
$s_{10} s_9 s_8 s_7 s_6 s_5 s_4 s_3 s_2 (s_1 C_3 \wedge s_1 G_4 \wedge   s_1 C_6)$
& $1$
\\
\hline
\end{tabular} 
}

\vspace{4mm} 
\noindent Here all the indices
satisfy $ i < j< k < m < n < r < s < t$, except that in the second-to-last line, the index $i$ is the least of the three indices different from the seven indices $j < k < m < n < r < s < t$. Here we also assume $s_l s_i := - s_i s_l$ whenever $i > l$.
Note that, using the interpretation in \cref{subsec-EMdual} of sums in the Picard group as 
corresponding to (wedge) products of fields, we have split
the last two terms as: 
$$
3\epsilon_0 - 2\epsilon_i - \epsilon_j - \epsilon_k- \epsilon_m - \epsilon_n - \epsilon_r -\epsilon_s -\epsilon_t =
(\epsilon_0 - \epsilon_i) + (2\epsilon_0 - \epsilon_i) - (\epsilon_j + \epsilon_k + \epsilon_m + \epsilon_n + \epsilon_r + \epsilon_s + \epsilon_t)
$$ 

\vspace{-2mm} 
\noindent and 
\vspace{-2mm} 
$$
\;\;\;
4\epsilon_0 - 3\epsilon_1 - \epsilon_2 - \cdots -\epsilon_{10}=
(\epsilon_0 - \epsilon_1) + 
(\epsilon_0 - \epsilon_1 ) + 
(2\epsilon_0 - \epsilon_1) -  (\epsilon_2 + \epsilon_3 + \epsilon_4 + \epsilon_5+  \epsilon_6 + \epsilon_7 + \epsilon_8 +\epsilon_9 +\epsilon_{10}) \;.
$$ 

The above can be viewed as a reduction from infinite 
to finite number of fields. 

\subsection{Cyclifying $\ge 11$ times and blowing up $\ge 11$ points}
\label{sec-blow11}

\paragraph{Reduction on $T^{11}$ and $E_{11}$-symmetry.} 
The appearance of an $E_{11}$ symmetry in physics has been originally advocated
to be already in 11 dimensions rather than through dimensional reduction 
(see \cite{We03}\cite{We12}). 
However, the reduction on $T^{11}$ to dimension zero may also lead to possible emergence of an $E_{11}$ symmetry (see \cite{Ch04}\cite{FK12}).

\paragraph{Fields associated to $E_{11}$.} 
Decomposing  the adjoint representation of the Lie algebra $\mathfrak{e}_{11}$ 
 under $\mathfrak{gl}(11)$ gives the representations in which the charges of the M-branes and other BPS states transform.
Levels 0, 1, 2, and 3, correspond, respectively, to the vielbein, the 3-form or M2-brane charge, the 
6-form or M5-brane charge, and an $(8,1)$-tensor or the dual graviton charge, 
the latter treated as part of the gravity sector
\cite{KSW04}\cite{We12}. 
Considering a tensor hierarchy algebra 
 introduces new generators starting at level
three, the first of which is a 9-form, but it is still associated with dual
diffeomorphisms
 \cite{BKPPS}.

\paragraph{The composite fields $G_{10}$ and $G_{11}$?} We have seen indications for the 
existence of fields of higher degree,  corresponding to our interpretation of 
the anticanonical class $-\cK_k$ in \cref{subsec-EMdual}.  
In the perspective corresponding to the Sullivan model, we would have a class,
call it $G_{11}$, that is the wedge product of $G_4$ with $G_7$ or the usual Chern-Simons triple wedge
$CS_{11}$ involving only $C_3$ and $G_4$.
From this
point of view, we stop at $E_{11}$
because of the maximal contractions
$
s_{11} \cdots s_{1}G_{11}
$, giving scalar field strengths in zero dimensions. 
From the  perspective of the Quillen model, 
there is a shift down in degree so that we end up with $G_{!0}$
as a combination in M-theory, so that we stop at $E_{10}$ because 
of the maximal contractions 
$
s_{10} \cdots s_{1}G_{10}
$,
reflecting $(-1)$-forms as potentials.

\paragraph{Beyond $k=11$?} 
For $k \ge 11$,
$\mathbb{E}_k$
is the unique odd unimodular lattice of signature $(1, k-1)$.
Without any further conditions,  this works for every $k \ge 11$. 
This shows that mathematically one can continue with the patterns, through 
$k \ge 11$, so that the duality between algebraic geometry and algebraic
topology in our Triality in Figure \ref{MT} 
seems to persist.  Now extending to the physics node takes us outside 
the traditional M-theory range. However, higher degree fields do show up in the 
infinite-dimensional setting in this 
context, and one would need to view spacetime in an extended sense. 
On the side of cyclifications of $S^4$ and algebraic surfaces, there are some dramatic changes happening at the borderline $k = 11$:

\vspace{-1mm} 
\begin{enumerate}[{\bf (i)}]  
\setlength{\itemsep}{-2pt}

\item
For $k \ge 11$, the Weyl group $W_k \subset \operatorname{O}(K_k^\perp)$ has infinite index in $\operatorname{O}(K_k^\perp)$; see \cite{CD12} in the algebraic geometric context.

\item The rank of the Picard group for a rational surface whose automorphism group is large in a certain sense must be 10 or 11.
An arbitrary projective surface with a large automorphism group, which contains a sufficiently large nonabelian part, must have the Picard number at most 11 \cite[Corollary 7.3]{CD12}.

\end{enumerate} 

\vspace{-1mm} 
\noindent It is remarkable that there seems to be a correlation on the limit of 
dimension as well, in the sense that this seems to terminate around the dimension of M-theory.
This is perhaps analogous to the situation of
supersymmetry, where one 
has a natural transition or upper limit of 11 dimensions, 
 beyond which one can extend but at the expense of some desirable properties (there:
super-Poincar\'e symmetry) being relaxed or lost.


\noindent  Hisham Sati, {\it Mathematics, Division of Science, and 
\\
\indent Center for Quantum and Topological Systems (CQTS),  NYUAD Research Institute, 
\\
\indent New York University Abu Dhabi, UAE.}
\\
{\tt hsati@nyu.edu}
\\
\\
\noindent  Alexander A. Voronov, {\it School of Mathematics, University of Minnesota, Minneapolis, MN 55455, USA, and
\\
\indent Kavli IPMU (WPI), UTIAS, University of Tokyo, Kashiwa, Chiba 277-8583, Japan.}
\\
{\tt voronov@umn.edu}

\end{document}